\documentclass[11pt,a4paper]{article}
\usepackage{jstyle}
\usepackage{slashed,framed}
\usepackage{empheq}
\usepackage[bbgreekl]{mathbbol}
\usepackage{caption}
\usepackage{subcaption}
\usepackage{multirow}
\usetikzlibrary{calc,decorations.markings}

\def\cf{{\mathcal F}}

\def\cfhat{\hat{\cf}}
\def\cfmus{{\mathcal F}_{\mu_1\ldots\mu_s}}
\def\cfhmus{{\cfhat}_{\mu_1\ldots\mu_s}}
\def\gsym[#1]{g_{(\mu_1\mu_2}{{#1}_{\mu_3\ldots\mu_s)}}}
\def\ZZ{\mathcal{Z}}
\author[a]{Jin-Beom BAE}
\author[b]{\quad Euihun JOUNG}
\author[c]{\quad Shailesh LAL}

\affiliation[a]{Korea Institute for Advanced Study, 85 Hoegiro, Dongdaemun-Gu, Seoul 02455, Korea}
\affiliation[b]{Department of Physics and Research Institute of Basic Science, Kyung Hee University, Seoul 02447, Korea}
\affiliation[c]{LPTHE -- UMR 7589, UPMC Paris 06, Sorbonne Universit{\'e}s,  Paris 75005, France}

\emailAdd{jinbeom@kias.re.kr}
\emailAdd{euihun.joung@khu.ac.kr}
\emailAdd{shailesh@lpthe.jussieu.fr}

\begin{document}

\title{\centering 
Exploring Free Matrix CFT Holographies \\ at One-Loop}

\abstract{We extend our recent study on the duality between 
stringy higher spin theories and free CFTs in the $SU(N)$ adjoint representation
to other matrix models namely the free $SO(N)$ and $Sp(N)$ adjoint models
as well as the free $U(N)\times U(M)$ bi-fundamental  and $O(N)\times O(M)$ bi-vector models. 
After determining the spectrum of the theories in the planar limit by Polya counting, we compute
the one loop vacuum energy and Casimir energy for their respective bulk duals by means of the CIRZ method 
that we have introduced recently.
We also elaborate on possible ambiguities in the application of this method.
}


\maketitle

\section{Introduction}

The AdS/CFT duality \cite{Maldacena:1997re} is a remarkable conjecture proposing the equivalence between a quantum gravity in Anti de Sitter (AdS) space and a conformal field theory (CFT) defined on the boundary of the same AdS space (see \cite{Aharony:1999ti} for a review).

These dualities were observed in string theory, building on the observation \cite{Polchinski:1995mt} that the D-branes of string theory and the black branes of supergravity are essentially complementary descriptions of the same system, being valid respectively at weak and strong string coupling.
The AdS theory is the closed string theory
--- a theory of quantum gravity --- that the black branes are embedded in, and the CFT is the field theory which describes the low energy dynamics of the world volume of the D-branes.
We therefore expect that the AdS/CFT dualities would share two very common features. 
First, fields in CFT should be matrix valued because of the CFT is the low energy effective theory of a stack of D-branes. Second, since closed string theories contain supergravity as a low-energy limit, there should be a regime in the parameter space of the duality where the AdS theory is described well by supergravity. 
It turns out that in the field theory this corresponds to taking the strongly coupled limit.
We also observe that supersymmetry is almost ubiquitous in these dualities, and is an important ingredient for ensuring that the weak and strong string coupling descriptions can be extrapolated to each other to lead to the duality.

It is interesting to contrast this situation with the case of AdS/CFT dualities involving higher-spin theories and  vector model CFTs \cite{Klebanov:2002ja,Gaberdiel:2010pz}. These dualities are counterexamples to the above expectations in almost every way.
Firstly, they are non-supersymmetric, or at least there is no apparent benefit in working with their supersymmetric extensions.
Secondly, the CFT is a typically a  vector model rather than a matrix model, this leads to important simplifications in the spectrum of the bulk and boundary theories and may also have important implications on black hole physics in these theories \cite{Shenker:2011zf}.
Thirdly, there is no obvious point in the parameter space of the duality where we obtain bulk GR.

Nonetheless the study of these dualities might have important insights into the physics of `stringy' AdS/CFT dualities.
This turns out to be the case from two \textit{a priori} distinct motivations. Firstly, while it is clearly desirable to use AdS/CFT dualities to probe bulk quantum gravity using the dual CFT, in practice it is somewhat more difficult as the dynamics of a CFT at a generic coupling is extremely complicated in itself. 
From this point of view, it is natural to consider AdS/CFT dualities involving \emph{free} CFTs to examine how the CFT repackages itself into a theory of quantum gravity\footnote{We consider `free' CFTs as being obtained from a zero `t-Hooft coupling limit of the large-$N$ expansion of a given CFT. Hence the bulk theory still admits a semi-classical expansion, identified to the 't-Hooft expansion of the dual CFT, and single trace conformal primaries in the CFT correspond to fields in the bulk.}. 
Secondly, taking the free `t-Hooft coupling limit on the CFT side corresponds to a particularly interesting limit on the AdS side as well \cite{Sundborg:1999ue,wittenHS}.

For definiteness, let us consider the case of the duality between Type IIB superstrings on AdS$_5\times S^5$ and $\cN=4$ super Yang-Mills, where the dictionary between the bulk and boundary parameters reads as
\begin{equation}
N^2 \sim \left(\ell_{AdS}\over\ell_{P}\right)^8,
\qquad \lambda\sim{\ell_{AdS}^4\over \left(\alpha'\right)^2}\,.
\end{equation}
This dictionary indicates, as is familiar, that taking the planar limit of $\cN=4$ super Yang-Mills corresponds to taking the classical limit in the bulk, where the radii of AdS$_5$ and $S^5$ are much larger than the Planck length. 
Further, now setting $\lambda$ to zero corresponds to setting the string tension $\alpha'^{-1}$ to zero or, equivalently, taking the string length to be much larger than the AdS$_5$ radius. 
In either way of thinking about this limit in the bulk, it should be clear that this is a very stringy limit as it corresponds to working at an energy scale much larger than the string tension, at which point the string no longer looks like a point object as it would to a low energy observer, which is essentially the supergravity approximation, corresponding to taking $\lambda$ to infinity.

Moreover, the tensionless limit is a window of string theory about which much remains to be understood, however there are important hints that new symmetries should manifest themselves in this phase \cite{Gross:1988ue,Gross:1988ib,Beisert:2003te,Bianchi:2003wx,Beisert:2004di,Bianchi:2005ze} and indeed that higher-spin symmetry may be one such symmetry \cite{Bianchi:2003wx,Beisert:2004di,Bianchi:2005ze}. 
It is therefore natural to explore this window of AdS/CFT duality both for gaining a foothold into tensionless string theory and also for a more general program of extracting bulk physics from CFT data.

The approach we adopt in this paper is to assume that a CFT with an 't-Hooft expansion admits an AdS dual in the planar limit, and then compute $1/N$ corrections in the duality. 
This approach also provides an interesting point of view into a different but related question. 
In particular, how does one couple massive representations of the higher-spin algebra to the Vasiliev system? 
Though the coupling of massive and massless higher-spin fields in AdS has been studied at the cubic level in \cite{Joung:2012fv,Joung:2012rv,Joung:2012hz}, directly constructing the bulk theory is still quite difficult. 
However, since the single-trace operator spectrum of a free matrix model CFT contains the conserved currents found in the vector model along with conformal primaries lying \textit{above} the unitarity bound, we expect its AdS dual to be a theory of massless higher spins coupled to massive higher spins. 
Further, by varying the content of the CFT, the operator spectrum can be quite easily varied.
Hence this setting is expectedly useful for generating a zoo of theories with massless and massive higher-spins coupled to each other in AdS.

As a preliminary exploration of 
the duality between tensionless strings in AdS and free matrix model CFTs, it is particularly appealing to focus on one-loop quantum effects in the bulk, especially the 
vacuum energy in AdS with sphere boundary
and thermal AdS with torus boundary.
We shall refer these two quantities as 
one-loop vacuum energy and Casimir energy.
The spectral problem for arbitrary spin tensor (and spinor) fields has been almost completely solved for Laplacians on hyperboloids \cite{Camporesi:1990wm,Camporesi:1993mz,Camporesi,Camporesi:1994ga},
and this provides the vacuum energy
of the corresponding particle.
The full result is expectedly determined by summing over contributions from every particles in the spectrum of the bulk theory. 
This was very explicitly carried out for the 
higher spin theories in \cite{Giombi:2012ms,Giombi:2013fka,Giombi:2014iua,Giombi:2014yra,Beccaria:2014jxa,Beccaria:2014xda,Beccaria:2014zma,Beccaria:2014qea,Giombi:2016pvg,Gunaydin:2016amv,Pang:2016ofv,Skvortsov:2017ldz} and the resulting computation matched. We refer the reader to \cite{Giombi:2016ejx} for a review higher spin holography in general including the one-loop computations mentioned here.

It is clearly of interest to explore how these computations can be extended to the case of the 
tensionless string, but there is an obvious complication. 
The higher-spin computations of one-loop free energies rely on an explicit knowledge of the bulk spectrum. 
Further, summing over free energy contributions of each particle leads to naively divergent sums that need to be regulated. 
Meanwhile, an independent formulation of even the classical bulk theory for the tensionless string is lacking. 
Even if we were to attempt to use the CFT data to reconstruct the bulk by identifying the CFT single trace operator spectrum to the spectrum of bulk particles, there is so far no simple closed form expression for the operator spectrum of a matrix valued CFT \cite{Barabanschikov:2005ri,Newton:2008au}. 

For these reasons, an alternate approach was adopted in \cite{Bae:2016rgm,Bae:2016hfy,Bae:2016xmv,Bae:2017spv} which bypasses both these problems by expressing the one-loop vacuum energy of a given field in terms of a linear operator acting on the conformal algebra character corresponding to the field. 
For technical reasons, this was referred to as the \textit{Character Integral Representation of the Zeta Function (CIRZ) method}. 
This method completely reproduces the previous results for Vasiliev's higher spin theory as well as readily extracts the answers for the tensionless string as well as its bosonic cousins, the bulk duals of the free $SU(N)$-adjoint scalar CFT and free $SU(N)$ Yang-Mills. 
In particular, it was found that the one-loop free energies 
of these bulk dual theories are non-zero, and equal to 
minus of the one-loop free energy of 
the corresponding boundary conformal field (scalar, spin-1 etc).
Further, the computations involved undergo simplifications for the \textit{maximally} supersymmetric case which are seemingly quite miraculous \cite{Bae:2017spv}. 

In this paper we shall discuss the extension of these results to the 
free CFTs in the adjoint representation 
of $SO(N)$ and $Sp(N)$,
as well as the bi-fundamental and 
the bi-vector representation
of $U(N)\times U(M)$ and $O(N)\times O(M)$,
respectively.
We concentrate our consideration on the AdS$_5$/CFT$_4$ dualities,
but all our analysis can be generalized to any even $d$ 
in a straightforward manner and
to odd $d$ with a bit more effort (see \cite{Bae:2016rgm}
for AdS$_4$/CFT$_3$ case, and \cite{Skvortsov:2017ldz} for a generalization to arbitrary dimensions.).
Another aim of the current paper is to provide 
a concise summary of the series of our recent works \cite{Bae:2016hfy,Bae:2016rgm,Bae:2016xmv}
and to append more details on the relevant technicalities
such as the spectral analysis of AdS space and 
the operator counting problem.

\subsection*{Organization of Paper}

A brief overview of this paper is as follows. In Section \ref{sec: Casimir} and Section \ref{CIRZ} we shall review the formalism for one-loop computations in AdS$_5$, recollecting the essential results for computing the Casimir Energy and vacuum energy at one-loop. Section \ref{matdual} provides a few more details about the duality between the tensionless string and free matrix models, focusing strongly on a pedagogical treatment of Polya counting, an essential tool for many of the computations presented here. Section \ref{stringyrev} then presents the applications of this formalism to adjoint $Sp(N)$ and $SO(N)$ CFTs, namely free scalar, Yang Mills and $\cN=4$ SYM, and bi-fundamental and bi-vector scalar and fermion models. Finally, some more technical details are reviewed in the Appendices. Appendix \ref{harmonic} contains a review of some facts of harmonic analysis on AdS spaces which are useful to these computations, while Appendix \ref{app: so24rep} reviews key features of unitary representations of $so(2,4)$. Finally, Appendix \ref{hsrev} contains an overview of the applications of the methods of Section \ref{sec: Casimir} and Section \ref{CIRZ} to the higher-spin/CFT dualities.

\section{Casimir Energy in Thermal AdS$_5$}\label{sec: Casimir}

We begin with how the one-loop AdS/CFT Casimir energy may be computed in thermal AdS$_5$. In particular, we will review the observation of \cite{Giombi:2014yra} that `naive' computation of the AdS/CFT Casimir energy for higher-spin theories yields a divergent answer which may be suitably regularized to obtain a result consistent with CFT expectations.  Importantly, 
the latter regularization also does not require us to know the precise spectrum of the theory, except in some implicit way through the thermal partition function of the theory, computed in the \textit{canonical ensemble}. This is discussed below. Equally importantly, the computations here contain the same key idea which is very useful for the analysis presented later, but in a simpler setting. 

We begin with the Vasiliev Type A theory in AdS$_5$. Its duality is discussed at somewhat greater length in Section \ref{hsrev} but for now it is sufficient to note that the \textit{non-minimal} Vasiliev theory contains massless spins from spin equal to $1$ to infinity appearing once each in the spectrum, along with a scalar with $\Delta=2$, and is dual to the $U(N)$ vector model. Further, there is a \textit{minimal} Vasiliev system arrived at by truncating the non-minimal one to even spins only, and this is dual to the $O(N)$ vector model. Next, we note that the Casimir energy of a massless spin-$s$ field in AdS$_5$ is given by \cite{Giombi:2014yra}
\begin{equation}\label{ecs}
    E_c^{(s)} = -{\frac{1}{1440}}\,s\left(s+1\right)\left[18s^2\,\left(s+1\right)^2-14s\left(s+1\right)-11\right].
\end{equation}
While the scalar of the theory is not massless, its Casimir energy can be determined from the formula \eqref{ecs} by setting $s=0$ in it. Therefore the Casimir energy for the \textit{non-minimal} AdS theory is given by
$E_c =  \sum_{s=0}^{\infty} E_c^{(s)}$\,,
which is clearly divergent.
This divergence can be regularized by means of an appropriate zeta function, or by inserting an exponential damping $e^{-\epsilon\left(s+\tfrac12\right)}$ when evaluating the sum and discarding all terms divergent in $\epsilon$ in the limit $\epsilon\rightarrow 0$. We thus obtain \cite{Giombi:2014yra}
\begin{equation}\label{echs}
    E_c =  \sum_{s=0}^{\infty}E_c^{(s)}\,e^{-\epsilon\left(s+\tfrac12\right)}\vert_{\text{finite}}= 0.
\end{equation}
As is apparent from the above analysis, carrying out this computation requires knowledge of the precise spectrum of the theory, along with a prescription for regulating the divergence for summing over the infinite number of fields in the spectrum of the theory. 
This data is unavailable for the bulk duals of matrix CFTs at present. 
We will now show in below how this requirement may be evaded \footnote{Somewhat related arguments are also implicit in some computations of \cite{Giombi:2014yra}. 
In particular, note their computations from Eq. (5.16) to Eq. (5.21) which are essentially a `one-shot' computation of the full bulk Casimir energy from the thermal partition function, much as we present here in \eqref{totecs}.}. 

Our starting point is the relation between the (blind) character
\be
    \chi_{\mathcal{V}}\!\left(\beta\right)=\tr\left(e^{-\beta H}\right)=\sum_{n}d_n\, e^{-\beta\,E_n}, \label{blind ch}
\ee
computed over the UIR $\mathcal{V}$ of $so(2,4)$ and the Casimir energy in AdS of the corresponding field. Here $(E_n,\,d_n)$ are the eigenvalues and degeneracies of the hamiltonian $H$. Given $\chi_{\mathcal{V}}\left(\beta\right)$ we may take its Mellin transform to obtain $\tilde{\chi}_{\mathcal{V}}\left(s\right)$ as
\begin{equation}\label{chi int}
\tilde{\chi}_\mathcal{V}\left(z\right)=\mathcal{L}_{\text{Mellin}}
\left[\chi_{\mathcal{V}};z\right]\equiv \int_0^\infty\,d\beta\,\frac{\beta^{z-1}}{\Gamma\left(z\right)}\,\chi_{\mathcal{V}}\left(\beta\right) = \sum_{n}d_n\,E_n^{-z},
\end{equation}
which implies
\begin{equation}\label{ecb}
\tilde{\chi}_\mathcal{V}\left(-1\right)= \sum_{n}d_n\,E_n = E_c\,,
\end{equation}
where $E_c$ is the Casimir energy \cite{Gibbons:2006ij}. Anticipating future developments, in the above we have defined a linear functional $\mathcal{L}_{\text{Mellin}}$ which acts on the character $\chi_{\mathcal{V}}$ to return the Mellin transform. 
Note that it is straightforward to apply \eqref{ecb} to the case where $\mathcal{V}$ is the short representation $\left(s+2,\tfrac{s}{2},\tfrac{s}{2}\right)$ to obtain the expression \eqref{ecs} for the Casimir energy of a massless spin-$s$ field. 
For fermions, the Casimir energy is defined with an overall minus sign, so we insert the fermion number operator into the character and define a 
partition function $Z_\mathcal{V}\left(\beta\right) = \tr_{\mathcal{V}}\left(\left(-1\right)^F e^{-\beta H}\right)$ in terms of which we obtain
$E_c$ as \eqref{chi int} and \eqref{ecb}.

Now we use the fact that the Hilbert space $\mathcal{H}$ of one-particle excitations of the AdS$_5$ theory decomposes by definition into UIRs of the conformal algebra $so(2,4)$. Further, again by definition
\begin{equation}
Z_\mathcal{H}\left(\beta\right) = \tr_\mathcal{H}\left(\left(-1\right)^F e^{-\beta H}\right) = \sum_{\lbrace b\rbrace}n_b\,\chi_{b}\!\left(\beta\right) - \sum_{\lbrace f\rbrace} n_f \,\chi_{f}\!\left(\beta\right),
\end{equation}
where $\lbrace b\rbrace$ and $\lbrace f\rbrace$ denote respectively the sets of bosonic and fermionic fields in the theory. Then we may use the linearity property of $\mathcal{L}_{\text{Mellin}}$ and act with it on the total partition function $\cZ_\mathcal{H}\left(\beta\right)$ to find the total Casimir energy
\begin{equation}\label{totecs}
E_c = \tilde{Z}_\mathcal{H}\left(-1\right);\quad \text{where} \quad  \tilde{Z}_\mathcal{H}\left(z\right) = \mathcal{L}_{\text{Mellin}}\left[Z_\mathcal{H};z\right].
\end{equation}
It turns out that in all cases of which we are aware, this definition of the Casimir energy perfectly reproduces the expressions found by the regularots such as \eqref{echs} that are used in the literature. Further, often it is possible to evaluate the full partition function $\chi_\mathcal{H}$ without knowing the explicit spectrum of the theory. Indeed matrix CFTs are an example of this possibility, as we review below. Therefore, the definition \eqref{totecs} is particularly useful to apply to the cases of matrix model CFTs which we encounter in `stringy' AdS/CFT dualities.

\begin{figure}[h]
\centering
\begin{tikzpicture}
\draw [help lines,->] (-2,0) -- (3.6,0);
\draw [help lines,->] (0,-1) -- (0,1);
\node at (4.2,0){Re$(\b)$};
\node at (-0.6,1) {Im$(\b)$};
\draw [line width =1.3pt, red] (0,0) to (3.2,0);
\draw [blue,
   decoration={markings,
  mark=at position 0.2 with {\arrow[line width=1.2pt]{>}}
  }
  ,postaction={decorate}]
  plot [smooth, tension=0.3]
  coordinates {(3.2,0.3) (0,0.3) (-0.3,0)};
  \draw [blue, postaction={decorate}]
  plot [smooth, tension=0.3]
  coordinates {(-0.3,0) (0,-0.3)  (3.2,-0.3)};
\end{tikzpicture}
\caption{Integration contour for the zeta function}
\label{fig: ct1}
\end{figure}
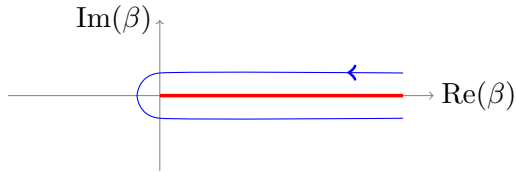

Finally, we also note that \eqref{chi int} may be efficiently evaluated by deforming the contour of $\beta$ integration, which originally stretches along the positive real $\beta$ axis from $0$ to $\infty$, to 
Figure\,\ref{fig: ct1} to get
\be
      \tilde Z_{\cH}(z)=\frac{i}{2\,\sin(\pi\,z)}\oint_C d\b\,\frac{\b^{z-1}}{\G(z)}\,
    Z_{\cH}(\b)\,.
\ee
Further, if the partition function $\chi_{\cH}(\b)$ has no singularities
on the positive axis of $\b$ except for poles at $\b=0$\,,
then the contour $C$ can be shrunk to a small circle around $\b=0$ to give
\begin{equation}\label{casimirexp}
    E_{c} =-\frac12\,\oint_C \frac{d\b}{2\,\pi\,i\,\b^{2}}\,
    Z_{\cH}(\b)\,,
\end{equation}
which may be evaluated by the residue theorem. 
We therefore find that the AdS Casimir energy is simply $-\tfrac12$ of the $\mathcal{O}\left(\b\right)$ term in the Laurent series expansion of
$Z_{\cH}(\b)$ about $\b=0$\,.

\section{Formalism for One-Loop Computations in AdS$_5$}\label{CIRZ}

In this section we review the formalism and techniques for carrying out one-loop computations in AdS$_5$. The techniques are more generally applicable and extend to arbitrary odd-dimensional AdS spaces straightforwardly and even-dimensional AdS spaces with a bit more effort. 

\subsection{Vacuum Energy in AdS$_5$}\label{zeta}

Evaluating the one-loop partition function of a quantum theory reduces to the problem of evaluating functional determinants:
\be
    Z^{\sst (1)}=\int \cD{\Psi}\,e^{-\frac12\la \Psi, \cK\,\Psi\ra}
    =\frac{1}{\sqrt{\det \cK}}\,,
\qquad
 \cK(x_1,x_2)=\frac{\delta^2 S[\Phi]}{\delta\Phi(x_1)\delta\Phi(x_2)}\Big|_{\Phi=\bar\Phi}\,,
\ee 
where $S[\Phi]$ is the classical action and 
$\Phi=\bar\Phi+\Psi$\,.
The $\bar\Phi$ and $\Psi$ are 
a background field  and the fluctuation over it, respectively.
The bracket $\la\cdot,\cdot\ra$ is the scalar product defined by
\be 
    \la \Psi_1, \Psi_2\ra =
    \int d^Dx\sqrt{g}\,\Psi_1(x)^*\cdot\Psi_2(x)\,.
\ee 
Here, we suppressed all the indices for simplicity
but one should understand that the fields
 $\Phi$ or $\Psi$ are tensors in general.
 `$\cdot$' is a Lorentz invariant scalar product that contracts the (suppressed) spin indices of the fields $\Psi$.
 
The one-loop free energy $\G^{\sst (1)}$ 
or the vacuum energy is simply 
\be 
    \G^{\sst (1)}=-\ln Z^{\sst (1)} 
    =\frac12\,\tr\ln\cK\,,
\ee
hence, we need to evaluate the $\tr\ln$ (or functional determinant) of the operator $\cK$\,.
If we treat the operator $\cK$
as if it is a finite dimensional diagonalizable matrix with eigenvalues 
$\k_n$, then we would get
\begin{equation}
    \tr\ln \cK= \sum_{n} d_n\ln \k_n\,,
\end{equation}
where $n$ parametrizes the eigenvalue
and $d_n$ is the degeneracy. 
Defining the zeta function $\zeta\!\left(z\right)$ to be
\begin{equation}\label{zetadef}
    \zeta\!\left(z\right) = \sum_n \frac{d_n}{{\k_n}^z},
\end{equation}
it is easy to see that
\begin{equation}
    \zeta'\!\left(0\right)=-\tr\ln\cK=-2\,\G^{\sst (1)}\,.
\end{equation}

However, expressions such as \eqref{zetadef} are not ideally suited for a direct evaluation in the case of a differential operator $\cK$\,. 
Typically, the naive degeneracy corresponding to a given eigenvalue is infinite. 
We shall therefore use the fact that given an orthonormal set of eigenvectors\footnote{The operators we are interested in will be of the form $-\square + c$ where $c$ is a constant and $-\square = g^{\mu\nu}\nabla_{\mu}\nabla_{\nu}$. 
The spectral problem for operators of this form has been explicitly solved for a wide class of spin fields in AdS space. 
In contrast, if we wish to compute the same determinants over quotients of AdS, in principle we have to impose quantization conditions over $\Psi_{m}^{(n)}$. This may prove easy or difficult depending on the orbifold at hand. Nonetheless, for the quotients we are interested in, it is possible to compute the determinants on the quotient space by the method of images. We review these facts in Appendix.} $\left\lbrace \Psi^{(n)}_m\right\rbrace$ belonging to the eigenvalue $\k_n$, the degeneracy may be defined as
\begin{equation}\label{degdef}
    d_{n} = \sum_{m} \la \Psi^{(n)}_m, \Psi^{(n)}_m\ra\,.
\end{equation}
We emphasize that though \eqref{degdef} is a tautology for compact spaces, for non-compact spaces it is essentially a non-trivial definition. When evaluated explicitly, the answer is still divergent but may be regulated in accordance with general principles of AdS/CFT. 
We shall be applying these methods to the typical kinetic operators 
$\cK=-\square + c$ in AdS$_5$, so it is useful to specialize a little to that case. 
It turns out that the spectrum of eigenvalues is continuous, labeled by a positive real number $u$, and is given by
\begin{equation}
    \kappa_{u} =u^2 + c'\,.
\end{equation}
Here, $c'$ is a constant number, essentially encoded in the parameter $c$ appearing in the bulk kinetic operator. For physical fields it will be related to $\Delta$, the conformal dimension of the dual operator on the boundary. The zeta function for the operator $\mathcal{K}$ is then given by
\begin{equation}\label{zetagen}
    \zeta\!\left(z\right) = \int d u\,\frac{\sum_{m} \la\Psi^{(u)}_m, \Psi^{(u)}_m\ra}{ \left(u^2 + c'\right)^z },
\end{equation}
where the wave functions $\Psi^{(u)}_m$ now obey the orthonormality conditions
\begin{equation}
    \la \Psi^{(u)}_m, \Psi^{(u')}_{m'}\ra = \delta(u-u') \,\delta_{m,m'}\,.
\end{equation}
We now specialize to the case of global AdS$_5$, to develop general expressions useful for the
forthcoming analysis. 
In this case, for a wide class of fields, the eigenfunctions of the Laplace operator $\square = g^{\mu\nu}\nabla_\mu\nabla_\nu$ have been explicitly computed \cite{Camporesi:1990wm,Camporesi,Camporesi:1993mz,Camporesi:1994ga,Camporesi:1995fb}. 
Further, using the homogeneity of AdS, it follows that $\sum_{m} \Psi^{(u)}_m\!\left(x\right)^*\cdot \Psi^{(u)}_m\!\left(x\right)$ is independent of $x\in$ AdS$_5$,\footnote{This statement is a generalization of the addition theorem for spherical harmonics on $S^2$ to general spin fields on symmetric spaces, and is particularly transparent when the group theory underlying harmonic analysis on symmetric spaces is used. These facts are reviewed in Appendix \ref{harmonic}.} and we define
\begin{equation}\label{plancherel}
   \sum_{m} \Psi^{(u)}_m\left(x\right)^*\cdot \Psi^{(u)}_m\left(x\right)=\sum_{m} \Psi^{(u)}_m\left(0\right)^*\cdot \Psi^{(u)}_m\left(0\right) \equiv \, \mu\!\left(u\right).
\end{equation}
Here $x=0$ is a point on AdS$_5$ which may be arbitrarily chosen. In practice, it is chosen so that all but a finite number of eigenfunctions $\Psi^{(u)}_m\!\left(x\right)$ vanish at that point, and the sum over $m$ may be easily evaluated. Finally, we see that \eqref{zetagen} reduces to
\begin{equation}
    \zeta\!\left(z\right) = \text{Vol}_{\text{AdS}_5} \int d u\, {\mu\!\left(u\right)\over \left(u^2 + c'\right)^z }\,,
\end{equation}
where 
$\mu\!\left(u\right)$ plays 
the role of measure over the parameter $u$ which indexes the eigenvalues of the Laplacian. 
It is known as the Plancherel measure.

For the operator $\cK$ 
corresponding the irreducible representation
$\cD(\Delta,(\ell_1,\ell_2))$\,,
the constant $c'$ is given by
\be
    c'=(\D-2)^2\,,
\ee
and the measure $\mu(u)$ by
\be
    \mu\!\left(u\right) = {1\over 3\pi^2}\, 
    \frac{\ell_1+\ell_2+1}2\,\frac{\ell_1-\ell_2+1}2
	\left(u^2+(\ell_1+1)^2\right)\left(u^2+\ell_2^2\right).
	\label{S fn}
\ee
The derivation of the $c'$ and $\mu(u)$ 
is provided in Appendix \ref{harmonic}.
The factor of the volume $\rm{Vol}_{AdS_5}$ 
is infinity due to the non-compactness nature of AdS space. 
This IR divergence can be also regularized as 
\be
	{\rm Vol}_{AdS_5}=\pi^2\,\log(\mu\,R)\,,
\ee
where $R$ is the raduis of AdS space
and $\mu$ the renormalization scale.
See \cite{Diaz:2007an} for the details and discussions. 
With the above result, suppressing the $\mu$ dependence, 
the AdS$_5$ vacuum energy 
is given always proportional to $\log R$\,.

\subsection{Character Integral Representation of Zeta function}

We have seen in Section \ref{sec: Casimir} that the Casimir energy for a theory in thermal AdS$_5$ is naturally encoded in the thermal partition function, or the blind character in other words, \textit{via} a linear operator acting on it. 
This provides a natural resummation of the Casimir energies of the individual fields in the spectrum. 
In \cite{Bae:2016rgm}, we have shown how the character may be similarly used to resum the one-loop free energies in \textit{global} AdS$_5$.
That is, there exists a linear operator $\mathcal{L}$ which, like $\mathcal{L}_{\text{Mellin}}$ of Section \ref{sec: Casimir}, acts on the character over a UIR $\mathcal{V}$ of $so(2,4)$ and returns the one-loop vacuum energy of the corresponding field, now in global AdS. 
 $\mathcal{L}$ again takes the form of a $\beta$ integral over the character, now with additional operations included, and returns the zeta function corresponding to the one-loop determinant, as defined in Section \ref{zeta}. 
For this reason, we refer to this method  as Character Integral Representation of Zeta function (CIRZ).

Let us provide a brief summary of the result of \cite{Bae:2016rgm}.
The zeta function for a 
Hilbert space $\cH$ 
--- which might be a single UIR space
or any collection of them ---
can be written as
the sum of three pieces:
\begin{equation}
\zeta_{\cH}\!\left(z\right)=
\zeta_{\cH|2}\!\left(z\right)+\zeta_{\cH|1}\!\left(z\right)+\zeta_{\cH|0}\!\left(z\right)\,,
\label{zeta sum}
\end{equation}
where $\zeta_{\cH|n}$ are the Mellin transforms,
\be
\frac{\Gamma(z)\,\zeta_{\cH|n}\!\left(z\right)}{\log R}=\int_0^\infty 
d\beta\,\frac{\big(\frac\beta2\big)^{2(z-1-n)}}{\Gamma\!\left(z-n\right)}\,
f_{\cH|n}(\b)\,,
\label{zeta f}
\ee
of the functions $f_{\cH|n}(\b)$ given by
\begin{equation}\label{fhn}
\begin{split}
f_{\mathcal{H}|2}(\b)&= {\sinh^4{\tfrac\beta2}\over 2}\,\chi_{\mathcal{H}}\!\left(\beta,0,0\right),\\
f_{\mathcal{H}|1}(\b)&= \sinh^2{\tfrac\beta2}\left[{\sinh^2{\tfrac\beta2}\over 3}-1-\sinh^2{\tfrac\beta2}\left(\partial_{\alpha_1}^2 +\partial_{\alpha_2}^2\right)\right]\chi_{\mathcal{H}}\!\left(\beta,\alpha_{1},\alpha_{2}\right)\bigg|_{\alpha_{i}=0},\\
f_{\mathcal{H}|0}(\b)&=\left[1 + {\sinh^2\tfrac\beta2\left(3-\sinh^2\tfrac\beta2\right)\over 3} \left(\partial_{\alpha_1}^2 +\partial_{\alpha_2}^2\right)\right.\\&\qquad\left. -{\sinh^4\tfrac\beta2\over 3}\left(\partial_{\alpha_1}^4-12\,\partial_{\alpha_1}^2\partial_{\alpha_2}^2+\partial_{\alpha_2}^4\right)\right]\chi_{\mathcal{H}}\!\left(\beta,\alpha_{1},\alpha_{2}\right)\bigg|_{\alpha_{i}=0}.
\end{split}
\end{equation}
Here $\chi_\cH$ is the character defined by
\be 
\chi_\cH\!\left(\beta,\alpha_1,\a_2\right) = \tr_{\mathcal{H}}\left( e^{-\beta\,H+i\,\alpha_1\,M_{12}
+i\,\alpha_2\,M_{34}}\right).
\ee
Since both of the relations \eqref{zeta f}
and \eqref{fhn} are linear,
they define a linear map $\mathcal{L}$ between the zeta function and the character:
$\zeta_{\cH}(z)=\cL[\chi_\cH;z]$.

One can recast the $\beta$ integral
\eqref{zeta f} with sufficently large ${\rm Re}(z)$ into an integral
 over the contour 
 which
 runs from the positive real infinity and encircles the branch cut generated by $\b^{2(z-1-n)}$
in the counter-clockwise direction (see Fig.\,\ref{fig: ct1}) as
\be
  \frac{\Gamma(z)\,\zeta_{\cH|n}\!\left(z\right)}{\log R}= \frac {i\,(-2)^{2(n+1-z)}}{2\,\sin (2\pi z)\,\G(z-n)}\,\oint_C d\b\,
	\frac{f_{\cH|n}(\b)}{\b^{2(n+1-z)}}\,.
	\label{f contour}
\ee 
Now the right hand side of the above
equation is well defined in
the $z\to 0$ limit. 
Defining 
\be 
	\g_{\cH|n}
	=-(-4)^{n}\,n!\oint \frac{d\b}{2\,\pi\,i}\,
	\frac{f_{\cH|n}(\b)}{\b^{2(n+1)}}\,,
	\label{g H n}
\ee
the total one-loop vacuum energy of the AdS$_5$ theory is given by the sum 
\be
	\Gamma^{\sst (1)}_{\cH}=\log R \left(\gamma_{\cH|2}
	+\gamma_{\cH|1}+\gamma_{\cH|0}\right).
	\label{fin G g}
\ee
When the function $f_{\cH|n}$ does not
have any singularities in positive
real axis of $\b$\,
the contour can be eventually shrunken
to a small circle around $\b=0$. 
The functions $f_{\cH|n}$ for any one particle state in AdS$_5$ as well as for the spectrum of Vasiliev's theory indeed satisfy this property.
However, quite generically, the functions $f_{\cH|n}$ for the AdS dual to a matrix model CFT do have additional poles
or branch cuts. Physically, this corresponds to the fact that higher-spin theories do not have a Hagedorn transition \cite{Shenker:2011zf} while string theory does \cite{Sundborg:1999ue}.

When there are bulk fermionic degrees of freedom to be summed over, as in the case of supersymmetric theories, it is sufficient  to use the CIRZ method as presented for bosons, but instead of using the thermal partition function, use the weighted partition function 
\begin{equation}\label{index}
    Z_\cH\!\left(\beta,\alpha_1,\a_2\right) = \tr_{\mathcal{H}}\left(\left(-1\right)^F\, e^{-\beta\,H
    +i\,\a_1\,M_{12}+i\,\a_2\,M_{34}}\right),
\end{equation}
introduced in \cite{Beisert:2003te,Bianchi:2006ti}.

\section{Computing Partition Functions by Polya Counting}\label{matdual}

The large $N$ expansion qualitatively works in the same way in the vector model as the matrix one: 
all the correlation functions can be organized in terms of the color loop number.
At the leading order of $N$, it is sufficient to consider the single trace operators i.e. those made with a single color loop.
On these single trace operators, there is important difference between vector models and matrix models.
In vector models, the single trace operators are the scalar product of two fields in vector representation,
\be 
    \partial \partial \cdots \partial \vec \phi_1 \cdot
    \partial\partial \cdots \vec \phi_2\,,
    \label{vec st}
\ee 
whereas in matrix models they are the traces of 
arbitrary number of fields in a matrix valued  representation,
\begin{equation}
\tr\Big[ 
\big(\partial \partial\cdots \partial 
\bm{\phi}_1 \big)
\big(
\partial \partial \cdots \partial
\bm\phi_2\big)
\cdots 
\big(
\partial \partial \cdots \partial 
\bm\phi_n\big)\Big]\,.
\label{ST operators}
\end{equation}
In \eqref{vec st} and \eqref{ST operators}, we suppressed all the indices
and the field operators $\phi_n$ can be either scalar, spinor or vector in four dimensions.
Therefore, even though the number of single trace operators is infinite in both cases, 
the number is infinitely larger in matrix models than vector models.

One of difficulties in matrix models is the control or organization of 
infinitely many single trace operators. For this reason, one often focuses on a certain class
of single trace operators, such as BPS operators, whose study does not invoke the knowledge of the rest of operators.
However, in studying the total one-loop Casimir or vacuum energy, we need the full operator spectrum of the theory.
This can in principle be identified by decomposing the operators \eqref{ST operators} into 
irreducible $so(2,4)$ representations.
The decomposition requires a particular symmetrization of indices
 which projects the operator \eqref{ST operators} to the irreducible representation. Finding out the exact forms of these projections is not easy, but this process can be cast as a standard group theoretical problem. 
For the scalar product or inside of a trace, we put derivatives of field operator up to its equation.
They form a basis for the Hilbert space $V$ of the conformal field $\phi$ carrying a short-representation of $so(2,4)$\,:
\be
	V={\rm Span}\{\,\phi,\, \partial\phi,\, \partial\partial\phi, \ldots\,\}
	={\rm Span}\{v_i\}\,,
\ee
where the $i$ is the index indicating one of descedant (or primary) states of $\phi$\,, hence it is infinite dimensional.

In vector models, we construct single trace operators by using two elements of $V$ as \eqref{vec st}, 
hence the vector space of single trace operators is the tensor product $V\otimes V$\,.
When the field $\vec \phi_1$ and $\vec \phi_2$ are the same, which is the case in the $O(N)$ vector model,
single trace operators are symmetric in the exchange of 1 and 2 as the latter 
label is dummy. Then the corresponding vector space of single trace operators are the symmetrized 
tensor product of two $V$'s, denoted by $V\vee V$\,.
In order to find out the operator spectrum, we need to decompose $V\vee V$ into $so(2,4)$ UIRs
and it can be conveniently done in terms of the $so(2,4)$ characters.
In addition, the symmetrizations of the tensor product, or the plethysm, can be also handily treated
at the level of characters.
Suppose that for an element $g\in SO(2,4)$, $V$ has eigenvalues $\{\lambda_i \}$
(since only the conjugacy class of $g$ matters due to the trace, we can focus on 
the Cartan subgroup as in \eqref{char cartan}),
then the character reads $\chi_V(g)=\tr_V(g)=\sum_i \lambda_i$\,.
Then, the character for $V\vee V$ is
\be
	\chi_{V\vee V}(g)=\sum_{i\le j}\,\lambda_i\,\lambda_j
	=\frac{\left(\sum_i\l_i\right)^2+\sum_i\,\l_i^2}2\,.
\ee
Since $\sum_i\,\l_i^2=\tr_V(g^2)$\,, we get the relation
\be
	\chi_{V\vee V}(g)=\frac{\chi_V(g)^2+\chi_V(g^2)}2\,.
	\label{sym 2}
\ee
In the case of matrix models, we need to consider $n$ tensor product of $V$ and 
impose appropriate symmetrization compatible with trace and also the gauge group.

\subsection{$SU(N)$ Adjoint Models}

\begin{figure}
\centering
\begin{tikzpicture}
	\draw (-50:1.5) arc (-50:210:1.5) ;
	\foreach \n in {0,...,8}
      \node at (220+10*\n: 1.5) {$\cdot$};
      \foreach \n in {0,1,...,4}
      \draw [semithick,fill=white] (180-50*\n: 1.5) circle [radius=0.3];
      \foreach \n in {1,...,4}
      \node at (180-50*\n: 1.5) {$i_{\n}$};
      \node at (180: 1.5) {$i_n$};
\end{tikzpicture}
\caption{Necklace with $n$ beads}
\label{nec}
\end{figure}
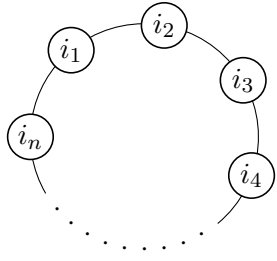

When the gauge group is $SU(N)$, the fields $\partial\partial\cdots \partial \bm\phi_p$
have the cyclic symmetry in $p \to p+1$ ($p=1,\ldots n, p+1\equiv 1$) due to the trace operation.
Then, the projection to the cyclic invariant requires only some combinatorial consideration.
For intuitive understanding let us consider a few lower $n$'s where $n$ is the number of operators in the trace.
First, the $n=2$ cyclic symmetry is nothing but the permutation symmetry.
For $n=3$, the character of cyclic 3 tensor product of $V$, denoted by ${\rm Cyc}_3(V)$, is 
\ba
	\chi_{{\rm Cyc}_3(V)}(g)\eq \sum_{i=j=k\,\vee\, i=j \neq k \,\vee\, i<j<k
	\,\vee\,i<k<j}\,\lambda_i\,\lambda_j\,\lambda_k \nn
	\eq \sum_i\l_i^3+\sum_{i\neq j}\l_i^2\,\l_j
	+2\,\sum_{i<j<k}\l_i\,\l_j\,\l_k\,.
\ea
where the summation over $(i,j,k)$ is chosen for the proper counting of elements
with the cyclicity $(i,j,k)\equiv (j,k,i)$\,.
Since 
\be
	\left(\sum_i \lambda_i\right)^3=\sum_i \lambda_i^3
	+3\,\sum_{i\neq j} \lambda_i^2\, \lambda_j
	+6\,\sum_{i<j<k} \lambda_i\,\lambda_j\,\lambda_k\,,
	\label{l3}
\ee
we find that
\be
	\chi_{{\rm Cyc}_3(V)}(g)=
	\frac{\left(\sum_i \lambda_i\right)^3+2\,\sum_i (\lambda_i)^3}3
	=\frac{\chi_V(g)^3+2\,\chi_V(g^3)}3\,.
\ee
As one can see from this example, the point is the
counting of $(i_1,\ldots, i_n)$ taking into account the cyclic equivalence.
This is well-know problem of counting inequivalent necklaces with $n$ beads
(see Fig.\ref{nec}).
The index $i_p$ indicates the type or color of beads (which, in our context, corresponds
to the descendant state of the conformal field $\phi$).
The solution to this problem is provided by
Polya's enumeration theorem as
\be
    \chi_{{\rm Cyc}_n(V)}(g)
    = \sum_{{\rm cyclic}\,(i_1,\ldots,i_n)} \l_{i_1}\cdots\l_{i_n}=
    \frac1{n}\,{\sum_{k=1}^n \left(\sum_i\l_i^\frac{n}{{\rm gcd}(k,n)}\right)^{{\rm gcd}(k,n)}}\,.
\ee
Here, ${\rm gcd}(k,n)$ is the greatest common divisor
of $k$ and $n$\,.
Alternatively, the above can be written as
\be
    \chi_{{\rm Cyc}_n(V)}(g)
    =
    \frac1{n}\,\sum_{k|n} 
    \varphi(k) \left(\sum_i\l_i^k\right)^{\frac{n}k}
    = 
    \frac1{n}\,\sum_{k|n} 
    \varphi(k)\,\chi_V(g^k)^{\frac{n}k},
    \label{cyc n}
\ee
where $k|p$ denotes the divisor $k$ of $p$
and the Euler totient function $\varphi(p)$
is the number of relative primes of $p$ in $\{1,\ldots,p\}$\,.

The total partition function is the sum of the above
from $n=2$ to infinity. 
Hence, we are temped to sum $\chi_{{\rm Cyc}_n(V)}$
over $n$.
It turns out that it is possible to at least partially sum over $n$ in \eqref{cyc n} \textit{via} \cite{Sundborg:1999ue,Polyakov:2001af,Aharony:2003sx}
\ba\label{cyc}
\chi_{{\rm Cyc}(V)}(g)\eq 
\sum_{n=2}^{\infty}\,\chi_{{\rm Cyc}_n(V)}(g)
=\sum_{n=2}^{\infty}\frac1n\sum_{k|n}\,\varphi(k)\,
[\chi_{V}(g^k)]^{\frac nk}
\nn 
\eq -\chi_{V}(g)
    +\sum_{k=1}^\infty\sum_{m=1}^\infty \frac1{m\,k}\,
    \varphi(k)\,
    [\chi_{V}(g^k)]^{m}\nn 
\eq -\chi_{V}(g)+\sum_{k=1}^\infty\,\frac{\varphi(k)}{k}\,
    \chi_{\text{log},k(V)}(g)\,,
\ea
where $\chi_{\text{log},k(V)}$ are given by
\begin{equation}\label{logpart}
\chi_{\text{log},k(V)}(\b,\a_1,\a_2)=
-\log\left[1-\chi_{V}\!\left(k\beta,k\alpha_1,k\alpha_2\right)\right].
\end{equation}

\subsection{$Sp(N)$ and $SO(N)$ Adjoint Models}

Now let us turn to the cases where the field $\bm\phi$ takes value in 
the adjoint representation $Sp(N)$.
Then the field 
is symmetric:
$\bm\phi^t=\bm\phi$\,.
Consequently,
the single-trace operators \eqref{ST operators}
admit the (anti-)symmetry,
\be
	\tr\left(\partial^{k_1}\bm\phi\,\partial^{k_2}\bm\phi\cdots \partial^{k_n}\bm\phi\right)
	=\tr\left(\partial^{k_n}\bm\phi\cdots \partial^{k_2}\bm\phi\,\partial^{k_1}\bm\phi\right),
\ee
under the flip of the $\partial^k\bm\phi$ ordering inside the trace.
In terms of indices,
\be
    (i_1,i_2,\ldots, i_n)
    \equiv (i_n,i_{n-1},
    \ldots, i_1)\,.
\ee
Hence, the space of independent single-trace operators
corresponds now to the subspace invariant under the actions of the dihedral group ${\rm Dih}_n$\,
(which includes also reflections on top of the cyclic rotations).

Let us again start the discussion with $n=3$ case:
\be
	\chi_{{\rm Dih}^+_3(V)}(g)= \sum_{i=j=k\,\vee\, i=j, k \,\vee\, i<j<k}\,\lambda_i\,\lambda_j\,\lambda_k = \sum_i\l_i^3+\sum_{i\neq j}\l_i^2\,\l_j
	+\sum_{i<j<k}\l_i\,\l_j\,\l_k\,.
\ee
where the summation over $(i,j,k)$ is chosen for the proper counting of elements
with the cyclicity $(i,j,k)\equiv (j,k,i)
\equiv (i,k,j)$\,.
Note that for $n=3$, the dihedral group coincides
with the symmetric group.
The above can be written as
\be 
    	\chi_{{\rm Dih}^+_3(V)}(g)
    	=\frac{\left(\sum_i \l_i\right)^3
    	+3\left(\sum_i \l_i^2\right)\left(\sum_j \l_j\right)
    	+2\sum_i \l_i^3}6\,,
\ee
using \eqref{l3} and 
\be 
    \left(\sum_i \l_i^2\right)\left(\sum_j \l_j\right)
    	=\sum_i\l^3+\sum_{i\neq j}\l_i^2\,\l_j\,.
\ee
In terms of the basic character $\chi_V$, it reads
\be
	\chi_{{\rm Dih}^+_3(V)}(g)
=\frac{\chi_V(g)^3+2\,\chi_V(g^3)+3\,\chi_V(g^2)
    	\,\chi_V(g)}6
    	=\frac12\,\chi_{{\rm cyc}^3(V)}(g)
    	+\frac12\,\chi_V(g)
    	\,\chi_V(g^2)\,.
    	\label{sp 3}
\ee
As one can see from this $n=3$ example,
the character of dihedrial tensor-product space ${\rm Dih}^+_n(V)$
is roughly half of the cyclic one, but not exactly. 
The precise formula is
\be\label{dih n}
		\chi_{{\rm Dih}^+_n(V)}(g)=
		\frac12\,\chi_{{\rm Cyc}_n(V)}(g)+
	\begin{cases} \frac{1}{2}\,
	\chi_V(g)\,\chi_V(g^2)^{\frac{n-1}2} & [n\ {\rm odd}] 
	\vspace{5pt}\\
	 \frac{1}{4} \left( \chi_V(g)^2\, \chi_V(g^2)^{\frac{n-2}2} + \chi_V(g^2)^{\frac{n}2} \right) & 
	  [n\ {\rm even}] \end{cases}.
\ee

If the field $\bm\phi$ takes value in the adjoint representation of $SO(N)$,
it is antisymmetric:
$\bm\phi^t=-\bm\phi$\,.
Then, the single trace 
operators have the following reflection 
property,
\be
    (i_1,i_2,\ldots, i_n)
    \equiv (-1)^n\,(i_n,i_{n-1},
    \ldots, i_1)\,,
\ee
in addition to the cyclicity. Due to the 
factor $(-1)^n$
we have less number of 
 operators in the $SO(N)$
 case compared to the $Sp(N)$ case.
 
 Again, let us consider
 the $n=3$ example.
 Since $(i,j,k)\equiv -(k,j,i)$, any repeated index vanish:
 $(i,i,j)\equiv (i,j,i)
 \equiv -(i,j,i)$.
 In the end, only
 strictly ordered 
 set $(i,j,k)$ with $i<j<k$
 survive. Hence, the character is
 \ba
	&& \chi_{{\rm Dih}^-_3(V)}(g)= 
	\sum_{i<j<k}\,\lambda_i\,\lambda_j\,\lambda_k =\frac{\left(\sum_i \l_i\right)^3
    	-3\left(\sum_i \l_i^2\right)\left(\sum_j \l_j\right)
    	+2\sum_i \l_i^3}6\nn
    &&=
    \frac{\chi_V(g)^3+2\,\chi_V(g^3)-3\,\chi_V(g^2)
    	\,\chi_V(g)}6
    =\frac12\,\chi_{{\rm cyc}^3(V)}(g)
    	-\frac12\,\chi_V(g)
    	\,\chi_V(g^2)\,.
\ea
One can notice that
compared to the $Sp(N)$ case of \eqref{sp 3},
we have minus sign
after the last equality.
This pattern extends
to an arbitrary odd $n$\,:
\be\label{so n}
		\chi_{{\rm Dih}^-_n(V)}(g)=
		\frac12\,\chi_{{\rm Cyc}_n(V)}(g)+
	\begin{cases} -\frac{1}{2}\,
	\chi_V(g)\,\chi_V(g^2)^{\frac{n-1}2} & [n\ {\rm odd}] 
	\vspace{5pt}\\
	 \frac{1}{4} \left( \chi_V(g)^2\, \chi_V(g^2)^{\frac{n-2}2} + \chi_V(g^2)^{\frac{n}2} \right) & 
	  [n\ {\rm even}] \end{cases},
\ee
because for odd $n$,
cyclic operators
can be split
into either symmetric or
anti-symmetric ones under
reflection. 

Finally, one may attempt to sum 
$\chi_{{\rm Dih}^\pm_n(V)}$ over $n$.
The term half of cyclic character can be treated 
as the cyclic case, whereas the additional
contributions can be summed as they are
geometric series. 
It will be useful to consider the two partition functions obtained by summing separately over the even and odd values of $n$. In particular
\begin{equation}\label{evenodd}
\begin{split}
\chi_{{\rm even}(V)}\left(g\right)& =\frac14\sum_{n=2,4,6, \cdots}^{\infty} \left(\chi_V(g)^2 \chi_V(g^2)^{\frac{n-2}{2}}+\chi_V(g^2)^{n/2} \right)=
\frac14\,\frac{\chi_V(g)^2+\chi_V(g^2)}{1-\chi_V(g^2)},\\
\chi_{{\rm odd}(V)}\left(g\right)& =\frac12\sum_{n=3,5,7,\cdots}^{\infty}\chi_V(g)\,\chi_V(g^2)^{\frac{n-1}2} =\frac12\,\frac{\chi_V(g)\,\chi_V(g^2)}{1-\chi_V(g^2)}.
\end{split}
\end{equation}
In the end we get
\be\label{dihso}
	\chi_{{\rm Dih}^\pm(V)}(g)=
	\frac12\,\chi_{{\rm Cyc}(V)}(g)+ \chi_{{\rm even}(V)}\left(g\right)\pm\chi_{{\rm odd}(V)}\left(g\right)\,.
\ee

\subsection{$U(N)\times U(M)$ Bi-Fundamental
and $O(N)\times O(M)$ Bi-Vector Models}

If the conformal fields $\bm\phi$  carry bi-fundamental
representations with respect to $U(N)$
and $U(M)$, hence taking value
in $M\times N$ complex matrix, then
the single trace operators will take the form of
\be 
    \tr\left(\partial^{k_1}\bm\phi\, \partial^{k_2}\bm\phi^\dagger\,
    \cdots\,\partial^{k_{2n}}\bm\phi^\dagger
    \right).
\ee
Note here that the operators involve always
even number of fields in $\bm\phi\,\bm\phi^\dagger$ form,
and the operators are invariant under
the cyclic rotation by 2.
This means that 
the basic vector space in this case is 
not $V$ but $V\otimes V$ and 
the single trace operators with $2n$ operators
is governed by the cyclic  group $C_n$\,.
The character can be constructed in 
an analogous manner and reads
\be\label{bf n}
		\chi_{{\rm Bf}_{2n}(V)}(g)=
		\frac1{n}\,\sum_{k|n} 
    \varphi(k)\,\chi_V(g^k)^{\frac{2n}k}\,.
\ee
We can collect the above for $n=1,\ldots, \infty$ to get
\be
    \chi_{{\rm Bf}(V)}(g)=\sum_{n=1}^\infty
    \chi_{{\rm Bf}_{2n}(V)}(g)
    = -\sum_{k=1}^\infty\frac{\varphi(k)}k\,
    \log\left[1-\chi_V(g^k)^2\right]\,.
    \label{bf}
\ee

The final case is the $O(N)\times O(M)$ bi-vector models
where the scalar fields $\bm\phi$ are real as opposed to the $U(N)\times U(M)$ bi-fundamental models. Hence, the space of its single-trace operators are spanned by
\begin{equation}
\mbox{Tr}\left(
\partial^{k_1} 
\bm\phi\,\partial^{k_2}\bm\phi^{t}
\cdots \partial^{k_{2n}}\bm\phi^{t}\right),
\label{O bi-vector}
\end{equation}
which has the reflection symmetry:
\be
	\mbox{Tr}\left(
\partial^{k_1} 
\bm\phi\,\partial^{k_2}\bm\phi^{t}
\cdots \partial^{k_{2n}}\bm\phi^{t}\right)
	=\mbox{Tr}\left(
\partial^{k_{2n}} 
\bm\phi
\cdots \partial^{k_2}\bm\phi\,\partial^{k_{1}}\bm\phi^{t}\right),
\ee
on top of the cyclic rotation by two.
Again, the character of the bi-vector models 
is the half of the bi-fundamental ones up to 
the contribution from the reflection symmetries. 
This time, the latter is simpler and we end up with
\be\label{bv n}
		\chi_{{\rm Bv}_{2n}(V)}(g)=
		\frac12\left(\chi_{{\rm Bf}_{2n}(V)}(g)
	 	+\chi_V(g^2)^{n} \right).
\ee
The character for all single-trace operator is again the sum
of the latter over all positive integer $n$ and reads
\be\label{bv}
	\chi_{{\rm Bv}(V)}(g)=\sum_{n=1}^\infty 	\chi_{{\rm Bv}_{2n}(V)}(g)=
	\frac12\left(\chi_{{\rm Bf}(V)}(g)+\frac{\chi_V(g^2)}{1-\chi_V(g^2)}\right).
\ee

\subsection{Symmetric Group}

Finally, we can consider the operators
made by conformal fields which are fully symmetric in any permutations.
This is the symmetric group and 
the corresponding character is 
given by
\be\label{sym n}
    \chi_{{\rm Sym}_{n}(V)}(g)
    =\sum_{j_1+2\,j_2+\cdots+n\,j_n=n}\,
    \prod_{k=1}^n
    \frac{{\chi_V(g^k)}^{j_k}}{k^{j_k}\,j_k!}\,,
\ee
or in terms of Bell polynomial as
\be 
      \chi_{{\rm Sym}_{n}(V)}(g)
    =\frac1{n!}\,B_n(0!\,\chi_V(g),1!\,\chi_V(g^2),\ldots,(n-1)!\,
    \chi_V(g^n))\,.
\ee
The generating function or
the full partition function has 
rather simple form,
\be\label{sym}
    \chi_{{\rm Sym}(V)}(g)
    =\exp\left(\sum_{k=1}^\infty\frac1k\, \chi_V(g^k)\right),
\ee
sometimes referred to as
plethystic exponential (PE).
Notice that here we do not have the notion
of large $N$ expansion (and single trace, multi trace etc) hence  we have also included the $n=1$ operator,
that is $\phi$ itself.

In order to see the implication of the above formula, let us consider 
a few toy examples. We first take the one-particle partition function of free scalar in two dimensions\,:
\be
    \chi_V(q,\bar q)=\frac q{1-q}+\frac{\bar q}{1-\bar q}\,.
\ee
Once can evaluate the sum over $k$ by expanding first $1/(1-q)$ and $1/(1-\bar q)$ as 
\be
    \sum_{k=1}^\infty\frac1k\,\frac{q^k}{1-q^k}
    =\sum_{k=1}^\infty\frac1k\,q^k\sum_{n=0}^\infty q^{k\,n}
    =\sum_{n=1}^\infty \log\left(\frac1{1-q^n}\right)\,,
\ee
hence we get
\be 
    \chi_{S(V)}(q,\bar q)
    =\left(\prod_{n=1}^\infty\frac1{1-q^n}\right)
    \left(\prod_{n=1}^\infty\frac1{1-\bar q^n}\right).
\ee 
This differs from the partition function of free boson
by $(q\,\bar q)^{-1/24}/\log(q\,\bar q)$\,.
The $(q\,\bar q)^{-1/24}$ factor is missing because
our character did not include the $q^{-c/24}$\,.
The $\log(q\,\bar q)$ factor is due to 
zero mode contribution.

Let us consider the following toy partition functions 
inspired by the two-dimensional free boson,
\be
    \chi_V(q)=\frac{q}{(1-q)^d}\,,
\ee 
which captures certain aspects of the scalar character
in higher dimensions.
By using
\be
    \frac{q}{(1-q)^d}=\frac1{(d-1)!}\left(\frac\partial{\partial s}\right)^{d-1}\,\frac{s^{d-1}\,q}{1-s\,q}\,\Big|_{s=1}\,,
\ee
we get
\ba 
    \log\chi_{S(V)}(q)
    \eq \frac1{(d-1)!}\left(\frac\partial{\partial s}\right)^{d-1}\,\sum_{n=1}^\infty\,s^{d+n-2} \log\left(\frac1{1-q^n}\right)\Big|_{s=1}\nn 
    \eq \sum_{n=1}^\infty \binom{d+n-2}{d-1}\,
    \log\left(\frac1{1-q^n}\right).
\ea
In the end, the full partition function,
\be 
    \chi_{S(V)}(q)
    =\prod_{n=1}^\infty\frac1{(1-q^n)^{\binom{d+n-2}{d-1}}}\,,
\ee
gives the MacMahon's 
(unsuccessful) guess formula
for the generating function of $d$-dimensional partitions.

\subsection{Fermions}

So far, our consideration was only on the 
Hilbert space $V$ of bosonic
conformal fields $\phi$. 
Let us now include the Hilbert space $W$ of fermionic fields $\psi$\,:
\be 
    W={\rm Span}\{\psi,\,\partial\psi,\,
    \partial\partial\psi,\ldots\}
    ={\rm Span}\{w_p\}\,.
\ee
The total Hilbert space of conformal fields
is then $H=V\oplus W$\,.
We generalize the character to cover the fermionic case
as
\be
    Z(g)=\tr\left((-1)^F\,g\right),
\ee
where $F$ is the fermionic number operator.

Let us reconsider the $SU(N)$ adjoint partition function for single trace operators of lower $n$'s. For $n=2$,
we have two additional class of operators. First, we have fermionic operators,
\be
    Z_{VW}(g)
    =-\tr_{V\otimes W}(g)
    =-\sum_{i,p}\l_i\,\l_p
    =Z_{V}(g)\,Z_W(g)\,.
\ee
Second, there is the bosonic one made by two fermions,
\be
    Z_{WW}(g)
    =\tr_{W\wedge W}(g)
    =\sum_{p<q}\l_p\,\l_q
    =\frac{\left(\sum_p \l_p\right)^2-\sum_p \l_p^2}2
    =\frac12\left(Z_W(g)^2+Z_W(g^2)\right).
\ee
In the end, we get the same form as \eqref{sym 2}:
\be 
     Z_{HH}(g)
     =\frac12\left(Z_H(g)^2+Z_H(g^2)\right).
\ee
Moving to $n=3$\,, we have three more classes 
of operators, $VVW$, $VWW$ and $WWW$\,.
The first and second are simply
\be 
    Z_{VVW}(g)=Z_V(g)^2\,Z_W(g)\,,
    \qquad 
    Z_{VWW}(g)=Z_V(g)\,Z_W(g)^2\,,
\ee
and the last is
\be
 Z_{WWW}(g)
 =\frac{Z_W(g)^3+2\,Z_W(g^3)}3\,.
\ee
Note that the fermionic nature does not play any role
in $WWW$ as the cyclic permutation can be viewed
as the commutation of bosonic $WW$ and fermionic $W$ space.
In the end, we get
\be
 Z_{HHH}(g)
 =\frac{Z_H(g)^3+2\,Z_H(g^3)}3\,.
\ee
In this way, one can convince her/himself
that the partition fuction of 
single trace operators made by both of
bosonic and fermionic conformal fields 
has the same form as the character in the pure bosonic case:
\be
    Z_{{\rm cyc}^n(H)}(g)
    =
    \frac1{n}\,\sum_{k|n} 
    \varphi(k)\,Z_H(g^k)^{\frac{n}k}.
\ee
One can include fermions
in the dihedral, bi-fundamental and bi-vector models
in the same way.

\section{One Loop Tests of 
Free Matrix CFT Holographies}\label{stringyrev}

In the previous section,
we have 
computed
the partition function 
of all single trace operators in
various free CFTs
with fields in various different representations of the internal symmetry group.
Using the AdS/CFT dictionary,
these operator spectrum can be identified to the spectrum of AdS fields in the dual theory.
Hence, the partition function for
single trace operators computed above can be
simply interpreted as 
the partition function of the dual AdS theory.
Then, the CIRZ formalism presented in Section \ref{CIRZ} 
can be readily applied to computing the one-loop vacuum energy. 
Analogously, the methods of Section \ref{sec: Casimir} can be used to compute the Casimir energy of such AdS theories. We now turn to these computations.

A natural starting point is to evaluate the one-loop vacuum and Casimir 
energies
of the AdS fields 
dual to the single trace operators
made by  $n$ boundary fields.
The corresponding 
partition functions are
given by $\chi_{{\rm Cyc}_n}$ \eqref{cyc n},
$\chi_{{\rm Dih}^+_n}$ \eqref{dih n},
$\chi_{{\rm Dih}^-_n}$ \eqref{so n},
$\chi_{{\rm Bf}_{n}}$ \eqref{bf n},
$\chi_{{\rm Bv}_{n}}$ \eqref{bv n} 
and 
$\chi_{{\rm Sym}_{n}}$ \eqref{sym n}.
The set of AdS fields dual to the single trace operators appearing in each tensor product above will be referred to as comprising the $(n-1)^{\mathrm{th}}$ order Regge trajectory, following
\cite{Bae:2016hfy,Bae:2016rgm,Bae:2017spv}.

In order to obtain the full vacuum energy
we need analytic expressions for
the vacuum energy of the fields in a given
 Regge trajectory.
As we shall show shortly,
this is not available except for the 
$\cN=4$ theory.
However, it is still possible 
to calculate these quantities
for $n$'s large enough 
to observe a certain pattern.
Either from the pattern or from
the analytic expression in the $\cN=4$ case, we can see that 
 the one-loop vacuum and Casimir energies seem to diverge with increasing $n$.
 
Given this, we need an alternative
prescription to sum over fields and
one-loop energies to cure this
divergence. 
For intution, let us consider the
corresponding computations for the
higher-spin theories dual to vector models
\cite{Giombi:2013fka,Giombi:2014iua,Giombi:2014yra,Beccaria:2014zma}. In that
case, divergence in the total 
energy can be traced back to the fact
that contributions from individual fields are computed \textit{first} and summed
over \textit{second}. Indeed, our
computations of
\cite{Bae:2016hfy,Bae:2016rgm} reviewed
in Section \ref{hsrev} show that this
divergence is cured by summing over
states first, by computing the thermal
partition function, and evaluating the
vacuum energy second. Hence it is natural
to attempt to cure the divergence arising
in matrix model CFTs 
by applying the CIRZ technique directly
to the full partition functions ---
$\chi_{{\rm Cyc}}$ \eqref{cyc},
$\chi_{{\rm Dih}^\pm}$ \eqref{dihso},
$\chi_{{\rm Bf}}$ \eqref{bf},
$\chi_{{\rm Bv}}$ \eqref{bv} 
and 
$\chi_{{\rm Sym}}$ \eqref{sym}. We shall also carry out these computations here.

The rest of the section is organized as follows. Section \ref{subsec: scalym} contains the computation of one-loop vacuum and casimir energies for the bulk duals of the free $Sp(N)$ and $SO(N)$ scalar matrix models as well as the free Yang Mills theories. Next, in Section \ref{subsec: neq4} we turn to the corresponding computations for the bulk dual of $\cN=4$ super Yang-Mills. Finally in Section \ref{subsec: bifund} we study the bulk duals of CFTs with bifundamental matter, in particular, scalars and Majorana fermions.

Finally, some reminders of notation. In what follows, the partition function of the boundary scalar, spin-$\tfrac12$, and spin-1 fields is respectively denoted by $\chi_0$, $\chi_{\tfrac12}$ and $\chi_1$,
and their explicit forms
are given in \eqref{shortscalar}, \eqref{shortmajo} and \eqref{chi1}.
The Casimir energy is denoted as $E$ and the one-loop vacuum energy is denoted as $\G^{(1)}$. Often these quantities will have subscripts which indicate which fields or set of fields they correspond to. For example, $E_0$ is the Casimir energy for a boundary scalar, while $\G^{(1)}_{{\rm Dih}^+_n}$ is the vacuum energy summed over all fields contained in the cyclic character for $Sp(N)$ at some fixed $n$.

\subsection{Non-Supersymmetric $Sp(N)$ and $SO(N)$ Adjoint Models} \label{subsec: scalym}

\subsubsection{Bulk Dual of Free Scalar}

\begin{figure}[h]
\begin{subfigure}{.5\textwidth}
  \centering
  \includegraphics[width=0.9\linewidth]{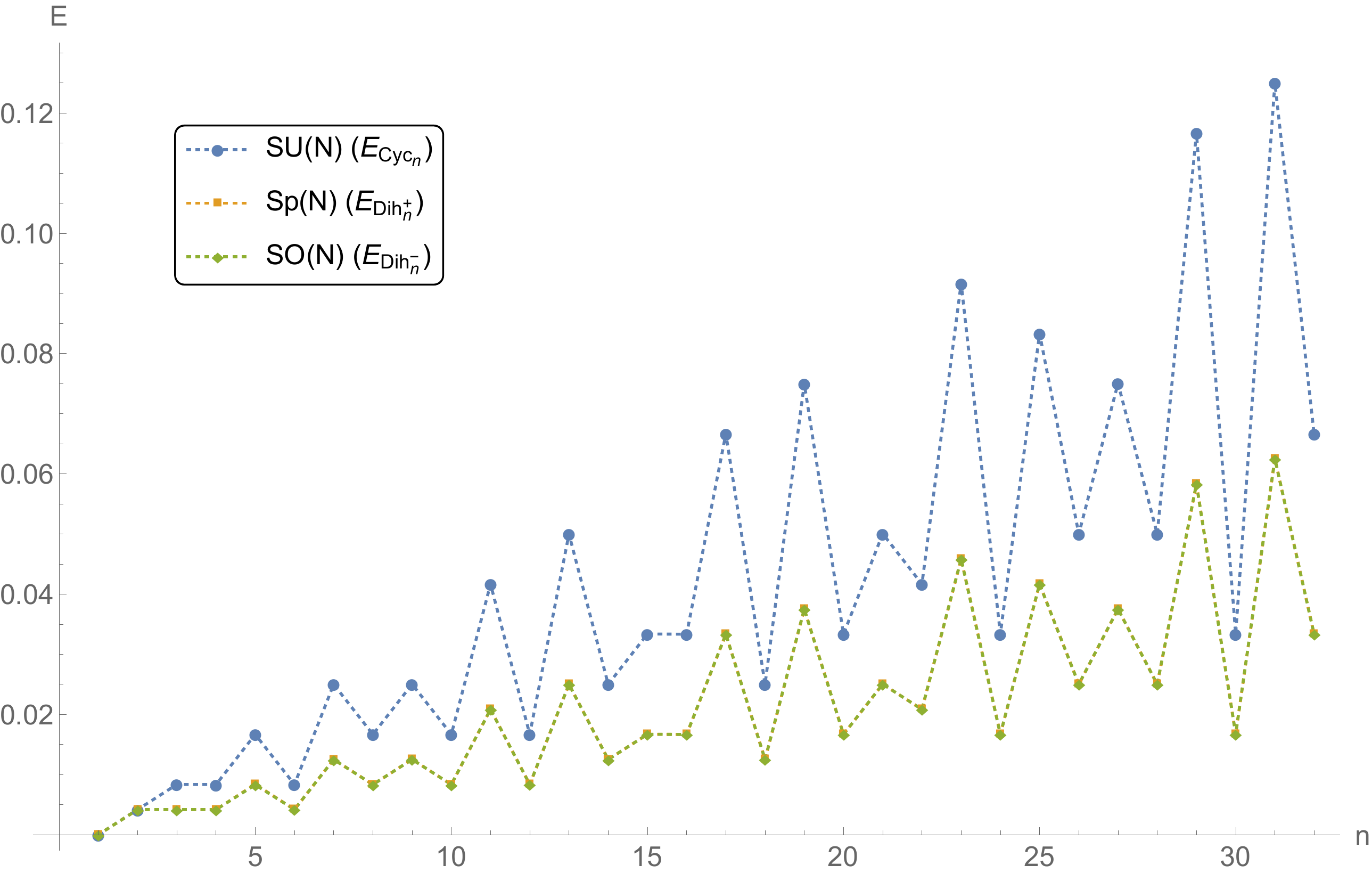}
  \caption{Casimir Energy}
  \label{fig:cas_scal}
\end{subfigure}%
\begin{subfigure}{.5\textwidth}
  \centering
  \includegraphics[width=0.9\linewidth]{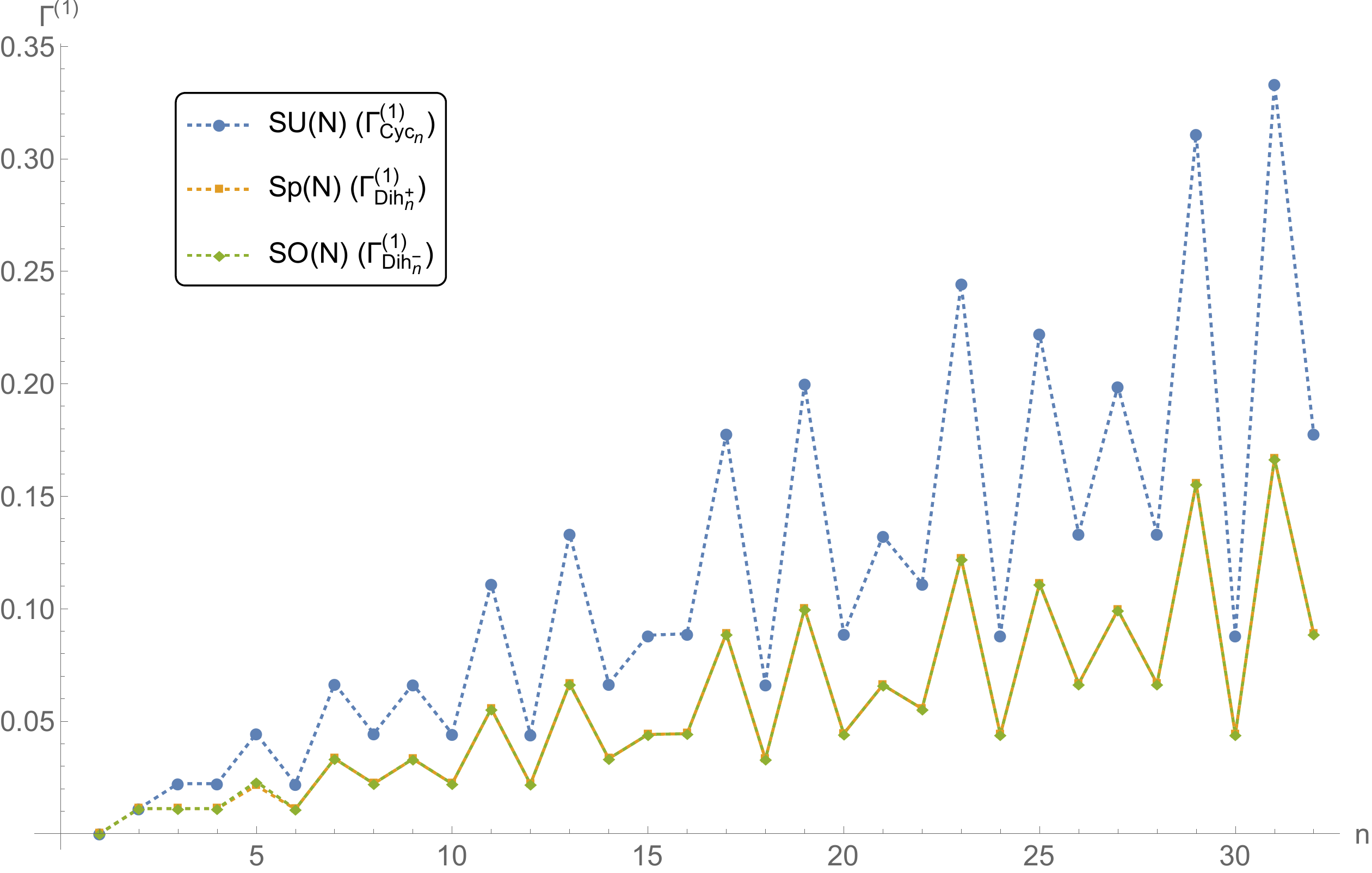}
  \caption{Vacuum Energy}
  \label{fig:vac_scal}
\end{subfigure}%
\caption{Plot
of the first 32 results for
$E_{{\rm Dih}^\pm_n}$
and $\G^{\sst (1)}_{{\rm Dih}^\pm_n}$
(in the unit of $\log R$)}
\label{fig:scalar}
\end{figure}

\begin{figure}[h]
\begin{subfigure}{.5\textwidth}
  \centering
  \includegraphics[width=0.9\linewidth]{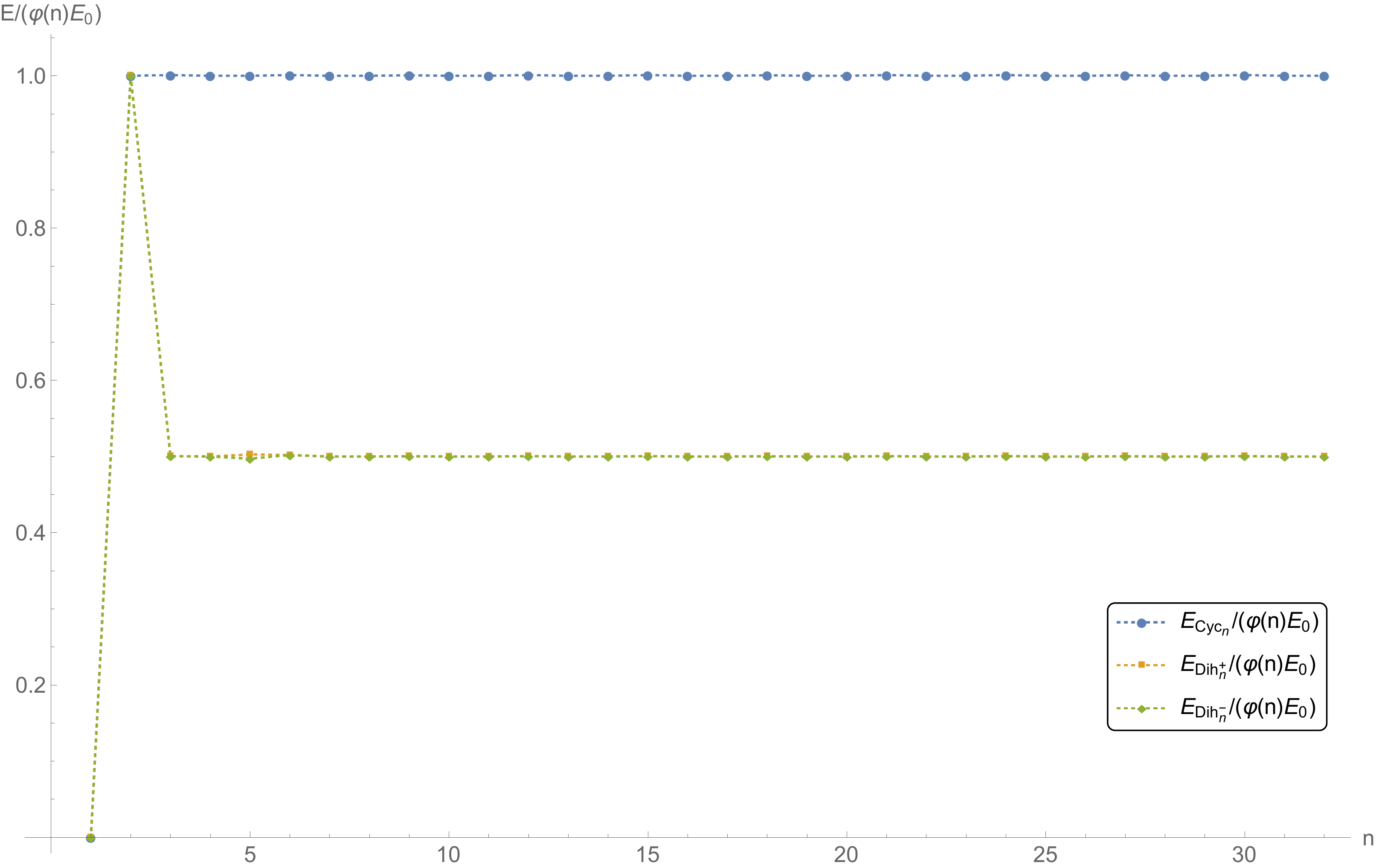}
  \caption{Casimir Energy}
 \end{subfigure} 
\begin{subfigure}{.5\textwidth}
  \centering
  \includegraphics[width=0.9\linewidth]{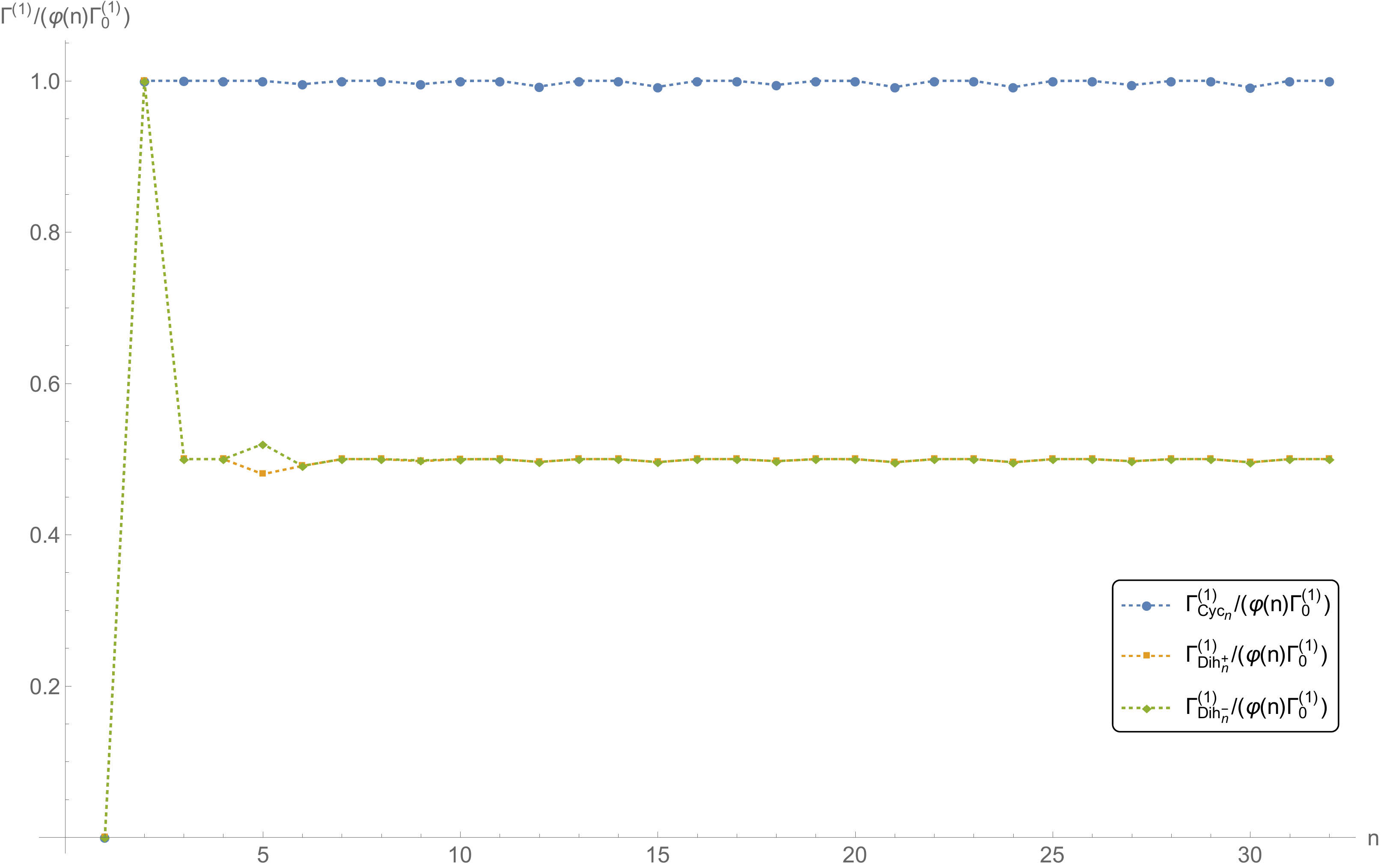}
  \caption{Vacuum Energy}
 \end{subfigure} 
 \caption{Plot of $E_{{\rm Dih}^\pm_n}/(\varphi(n)\,E_0)$
 and $\G^{\sst (1)}_{{\rm Dih}^\pm_n}/(\varphi(n)\,\G^{\sst(1)}_0)$}
  \label{fig: totient}
\end{figure}

At fixed values of $n$, the Casimir energies $E_{{\rm Dih}^\pm_n}$
and vacuum energy $\G^{\sst (1)}_{{\rm Dih}^\pm_n}$ for the  free $Sp(N)$ and $SO(N)$ adjoint scalar CFTs can be obtained by applying CIRZ methods to \eqref{dih n} and \eqref{so n}, where $\chi_{V}$ is taken to be $\chi_0$. The results obtained up to $n=32$ are exhibited graphically in Figure \ref{fig:scalar}.
We make the following observations at this stage. Firstly, we note that the $Sp(N)$ and $SO(N)$ plots almost overlap each other. Secondly, upon normalizing by $\varphi(n)\,E_0$ or $\varphi(n)\,\G^{\sst (1)}_0$, the chaotic pattern of the results in Figure \ref{fig:scalar} maps to the constant $\frac12$ with very tiny fluctuations: see Figure \ref{fig: totient}.
This is because the correction terms in \eqref{dih n} and \eqref{so n} very quickly decay as $n$ increases.
Further, it was also observed for the $SU(N)$ adjoint model in \cite{Bae:2016rgm} that when $n=2^m$ for an integer $m$, then the fluctuations
exactly vanish. This turns out to be no longer true for the dihedral models due to the presence of the correction terms \eqref{dih n} and \eqref{so n}.

In order to obtain the full 
Casimir or vacuum energy, we need to sum
these results over $n$\,.
In the $N\to\infty$ limit,
the summation is from $n=2$ to $\infty$.
Since we do not have 
an analytic expression for $n$ 
at our disposal, we cannot evaluate
this sum.
However, if the pattern of Figure \ref{fig: totient} persists,
one can expect from the pattern
that the total results are divergent. 

We will now examine if the CIRZ method can again be used to regulate this divergence by summing over the spectrum first and evaluating the free energy afterwards. It is quickly apparent that the CIRZ method, when applied to \eqref{dihso}, returns a finite value for the Casimir energy for both $Sp(N)$ and $SO(N)$ models. In particular, 
\begin{equation}\label{fullcas2scalar}
E_{{\rm Dih}^{+}} = {27\over 240}\,,
\qquad
E_{{\rm Dih}^{-}}={28\over 240}  \qquad \left(E_0 = {1\over 240}\right),
\end{equation}
where we have also presented the Casimir energy of the boundary scalar for comparison.
We now finally turn to the one-loop vacuum energy computation in the $N\rightarrow\infty$ case, where we provide a few more details. Firstly, again from examining \eqref{dihso} we see that it is useful to focus on the correction terms $\chi_{\mathrm{even}}$ and $\chi_{\mathrm{odd}}$ computed in \eqref{evenodd}. Applying the CIRZ formalism, we see that the one-loop vacuum energies receive the following contributions
\be 
 \G^{(1)}_{\mathrm{odd}} = -\frac{1}{180}\log R,\qquad  \G^{(1)}_{\mathrm{even}} = \frac{101}{180}\log R ,
\ee 
Using these results, as well as the result \cite{Bae:2016rgm}
\begin{equation}
 \G^{\sst (1)}_{{\rm Cyc}} = -{1\over 90}\log R,
\end{equation}
we see that for the bulk dual of the scalar matrix model
\begin{equation}
 \G^{\sst (1)}_{{\rm Dih}^+} =
 {11\over 20}\,\log R\,,
 \qquad 
 \G^{\sst (1)}_{{\rm Dih}^-} ={101\over 180}
 \log R
 \qquad \left(\G^{\sst (1)}_0=\frac1{90}\,\log R\right)\,.
\end{equation}
The vacuum energy for the boundary scalar is also presented above for comparison.
We postpone discussions of how to interpret these results to the conclusions.

\subsubsection{Bulk Dual of Free Yang Mills}

\begin{figure}[h]
\begin{subfigure}{.5\textwidth}
  \centering
  \includegraphics[width=0.8\linewidth]{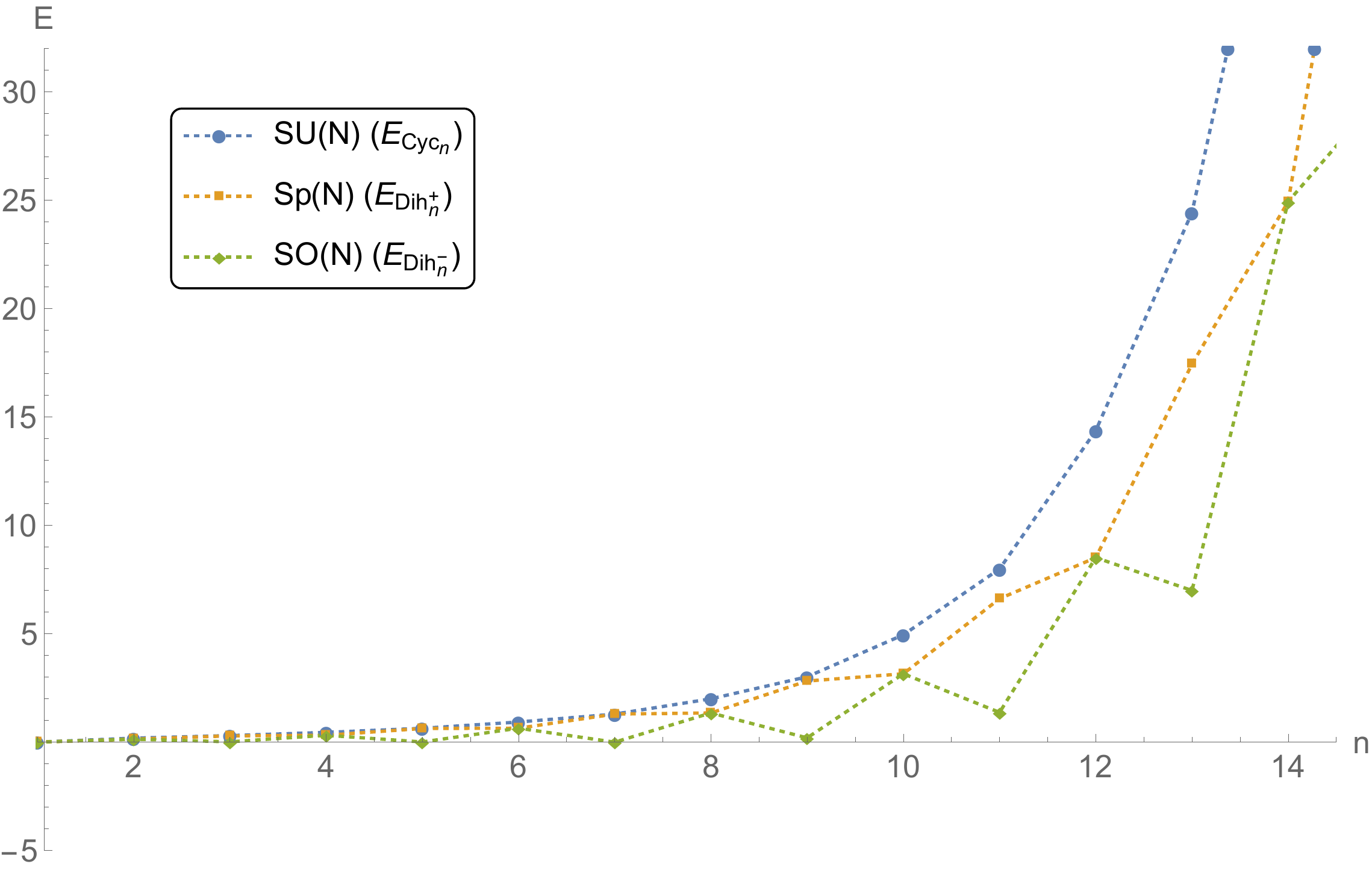}
  \caption{Casimir Energy for free Yang-Mills}
 \label{fig:cas_ym}
\end{subfigure}
\begin{subfigure}{.5\textwidth}
  \centering
  \includegraphics[width=0.8\linewidth]{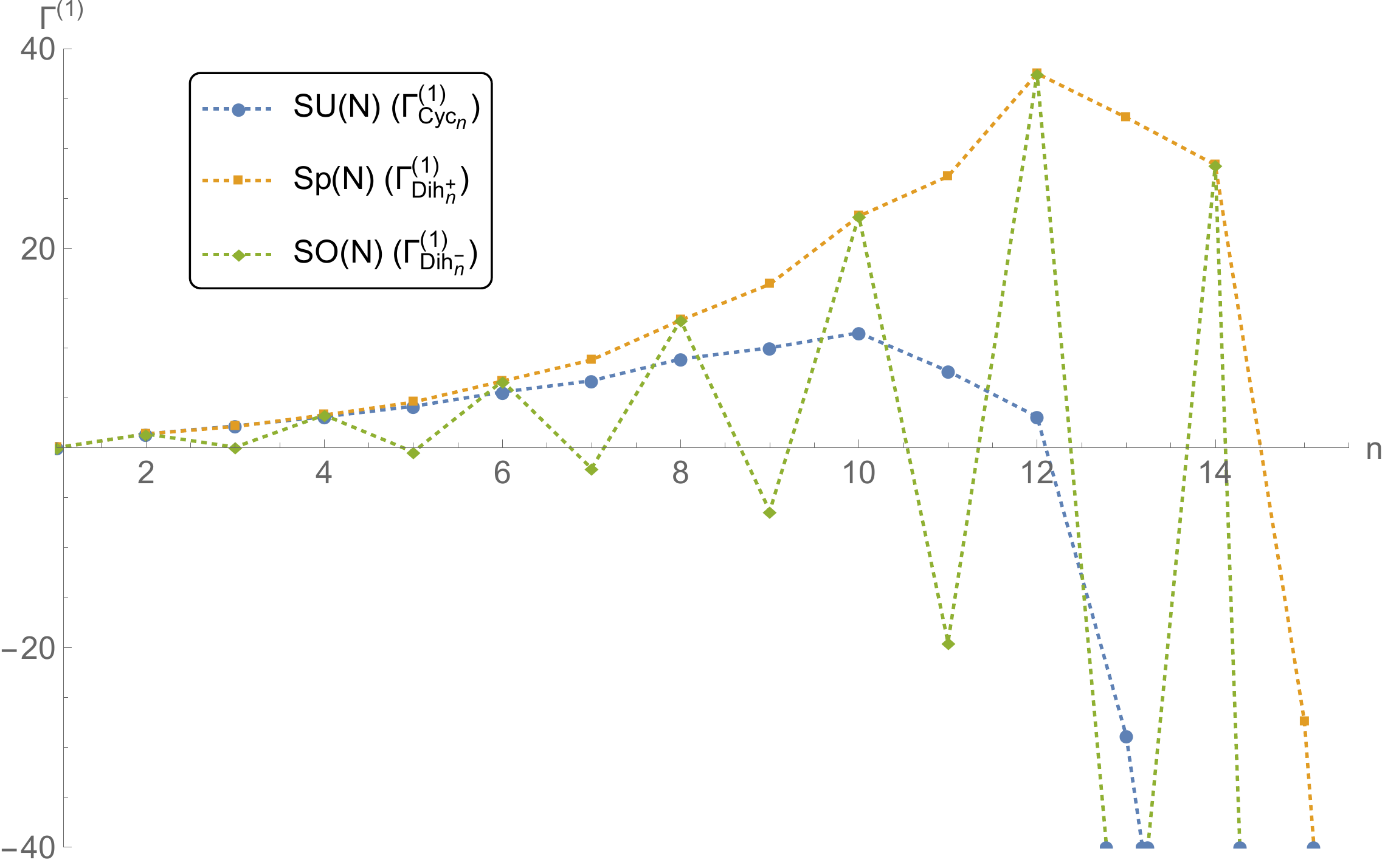}
  \caption{Vacuum Energy for free Yang-Mills}
 \label{fig:vac_ym}
\end{subfigure} %
\begin{subfigure}{.5\textwidth}
  \centering
  \includegraphics[width=0.8\linewidth]{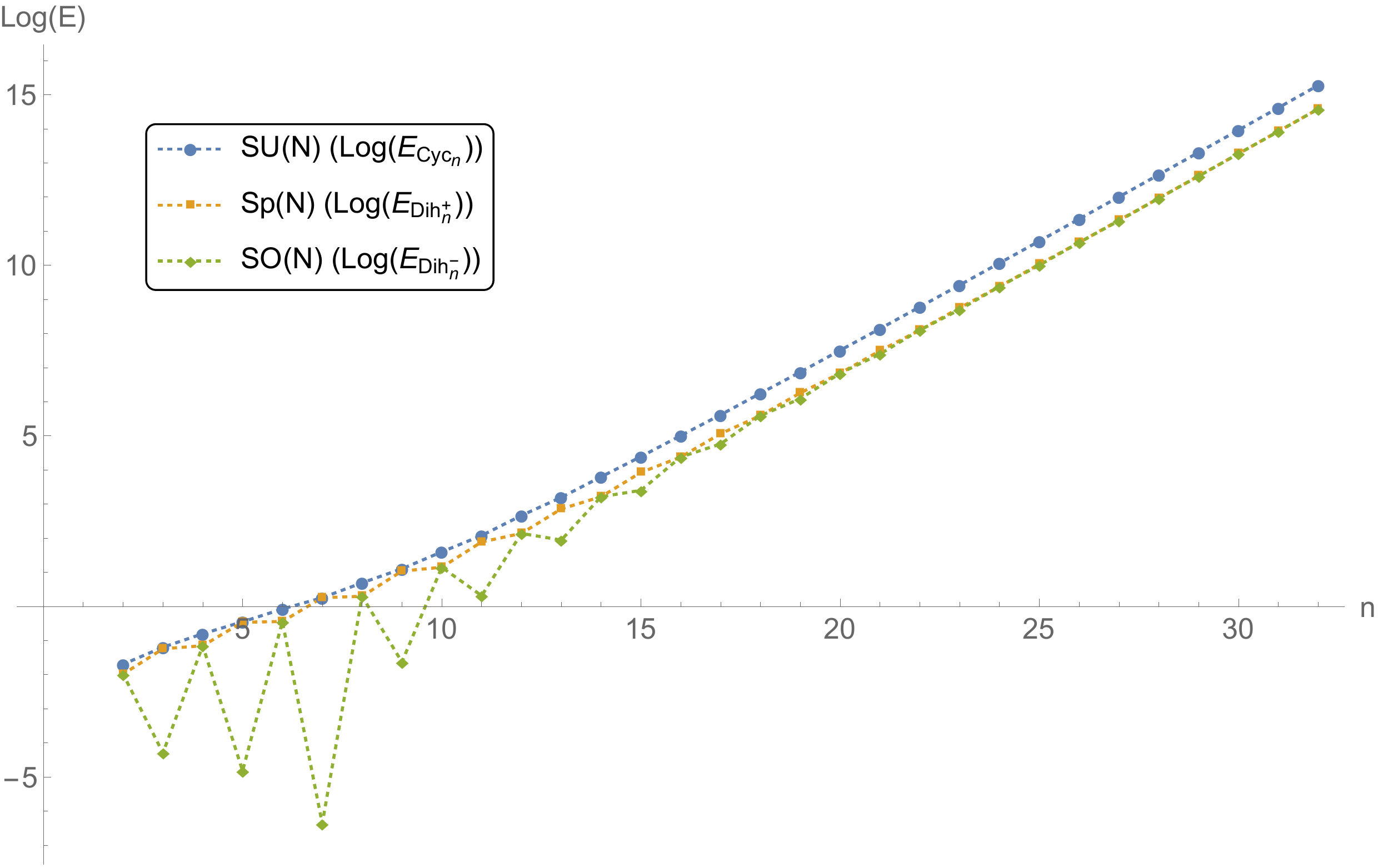}
  \caption{Log plot of the Casimir Energy}
 \label{fig:cas_ym_log}
\end{subfigure}
\begin{subfigure}{.5\textwidth}
  \centering
  \includegraphics[width=0.8\linewidth]{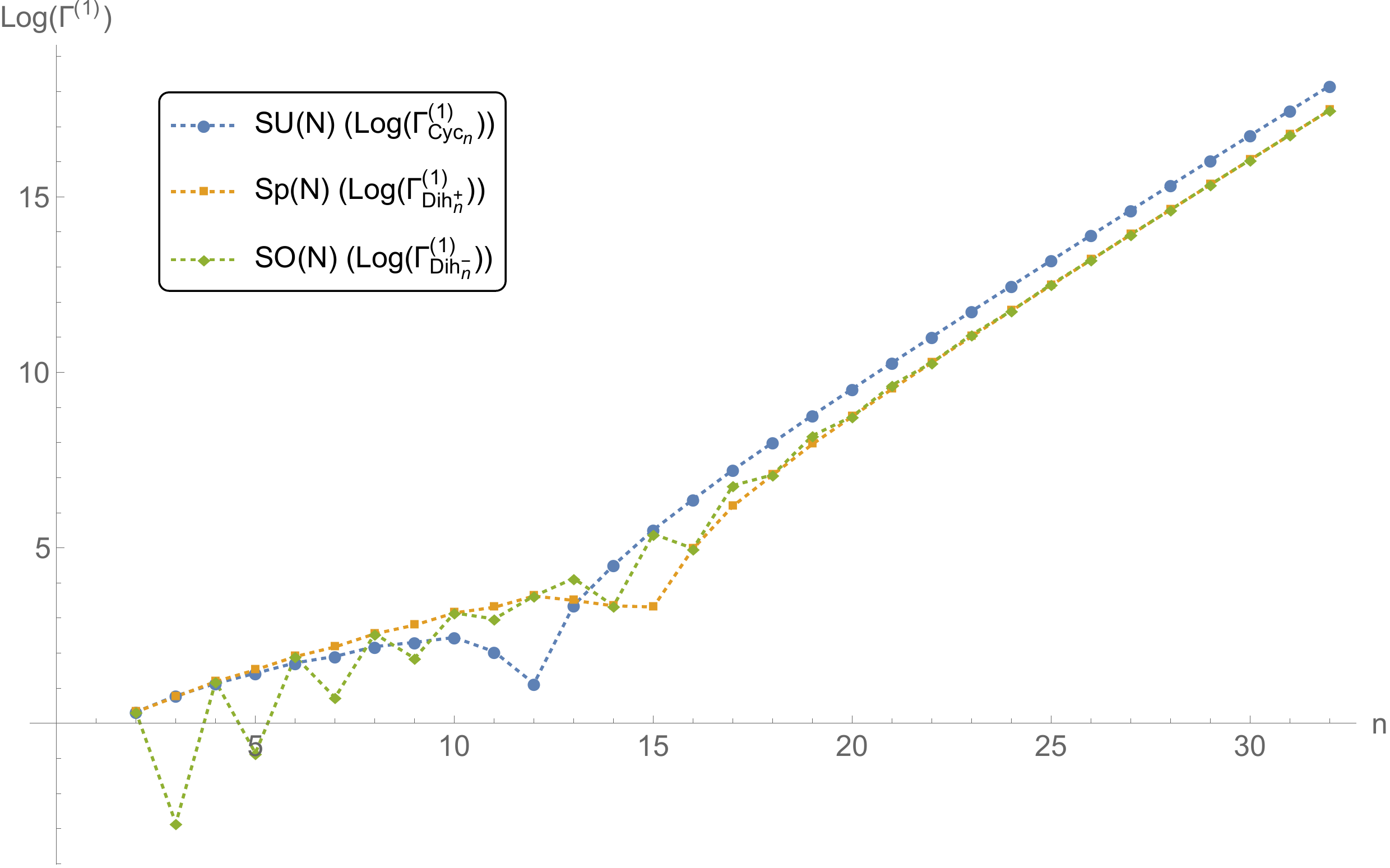}
  \caption{Log plot of the Vacuum Energy}
 \label{fig:vac_ym_log}
\end{subfigure}
\caption{Plots of the Casimir energy and the one-loop vacuum energy of the bulk dual of free Yang-Mills theory in $SO(N)$ and $Sp(N)$ adjoint representation up to $n=32$.}
\end{figure}

In contrast to the free scalar case, 
for free Yang Mills 
--- we take $\chi_V=\chi_1$ --- we see that the one-loop vacuum and Casimir energies plotted in Figures \ref{fig:cas_ym} and \ref{fig:vac_ym} show a runaway behavior.
Further, though the scale of the graph hides it, the contribution to the vacuum energy flips sign as $n$ is increased. 
This may be readily inferred from the fact that for $n=2$ the $Sp(N)$ and $SO(N)$ partition functions \eqref{dih n} and \eqref{so n} are equal to the $SU(N)$ partition function, for which the vacuum energy contribution was computed in \cite{Bae:2016hfy} and found to be $+\tfrac{62}{45}\log R$, and the fact that for larger values of $n$ the vacuum energy contribution takes negative values. 
Hence, even in the Yang Mills case the one-loop free energies do not appear to converge as we increase the value of $n$. 
It is therefore again natural to 
regulate
the result by directly using
the character of the full 
partition function \eqref{dihso}. 
In this way we obtain the total Casimir 
energy as
\begin{equation}\label{fullcas2ym}
E_{{\rm Dih}^+} = \frac{1377}{120}\,,
\qquad 
E_{{\rm Dih}^-} =\frac{1388}{120}
\qquad \left(E_1 = {11\over 120}\right),
\end{equation}
while for the total vacuum energy we find
\begin{equation}
 \G^{\sst (1)}_{{\rm Dih}^+} =  \frac{791}{10}\,\log R\,,\qquad 
 \G^{\sst (1)}_{{\rm Dih}^-} =  \frac{7181}{90}\,\log R\qquad
 \left(\G^{\sst (1)}_1=\frac{31}{45}\,\log R\right).
\end{equation}

\subsection{Bulk Dual of Free $\cN=4$ SYM}\label{subsec: neq4}

We now turn to the maximally supersymmetric case of the AdS$_5$ dual of free planar $\cN=4$ super Yang-Mills with gauge group $SO(N)$ or $Sp(N)$. 
The partition function to use
is $Z_V=\chi_1-4\,\chi_{\frac12}+6\,\chi_0$,
as in the $SU(N)$ case studied in \cite{Bae:2017spv} we again find that the computation can be analytically carried out for arbitrary values of $n$, in contrast to the non-supersymmetric cases studied above.

We remind the reader here that the result for the one-loop vacuum and Casimir energies from the $n$-th order Regge Trajectory in the $SU(N)$ (i.e. cyclic) case was given by 
\begin{equation}
\G^{\sst (1)}_{{\rm Cyc}_n} = n\,\log R,
\qquad E_{{\rm Cyc}_n} = {3\over 16} n\,.
\end{equation}
The partition function for the $n$-th order Regge Trajectories in the $Sp(N)$ and $SO(N)$ adjoint models is given in \eqref{dih n} and \eqref{so n} respectively. We focus on the correction term in both expressions. The contributions to the one-loop free energies from these terms readily be evaluated and summarized as 
\be
		\G^{\sst (1)}_{{\rm Dih}^\pm_n}-
		\frac12\,\G^{\sst (1)}_{{\rm Cyc}_n}
		=\log R\,
			\begin{cases} \pm\,\frac{n}{2} & [n\ {\rm odd}] 
	\vspace{5pt}\\
	\frac{n}{2} & [n\ {\rm even}] \end{cases},
    \quad
	  E_{{\rm Dih}^\pm_n}-
	  \frac12\,E_{{\rm Cyc}_n}=
			\begin{cases} \pm\,\frac{3\,n}{32} & [n\ {\rm odd}] 
	\vspace{5pt}\\
	\frac{3\,n}{32} & 
	  [n\ {\rm even}] \end{cases}.
\ee
Combining with the cyclic result,
we obtain the one-loop free energies for the $n$-th order Regge Trajectory in $Sp(N)$ as
\begin{equation}\label{vacspn}
\G^{\sst (1)}_{{\rm Dih}^+_{n}} = n\,\log R\,,\qquad  E_{{\rm Dih}^+_{n}} =\frac{3\,n}{16},
\end{equation}
while for $SO(N)$ they are given by
\be\label{vacson}
		\G^{\sst (1)}_{{\rm Dih}^-_{n}}=
			\begin{cases} 0 & [n\ {\rm odd}] 
	\vspace{5pt}\\
	n\,\log R & 
	  [n\ {\rm even}] \end{cases},\qquad 
	  E_{{\rm Dih}^-_{n}}=
			\begin{cases} 0 & [n\ {\rm odd}] 
	\vspace{5pt}\\
	\frac{3\,n}{16} & 
	  [n\ {\rm even}] \end{cases}.
\ee

The total one-loop free energies are given 
formally as
\ba 
    &&\G^{\sst (1)}_{{\rm Dih}^+}=
    \log R\sum_{n=2}^\infty n\,,
    \qquad 
    E_{{\rm Dih}^+}=\frac{3}{16}\,
    \sum_{n=2}^\infty n\,,
    \nn
    && \G^{\sst (1)}_{{\rm Dih}^-}=
    2\log R\sum_{p=1}^\infty p\,,
    \qquad 
    E_{{\rm Dih}^-}=\frac{3}{8}\,
    \sum_{p=1}^\infty p\,,
\ea
where for the $SO(N)$ case we used $n=2p$\,.
The above involve clearly divergent sum
$\sum_{p=1}^\infty p$\,,
which has been regularized to zero 
in \cite{Beccaria:2014xda}.
Hence, if we use the same
regularization scheme, 
we would obtain
\begin{equation}\label{result}
 \G^{\sst (1)}_{{\rm Dih}^+}=-\log R \,,
 \qquad
 \G^{\sst (1)}_{{\rm Dih}^-} = 0\,,
 \qquad
 E_{{\rm Dih}^+} = 
-{3\over 16}\,,
\qquad 
E_{{\rm Dih}^-} = 0\,.
\end{equation}
We can rederive the same result
applying the CIRZ directly to the full
partition functions 
$\chi_{{\rm Dih}^\pm}$ \eqref{dihso}.
 However, we see that at $\beta=0$, the $\cN=4$ singleton partition function $Z_V$ equals 1. As a result, the 
 geometric series in \eqref{evenodd} are 
 divergent at that point. 
 To avoid this, we 
 introduce a factor $r^{n-2}$
 in summing $\chi_{{\rm Dih}^\pm_n}$
 from $n=2$ to $\infty$\,.
 With the regulator $r$, we find that
 the correction terms \eqref{evenodd} give
\ba
  & \G^{(1)}_{\mathrm{even}}=
  \frac12\,q(r)\,\log R\,,
  \qquad 
   & \G^{(1)}_{\mathrm{odd}} = \frac12
  \,(q(r)-1)\,\log R\,, \\
    & E_{\mathrm{even}} = \frac{3}{32}\,q(r)\,,
    \qquad 
    & E_{\mathrm{odd}} = \frac{3}{32} \left(q(r)-1\right),
\ea
with 
\be
    q(r)=\frac{r+1}{(r-1)^2}\,.
\ee 
This immediately yields \eqref{result}
in the $SO(N)$ case as the
function $q(r)$ simply cancels out.
In the $Sp(N)$ case, 
this function survives
and becomes singular in the $r\to 1$ limit.
By adopting the scheme `$q(1)=0$' 
we can again recover \eqref{result}\,.
We would however like to emphasize that the result obtained by naively using \eqref{evenodd} as the partition functions for $\cN=4$ is finite and different from the above. 
In particular, for the $Sp(N)$ matrix model the Casimir and vacuum energies are $-\frac{371}{1152}$ and 
$-\tfrac{1049}{648}\log R$ respectively, while for the $SO(N)$ matrix model 
they are $-\tfrac{3}{128}$
and $-\tfrac18\log R$.

\subsection{$U(N)\times U(M)$ Bi-Fundamental
and $O(N)\times O(M)$ Bi-Vector Models}\label{subsec: bifund}

We now turn to a computation of the one-loop vacuum and Casimir energies for the bulk duals of the free $U(N)\times U(M)$ bi-fundamental model, and next, the $O(N)\times O(M)$ bi-vector model. We will consider the cases where the fundamental field is either a scalar ($\chi_V=\chi_0$)
or a Majorana fermion ($Z_V=-\chi_{\frac12}$). 

For the case of the $U(N)\times U(M)$ bi-fundamental model, it turns out that the one-loop vacuum and Casimir energies are almost trivially zero. This is because  both $\chi_0(g)$ and $\chi_{\tfrac12}(g)$ are odd in $\b$, and hence $\chi_0(g^k)^2$ and $\chi_{\tfrac12}(g^k)^2$ are even in $\b$. Hence the functions $f_{\mathcal{H}|0}, f_{\mathcal{H}|1}$ and $f_{\mathcal{H}|2}$ are also even in $\beta$ and therefore possess no odd powers of $\beta$ in the small $\beta$ expansion. Hence the one-loop vacuum energy for the corresponding AdS theories are trivially zero. By a similar reasoning, the one-loop Casimir energy also vanishes.

\begin{figure}[h]
\centering
\begin{subfigure}{.5\textwidth}
  \centering
  \includegraphics[width=0.8\linewidth]{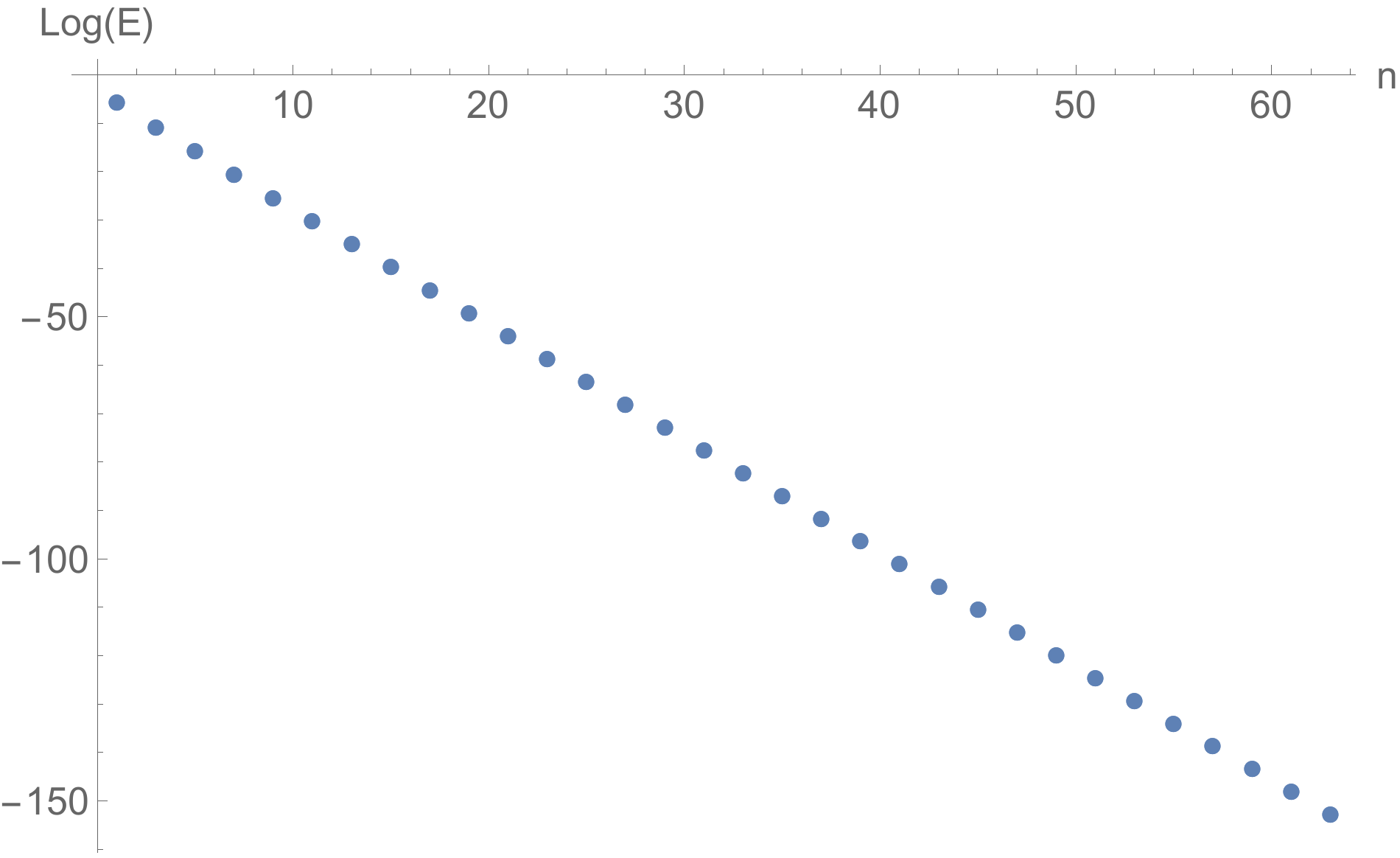}
 \label{fig:cas_scal_BV}
\end{subfigure}%
\begin{subfigure}{.5\textwidth}
  \centering
  \includegraphics[width=0.8\linewidth]{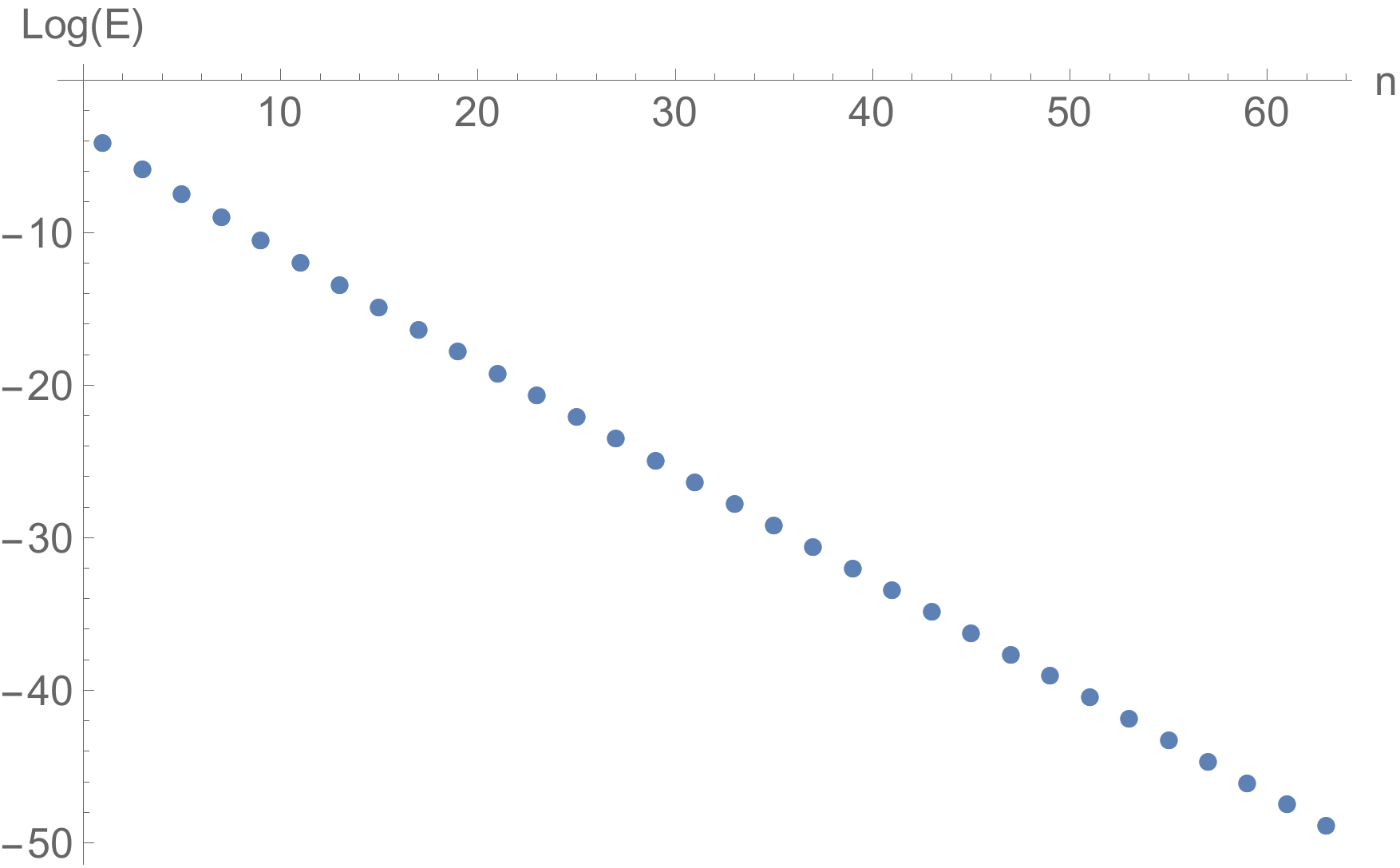}
 \label{fig:cas_ferm_BV}
\end{subfigure}
\begin{subfigure}{.5\textwidth}
  \centering
  \includegraphics[width=0.8\linewidth]{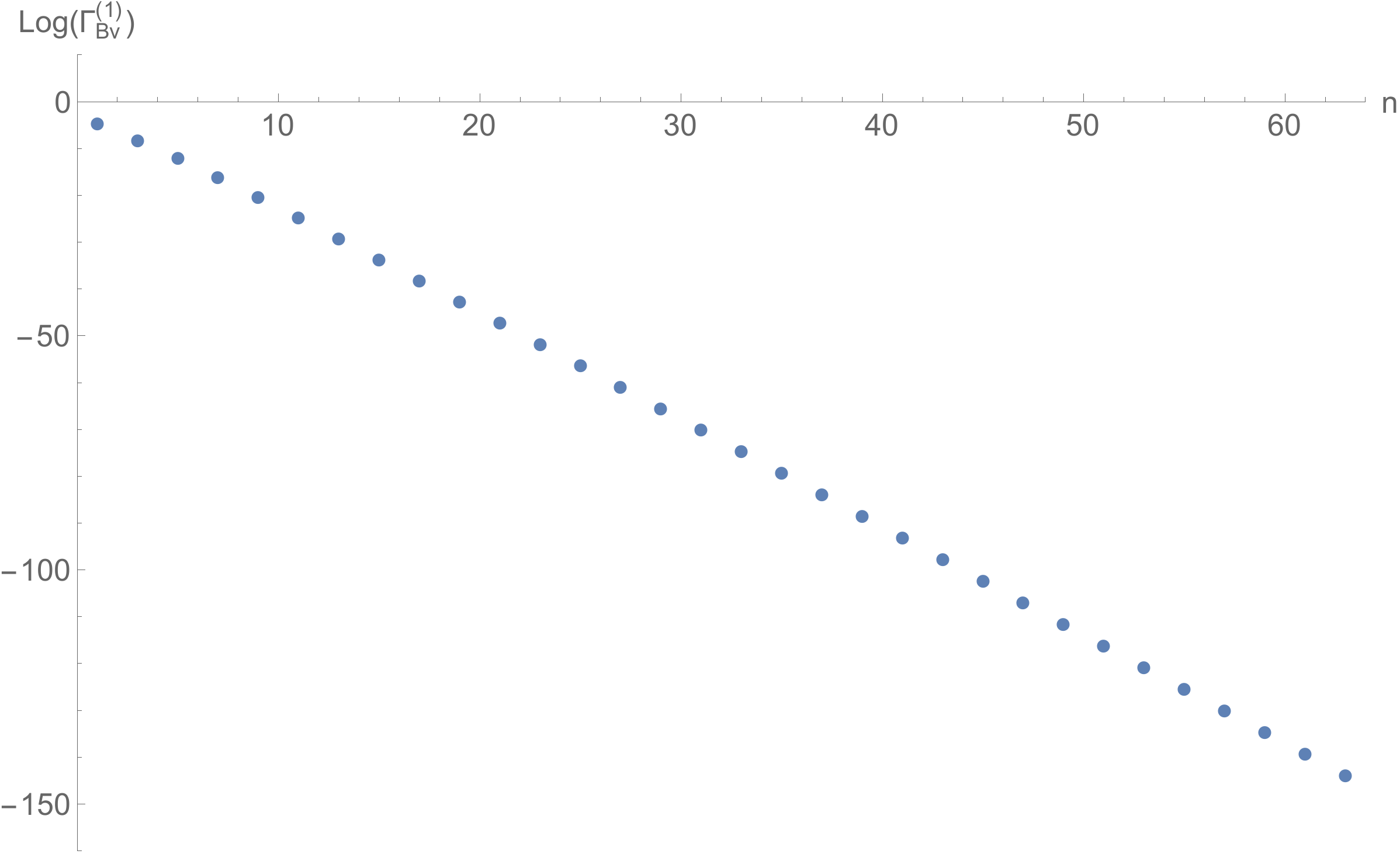}
  \label{fig:vac_scal_BV}
\end{subfigure}%
\begin{subfigure}{.5\textwidth}
  \centering
  \includegraphics[width=0.8\linewidth]{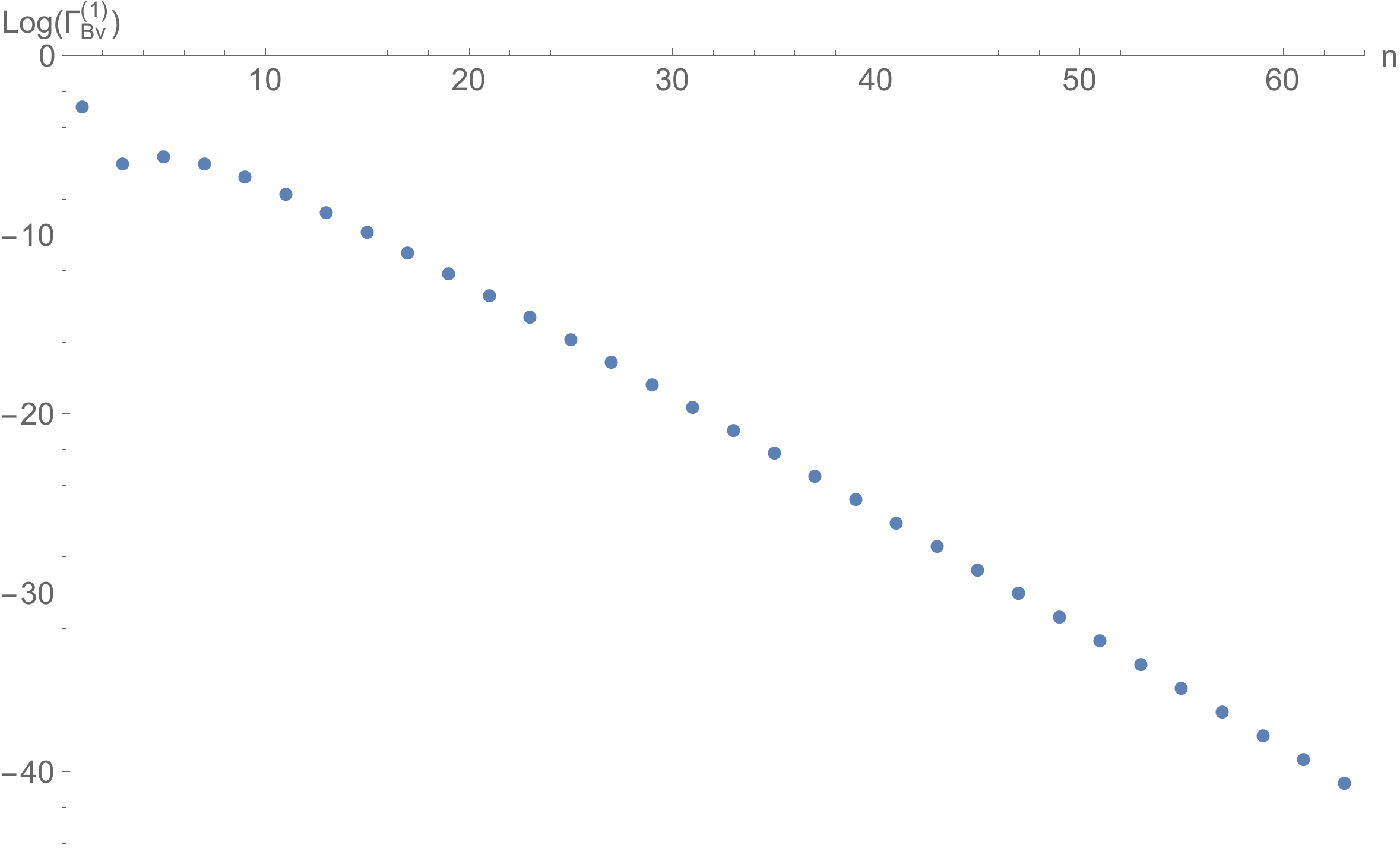}
 \label{fig:vac_ferm_BV}
\end{subfigure} 
\caption{The vacuum energy of the scalar bivector model (lower left) and fermion bivector model(lower right), and the Casimir energy of the scalar bivector model(upper left) and fermion bivector model(upper right).}
\label{fig:BV}
\end{figure}

Let us turn to the computation of the vacuum and Casimir energies for $O(N) \times O(M)$ bivector model. Working first at fixed values of $n$ (here $2n$ is the number of fields in single trace operators), we observe that as $n$ grows, the absolute value of the one-loop free energies
$E_{{\rm Bv}_n}$
and $\G^{\sst (1)}_{{\rm Bv}_n}$ rapidly decay for both scalar and fermion cases.
To understand better this decaying behavior,
we depict the log-plots of the Casimir and vacuum energies for scalar and fermion  bi-vector models in $n$ at Figure \ref{fig:BV}: the linear behavior
implies that the 
one-loop free energies exponentially decay to zero in $n$, for both scalar and fermion cases. 

When $N,M\to \infty$,
the full one-loop free energies
are the sum of all these results 
from $n=1$ to $\infty$.
But, due to the absence of analytic expressions,
 we cannot evaluate this sum.
 Instead, we can again apply the CIRZ method directly 
to the full partition
functions \eqref{bv}.
In the end, we find that the 
one-loop free energies of the bulk duals of $O(N) \times O(M)$ bivector model
are simply zero:
\be
    E_{\rm Bv}=0\,,
    \qquad \G^{\sst (1)}_{\rm Bv}=0\,,
\ee
both for the scalar and fermion cases.
As a final comment, we note that if 
only $N\to\infty$ while $M$ is kept finite,
the single trace operators
can involve only finite number of fields in a trace, hence possible value of $n$ is bounded above. An extreme case is 
the vector model with $M=1$ where
the only allowed value of $n$ is 2 
(remind that $2n$ is the number of fields
in a trace).
Therefore, for finite $M$,
the bulk dual theory will involve
only finite number of Regge trajectories
and the full one-loop free energies
will be a finite sum of $E_{{\rm Bv}_n}$
and $\G^{\sst (1)}_{{\rm Bv}_n}$.

\subsection{Symmetric Group}

\begin{figure}[h]
\centering
\begin{subfigure}{.5\textwidth}
  \centering
  \includegraphics[width=0.9\linewidth]{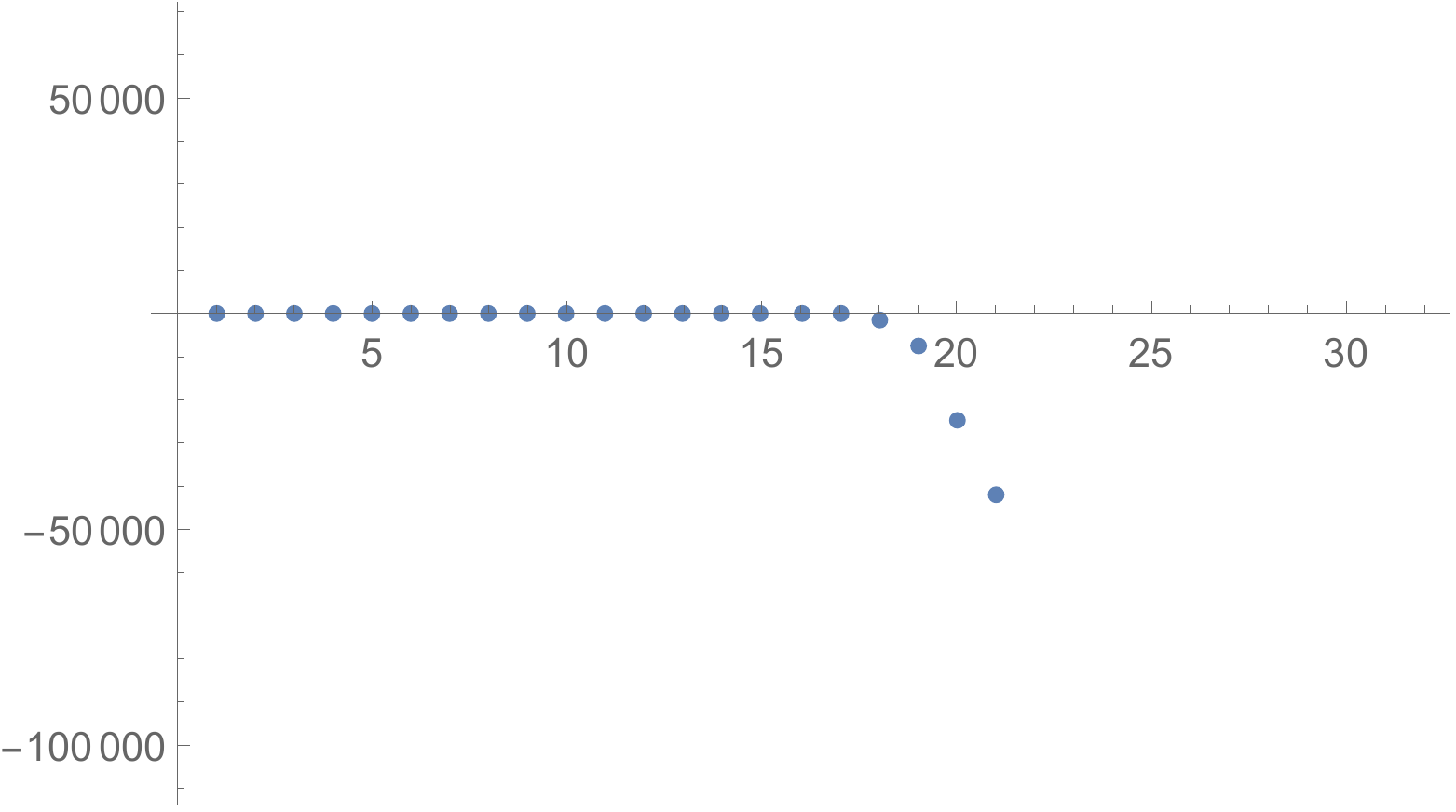}
 \caption{$E_{{\rm Sym}_n}$}
\end{subfigure}%
\begin{subfigure}{.5\textwidth}
  \centering
  \includegraphics[width=0.9\linewidth]{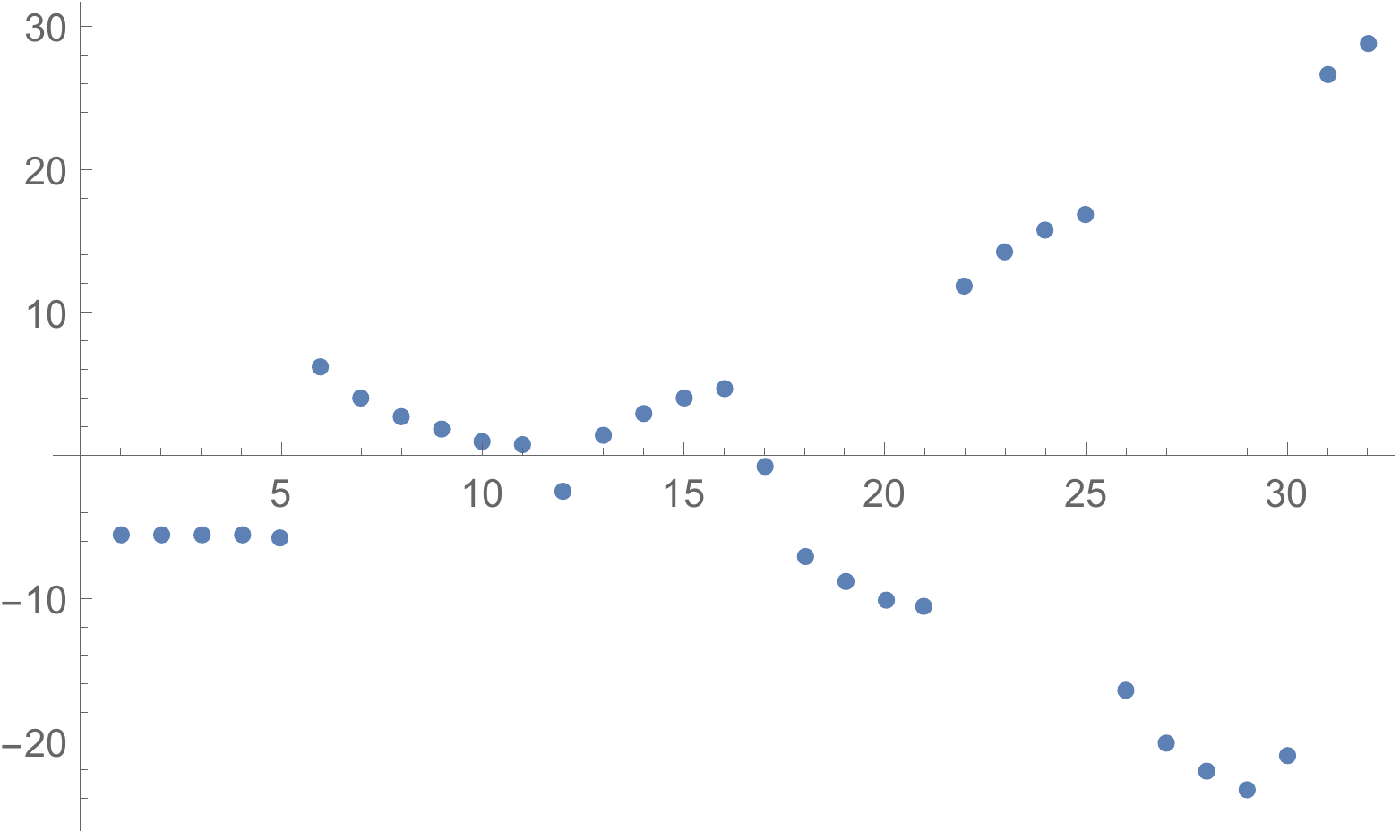}
  \caption{
 $f(E_{{\rm Sym}_n})$
 with $f(x)={\rm sign}(x)\,\log|x|$}
 \end{subfigure}
 \caption{Casimir energies
 for the first 32 results}
  \label{fig:sym}
\end{figure}

Let us consider
a toy AdS/CFT model
based on free scalar
with symmetric group,
even though it does not fit well in the standard picture on holography in many respects.
Putting the interpretation issues aside, let us simply provide
the result of Casimir energy
 computations. Using 
 the partition functions \eqref{sym n},
 we calculate first 32
Casimir energies
$E_{{\rm Sym}_n}$
and plot the result
in Figure \ref{fig:sym}.
We find the Casimir energy
has an oscillating behaviour
with 
exponentially growing oscillation amplitude. 

\section{Summary and Concluding Remarks}

In this paper we have computed the one-loop free energies for the holographic duals of a number of free CFTs. These results are summarized in Table \ref{table:1}. 
\begin{table}[h]
\centering
\begin{tabular}{||c | c |c| c||} 
 \hline
 Boundary CFT & Symmetry Group & Casimir Energy $E$ & Vacuum Energy $\G^{(1)}$ \\ [0.5ex] 
 \hline\hline
 \multirow{2}{8em}{Adjoint Scalar} & $Sp(N)$  & $\tfrac{27}{240}$ & $\tfrac{11}{20}$ \\ [1ex] \cline{2-4}
 & $SO(N)$  & $\tfrac{28}{240}$ & $\tfrac{101}{180}$ \\ [1ex]
 \hline
 \multirow{2}{8em}{Yang Mills} & $Sp(N)$  & $\frac{1377}{120}$ &  $\frac{791}{10}$ \\ [1ex] \cline{2-4} &
 $SO(N)$  & $\frac{1388}{120}$ &  $\frac{7181}{90}$ \\ [1ex]
 \hline \hline
 \multirow{2}{8em}{$\cN=4$ SYM} & $Sp(N)$  & $-\tfrac{3}{16}$ & $-1$ \\ [1ex] \cline{2-4} 
 &  $SO(N)$  & 0 & 0 \\
 \hline \hline
 Bi-fundamental Scalar & $U(N)\times U(M)$  & 0 & 0 \\ \hline
 Bi-vector Fermion & $U(N)\times U(M)$  & 0 & 0 \\ 
 \hline \hline
  Bi-vector Scalar  & $O(N)\times O(M)$  & 0 & 0 \\ \hline
 Bi-vector Fermion  & $O(N)\times O(M)$  & 0 & 0 \\
 \hline \hline
\end{tabular}
\caption{Summary of one-loop free energies computed for bulk duals of free matrix CFTs in the $N\to\infty$ limit. 
$\G^{\sst (1)}$ is in unit of $\log R$. The fermion is Majorana.}
\label{table:1}
\end{table}

We now briefly discuss the physical interpretation of these results in terms of matching the free energies across the bulk and the boundary theories. For definiteness, we will focus on vacuum energy in the scalar  $SO(N)$ adjoint model, though the discussion readily generalizes to the other dualities discussed above. The CFT free energy on $S^4$ has a logarithmic divergence corresponding to the $a$ anomaly, and is given by
\begin{equation}
 F_{{\rm CFT}} = {N\left(N-1\right)\over 2}\,{1\over 90}\,\log \Lambda\,,
\end{equation}
while the AdS free energy takes the form
\begin{equation}
 F_{{\rm AdS}} = \left(g^{-1}\mathcal{L}_0 +\frac{101}{180}+\mathcal{O}\!\left(g\right)\right)\log R\,,
\end{equation}
where $g$ is the bulk coupling constant. 
Then by  matching the free energies and using the correspondence between $\log \Lambda$ and $\log R$ we see that
\begin{equation}\label{ginvns4}
 g^{-1}  =  {N\left(N-1\right)\over 2}-{101\over 2}, \quad \mathcal{L}_0 = {1\over 90}\,.
\end{equation}
Hence, in contrast to the SU$(N)$ case, the one-loop shift in the definition of the coupling constant is by a half-integer, and not an integer. Further, if we apply a similar process to interpret the Casimir energy, we obtain
\begin{equation}\label{ginvns1s3}
 g^{-1}  =  {N\left(N-1\right)\over 2}+28.
\end{equation}
Note that this shift is different from that obtained for global
AdS$_5$ with $S^4$ boundary in
\eqref{ginvns4}. This is an interesting counterpoint to the
situation for higher-spin CFT
dualities as well as SU$(N)$ matrix CFT dualities the shift was the same in both backgrounds and was always by an integer amount.
It would be interesting to have a
better understanding of
\eqref{ginvns4} and
\eqref{ginvns1s3}. While it is
true that quantum effects are
sensitive to topologies, it is
puzzling that the AdS/CFT
dictionary itself can get altered
in a background dependent way. 
The results \eqref{ginvns4} and
\eqref{ginvns1s3} might be
indicating that putative AdS/CFT dualities involving the free $SO(N)$ and $Sp(N)$ adjoint scalar model
and free Yang Mills hold only in
the planar limit. It might also be possible that higher loop
corrections can alter this
discussion.

In this regard the situation for $\cN=4$ super Yang Mills is perhaps more satisfactory, once an appropriate regularization is adopted. In the $Sp(N)$ case we find the shift
\begin{equation}
 g^{-1} = {N(N+1)\over 2} -1
 =\frac{(N-1)(N+2)}2\,,
\end{equation}
while for the $SO(N)$ case it does not shift at all from the tree-level identification
\begin{equation}
 g^{-1} = {N(N-1)\over 2}.
\end{equation}
An additional curiosity regarding the dihedral matrix models is that for all the matter content that we have examined, be it the scalar, spin-1 or $\cN=4$ supersymmetric, the following relations hold:
\begin{equation}
 E_{{\rm Dih}^{-}} = E_{{\rm Dih}^{+}} + E_{{\rm singleton}},\qquad  \G^{(1)}_{{\rm Dih}^{-}} = \G^{(1)}_{{\rm Dih}^{+}} + \G^{(1)}_{{\rm singleton}}.
\end{equation}
Or in other words,  
the contribution of $\chi_{{\rm odd}(V)}$ to the one-loop free
energies 
is equal to  $-1/2$ times
the contribution of  $\chi_V$.

We finally turn to a summary of possible ambiguities in the application of the CIRZ method that we have so far only briefly discussed. These have to do with the presence of additional poles in the CIRZ integrands $f_{\cH|n}$ and the partition function $Z$ in the complex-$\beta$ plane when we work with the 
full partition function. Firstly, we note that due to the presence of terms of the form
\begin{equation}
 \log\left[1-\chi_V\left(g^k\right)\right], \quad {\rm and}\quad  \log\left[1-\chi_V\left(g^k\right)^2\right]
\end{equation}
the partition functions \eqref{dihso}, \eqref{bf} and \eqref{bv} contain branch points in $\beta$, one of which lies on the positive real axis\footnote{This branch point gives rise to the Hagedorn transition in CFTs with adjoint and bifundamental matter, see e.g. \cite{Sundborg:1999ue,Nishioka:2008gz}. This transition is absent in pure higher-spin/CFT dualities and is potentially an important distinguishing feature between higher-spin and tensionless string(y) theories \cite{Shenker:2011zf}.}. 

Since the reduction of the contours in \eqref{chi int} and \eqref{g H n} to small loops
around $\beta=0$ relies on the
integrands having no additional
singularities in the complex
$\beta$ plane,  the contribution
of these branch points 
to the $\beta$ integrals 
may need to be separately accounted for.
Next, as for the $SU(N)$ case we
also choose to apply the CIRZ
partition function on the
partition functions
$\chi_{\text{log},k}$ defined in
\eqref{logpart} \textit{at fixed k}. This is again because of the
singular points in $\beta$ contained in \eqref{logpart} at fixed $k$. In particular, if $\beta_c$ is a singular point for \eqref{logpart} at $k=1$, then $\tfrac{\beta_c}{k}$ is a singular point of $\chi_{{\rm Cyc}}$ in \eqref{dihso} at arbitrary integer values of $k$. 
Similar remarks apply to the singular points of \eqref{bf} and \eqref{bv}. These singularities would tend to cluster around $\beta=0$, making the partition function highly non-analytic in that neighbourhood.

In addition, the presence of this pole in the complex $\beta$ plane also introduces an additional ambiguity which we have so far not discussed in much detail. 
In particular, the summation of the geometric series carried out to evaluate \eqref{evenodd} and \eqref{bv} assumes that the absolute value of the character $\chi_V\left(g\right)$ is less than 1. If the absolute value is greater than 1, the summed expression may be used as an analytic continuation of the divergent geometric series.
However, this breaks down when $\vert\chi_V\left(g\right)\vert$ is equal to 1, and the resulting expressions for the partition functions \eqref{evenodd} such as diverge. 
This is what happens for the character of the $\cN=4$ Maxwell multiplet at $\beta=0$, and the
role of the regulation with $r$  that we carried out above is essentially to discard the contribution of this singularity to the contour integrals \eqref{chi int} and \eqref{g H n}.
Since the characters $\chi_0(g)$ and $\chi_1(g)$ become 1 at some positive real $\beta$, and not at $\beta=0$ we have implicitly carried out this prescription for the scalar matrix models and for free Yang Mills. 
The inclusion of this additional pole leads to contributions to the free energy which are not rational numbers.

\acknowledgments

We would like to thank Rajesh Gopakumar, Karapet Mkrtchyan, Zhenya Skvortsov and Mikhail Vasiliev for helpful discussions regarding various technical and conceptual points during the course of this and previous related work. We also thank the participants of the workshop ``New Ideas in Higher Spin Gravity and Holography" at Kyung-Hee University for helpful discussions and comments. SL thanks Kyung-Hee University for hospitality during the course of this work. 
The work of EJ was supported by the National Research Foundation of Korea through the grant NRF2014R1A6A3A04056670.
SL's research is supported by a Marie Sklodowska Curie Individual Fellowship 2014.

\appendix

\section{Harmonic Analysis on Spheres, Hyperboloids, and their Quotients}
\label{harmonic}
In this section we will review some essential facts about the spectrum of the Laplacian $\square = -g^{\mu\nu}\nabla_{\mu}\nabla_{\nu}$ on Anti-de Sitter space, focusing mostly on Euclidean AdS$_5$ and more generally AdS$_{2n+1}$. The extension to even dimensional subspaces has some subtleties which we mention below. Our starting point is the observation that Euclidean AdS$_5$ is the symmetric space $SO(5,1)/SO(5)$. From this it follows, see \cite{Camporesi:1990wm, Jaroszewicz:1991dd} for a review of these facts, that the spectrum of the Laplacian over arbitrary spin fields is determined in terms of the representation theory of $SO(5,1)$ and $SO(5)$. Further, since the results are more generally valid for all homogeneous spaces, we shall present the results for the general case, giving concrete examples along the way.

Firstly, given Lie groups $G$ and $H\subset G$, we define the coset space $G/H$ to be the set of equivalence classes of elements in $G$ obtained by the right action of the subgroup $H$, i.e.
\begin{equation}
 g_1 \equiv g_2\quad \text{if}\quad \exists\quad h\in H \quad\text{s.t.}\quad g_1 = g_2\cdot h.
\end{equation}
The set of all $g$ equivalent to an element $g_o$ under this relation is denoted by $g_oH$. Now $G$ is the principle bundle over $G/H$ with fibre isomorphic to $H$, and we can define the projection map from the bundle to the base space
\begin{equation}
  \pi\,:\,G\rightarrow G/H,\quad  \pi\left(g\right) = gH,\quad \forall\quad g\in G.
\end{equation}
Further we can also define a \textit{section} $\sigma\left(x\right)$, $x\in G/H$, through
\begin{equation}
 \sigma\,:\,G/H\rightarrow G,\quad \sigma\left(x\right) \in xH,
\end{equation}
i.e. $\sigma\left(x\right)$ is an element of the coset which contains $x$. Clearly, there is no canonical choice of section, and all sections are equivalent to each other upto right multiplication by $H$. Therefore, given two sections $\sigma_1$ and $\sigma_2$, at every $x_o\in G/H$, there exists an $h\in H$ such that $\sigma_1\left(x_o\right) = \sigma_2\left(x_o\right)\,h$. Then in this case the spin of a given field is fixed by specifying a UIR $S$ of $H$. Secondly, given a UIR $S$ of $H$, let us choose the set of all UIRs $R$ of $G$ that contain $S$. 
With these inputs, the eigenvalues of the Laplacian for a spin-$S$ field are given by
\begin{equation}\label{eigenv}
E_R^{(S)} = -{1\over a^2}\left(C_2\left(R\right)-C_2\left(S\right)\right),
\end{equation}
where $a$ is the AdS radius. The corresponding eigenfunctions are given by 
\begin{equation}\label{eigenf}
    \psi_a^I\left(x\right) = {1\over N_R^{(s)}}\left[\mathcal{U}^{R}\left(\sigma\left(x\right)^{-1}\right)\right]_{a}^{I}.
\end{equation}
Here $I$ is an index for the vector space carrying the representation $R$ and $a$ is an index for the vector space carrying the representation $S$. $N_R^{(s)}$ is a normalization constant, which we shall fix subsequently.
Notice that eigenfuctions carrying different values of $I$ for the same $R$ are necessarily degenerate. Hence the degeneracy of the eigenvalue $E_R^{(S)}$ is at least\footnote{The actual degeneracy can in principle be more as many representations $R$ can carry the same quadratic Casimir $C_2(R)$. We neglect this possibility below as it does not affect the subsequent analysis.} $d_R$. With these inputs, we may write the zeta function for the operator $-\square+\nu^2$ over a spin-$S$ field on the space $G/H$ as 
\begin{equation}\label{zetasps}
\zeta_{\nu,S}\left(z\right) = \sum_{R}\sum_{I}\sum_{a}{1\over \left(E_R^{(s)}+\nu^2\right)^z} \int_{G/H}\sqrt{g}\,d^dx\,\left[\psi_a^I\left(x\right)\right]^*\cdot \psi_a^I\left(x\right).
\end{equation}
Here the sum over $a$ is the dot product over local spin indices and the sum over $I$ is the sum over degenerate eigenvalues while the sum over $R$ is the sum over non-degenerate eigenvalues. Now, 
using \eqref{eigenf}, we have 
\begin{equation}\label{matmult}
\sum_{I}\left[\psi_b^I\left(x\right)\right]^*\cdot \psi_a^I\left(x\right) =\tfrac{1}{\vert N_R^{(s)}\vert^2}\delta_{a}^{b},
\end{equation}
which is independent of the point $x$ on the coset space\footnote{While we are working here with compact groups, this statement should hold equally well when we consider AdS$_5$ for which $G=SO(5,1)$. This is the group theoretic origin of the Equation \eqref{plancherel} which exploited the homogeneity of AdS$_5$ to define the Plancherel measure.}. Further, using the definition of degeneracy
\begin{equation}
\int\,d^{d}x\sqrt{g}\,\sum_{I,a}\left[\psi_a^I\left(x\right)\right]^*\cdot \psi_a^I\left(x\right) = d_R,
\end{equation}
we must have
\begin{equation}
{1\over\vert N_R^{(s)}\vert^2} = {d_R\over d_s}\,{1\over V_{G/H}}.
\end{equation}
The zeta function \eqref{zetasps} is therefore perfectly consistent with the original definition
\begin{equation}
\zeta_{\nu,S}\left(z\right) = \sum_{R} {d_R\over \left(E_R^{(s)}+\nu^2\right)^z}.
\end{equation}
It should be immediately apparent, however, that the above procedure is at first sight ill-defined for the case of hyperboloids like AdS$_5$. In this specific example $G$ is $SO(5,1)$, and its unitary representations are necessarily infinite dimensional. Further, the volume $V_{G/H}$ is infinite. For this reason, a slight modification of the above computation is adopted. We note from \eqref{matmult} that the quantity $\sum_{I}\left[\psi_b^I\left(x\right)\right]^*\cdot \psi_a^I\left(x\right)$ is independent of the point $x$. As a result, it is possible to define the \textit{coincident} zeta function
\begin{equation}
\zeta^{coin}_{\nu,S}\left(z\right) = \sum_{R}\sum_{I}\sum_{a}{1\over \left(E_R^{(s)}+\nu^2\right)^z} \left[\psi_a^I\left(x\right)\right]^*\cdot \psi_a^I\left(x\right),
\end{equation}
such that
\begin{equation}
\zeta_{\nu,S}\left(z\right) = V_{G/H}\,\zeta^{coin}_{\nu,S}\left(z\right).
\end{equation}
We shall now recapitulate the evaluation of the coincident zeta function of fields on AdS$_5$ by means of analytic continuation from S$^5$. For definiteness we shall focus on the Laplace operator $-\square$ but the discussion easily generalized to arbitrary $\nu$. Firstly, we specify the spin $S$ of the field by the UIR of $SO(5)$ that it carries. This representation is in turn specified by the quantum numbers $S = \left(s_1,s_2\right)$, where
\begin{equation}
s_1\geq s_2\geq 0.
\end{equation}
To solve for the zeta function on $S^5$, we consider all UIRs $R$ of $SO(6)$ that contain $S$ when restricted to $SO(5)$. Firstly, UIRs of $SO(6)$ are labelled by the triplet $\left(\ell,m_1,m_2\right)$ where
\begin{equation}
\ell\geq m_1\geq \vert m_2\vert,
\end{equation}
and $\ell,m_1,m_2$ are all simultaneously integers or half-integers. The dimension of such a representation $R$ is given by 
\begin{equation}
d_R = {\left(\ell+2\right)^2 - \left(m_1+1\right)^2\over 3}{\left(\ell+2\right)^2 - m_2^2\over 4}{\left(m_1+1\right)^2 - m_2^2\over 1}.
\end{equation}
This representation $R$ contains $S$ provided
\begin{equation}\label{branching}
\ell\geq s_1\geq m_1\geq s_2\geq \vert m_2\vert.
\end{equation}
We shall however consider the case that the given field satisfies irreducibility conditions, such as transversality and tracelessness, where some of the inequalities in \eqref{branching} saturate to yield 
\begin{equation}\label{sttgen}
\ell\geq s_1= m_1\geq s_2= \vert m_2\vert.
\end{equation}
Further, the eigenvalue of the Laplacian is determined from \eqref{eigenv} to be
\begin{equation}
E_{\ell}^{(s_1,s_2)}= \left(\ell+2\right)^2 -\left(s_1+s_2\right)-4,
\end{equation}
and as a result, the coincident zeta function on a five-sphere of unit radius is given by
\begin{equation}
\zeta^{coin}_{(s_1,s_2)} = \tfrac{1}{12\,V_{S^5}}\sum_{\ell\geq s_1} {\left[\left(\ell+2\right)^2 - \left(s_1+1\right)^2\right]\left[\left(\ell+2\right)^2 - s_2^2\right]\left[\left(s_1+1\right)^2 - s_2^2\right]\over \left(\left(\ell+2\right)^2 -\left(s_1+s_2\right)-4\right)^z}.
\end{equation}
Next, we move to the case of AdS$_5$ where to carry out the above procedure we need to enumerate all UIRs of $SO(5,1)$ which contain $S$ when restricted to $SO(5)$. UIRs of $SO(5,1)$ are labelled by the triplet $R=\left(i\lambda,m_1,m_2\right)$, where $\lambda\in \mathbb{R}_+$ and $m_1\geq \vert m_2\vert \geq 0$ and contain $S$ provided that
\begin{equation}\label{adsbranching}
s_1\geq m_1\geq s_2\geq \vert m_2\vert.
\end{equation}
Further, we shall apply the same irreducibility conditions on the field as we did on S$^5$ to saturate some inequalities in \eqref{adsbranching}. In particular, we take $m_1=s_1$ and $\vert m_2\vert =s_2$. It was explicitly demonstrated in \cite{Camporesi:1990wm,Camporesi:1993mz,Camporesi:1994ga,Camporesi:1995fb,Camporesi} that for a wide class of fields, the coincident zeta function in AdS$_5$ may be computed from the corresponding S$^5$ answer by means of the following analytic continutation
\begin{equation}
\ell +2 \mapsto i\lambda,\quad \lambda \in\mathbb{R}_+.
\end{equation}
As a result the coincident zeta function on AdS$_5$ is given by
\begin{equation}\label{ads5coin}
\zeta^{coin}_{(s_1,s_2)} = \tfrac{1}{12\,V_{S^5}}\int_0^\infty\,d\lambda\, {\left[\lambda^2 + \left(s_1+1\right)^2\right]\left[\lambda^2 + s_2^2\right]\left[\left(s_1+1\right)^2 - s_2^2\right]\over \left(\lambda^2 +\left(s_1+s_2\right)+4\right)^z}.
\end{equation}
We now provide an explicit example of computing the coincident zeta function on AdS$_5$ for a scalar field without using the analytic continuation proposed above. Also, as the analytic continuation as presented above seems somewhat abstract, we shall use this example as an explicit setting to demonstrate how this continuation works in practice.
\subsection*{The Scalar on AdS$_5$ and S$^5$}
We begin with noting that the metric on S$^N$ of unit radius in spherical polar coordinates
\begin{equation}
ds^2_{S^N} = d\chi^2 + \sin^2\chi\,ds^2_{S^{N-1}}
\end{equation}
is related to the metric on the corresponding hyperbolic space AdS$_N$ or $H^N$
\begin{equation}
ds^2_{H^N} = dy^2 + \sinh^2 y\,ds^2_{S^{N-1}}
\end{equation}
\textit{via} $\chi=iy$. The Laplace eigenvalue equation for scalar fields on $S^N$ is given by
\begin{equation}
    \square \,\varphi = \ell\left(\ell+N-1\right)\varphi,
\end{equation}
and its solutions are given in terms of hypergeometric functions as
\begin{equation}\label{sneigenf}
    \varphi_{\ell m\sigma} = \left(\sin\chi\right)^{m} {}_2F_1\left(\ell+m+N-1,m-\ell,m+\tfrac{N}{2};\sin^2\tfrac{\chi}{2}\right) Y_{m\sigma}.
\end{equation}
The hypergeometric function of the second kind is ruled out by requiring smoothness at $\chi=0$ while requiring the eigenfunctions to be smooth at $\chi=\pi$ restricts $m=0,1,\ldots,\ell$. The $Y_{m\sigma}$ solve the Laplace equation on $S^{N-1}$ with eigenvalue $m\left(m+N-2\right)$. Then the eigenfunctions of the scalar Laplace equation on $H^N$
\begin{equation}
 \square \,\varphi = \left(\lambda^2+\rho^2\right)\varphi,\quad \rho=\tfrac{N-1}{2},
\end{equation}
are obtained by making the replacements $\chi = iy$ and $\ell+\rho = i\lambda$ in \eqref{sneigenf}. We therefore have
\begin{equation}\label{hneigenf}
    \varphi_{\lambda m\sigma} = N_{\lambda m}\,\left(i\sinh y\right)^{m} {}_2F_1\left(i\lambda +\rho \ell+m,-i\lambda +m+\rho,m+\tfrac{N}{2};-\sinh^2\tfrac{y}{2}\right) Y_{m\sigma},
\end{equation}
where $N_{\lambda m}$ is an overall constant. Notice that these are purely local solutions on which the only boundary condition that has been imposed is regularity at $y=0$. Next, demanding that the eigenfunctions be square integrable and a complete set fixes $\lambda$ to be real and positive, while demanding that they be Dirac delta normalized fixes
\begin{equation}
    N_{\lambda m} = \left(2^{N-2}\over \pi\right)^{1/2}\left\vert \sqrt{\pi}\Gamma\left(i\lambda+(N-1)/2+m\right)\over 2^{N+m-2}\Gamma\left(i\lambda\right)\Gamma\left(m +N/2\right)\right\vert.
\end{equation}
Note that $\ell\mapsto i\lambda-\rho$ for $N=5$ is precisely the analytic continuation used above from AdS$_5$ to $S^5$. The coincident zeta function is therefore given by
\begin{equation}
\zeta^{coin}\left(z\right) = \int_0^{\infty}\,d\lambda\,\sum_{m\,\sigma} {\varphi_{\lambda m\sigma}^*\varphi_{\lambda m\sigma}\over \left(\lambda^2+\rho^2\right)^z}.
\end{equation}
We choose to evaluate this quantity at $y=0$ where the eigenfunction $\varphi_{\lambda m\sigma}$ vanishes unless $m=0$. Further, $Y_{m\sigma}$ for $m=0$ is just the constant mode on $S^{N-1}$, given by $\left\vert V_{S^{N-1}}\right\vert^{-1/2}$ for reasons of normalization, and the sum over $\sigma$ is also then trivial. We finally obtain
\begin{equation}
\sum_{m\,\sigma}\varphi_{\lambda m\sigma}^*\varphi_{\lambda m\sigma} = {1\over 2^{N-2}}\,\left\vert \Gamma\left(i\lambda+(N-1)/2\right)\over \Gamma\left(i\lambda\right)\Gamma\left(N/2\right)\right\vert^2\, \left\vert\frac{\Gamma(N/2)}{2\pi^{N/2}}\right\vert.
\end{equation}
For the case of $N=5$ we obtain
\begin{equation}
\sum_{m\,\sigma}\varphi_{\lambda m\sigma}^*\varphi_{\lambda m\sigma} = {1\over 12\pi^3} \lambda^2\left(\lambda^2+1\right).
\end{equation}
Hence the coincident zeta function for scalar fields on AdS$_5$ is given by 
\begin{equation}
\zeta^{coin}\left(z\right) ={1\over 12\pi^3} \int_0^{\infty}\,d\lambda\,{\lambda^2\left(\lambda^2+1\right)\over \left(\lambda^2+4\right)^z}.
\end{equation}
This matches perfectly with \eqref{ads5coin} which was obtained by analytic continuation if we set $s_1=0$ and $s_2=0$ there.
\subsection{Zeta Functions of  AdS$_5$ Fields}\label{fronsdaltads}
To evaluate the partition function of the bulk theory to one-loop order it is sufficient to consider the quadratic action for the fields about $AdS_5$. For the case of massless symmetric rank-$s$ fields, this is given by the Fronsdal action 
\begin{equation}
\label{action}
S\left[\phi_{\left(s\right)}\right]=\int\,d^Dx\sqrt{g} \phi^{\mu_1\ldots\mu_s}\left(\cfhat_{\mu_1,\ldots,\mu_s}-\frac{1}{2}{\gsym[\cfhat]}_\lambda\,^\lambda\right),
\end{equation}
where
\begin{equation}
\label{fhat}
\cfhmus=\cfmus-\frac{s^2+\left(D-6\right)s-2\left(D-3\right)}{\ell^2}\phi_{\mu_1\ldots\mu_s}-\frac{2}{\ell^2}{\gsym[\phi]}_\lambda\,^\lambda,
\end{equation}
and
\begin{equation}
\label{f}
\cfmus=\Delta\phi_{\mu_1\ldots\mu_s}-\nabla_{(\mu_1}\nabla^{\lambda}\phi_{\mu_2\ldots\mu_s)\lambda}+\frac{1}{2}\nabla_{(\mu_1}\nabla_{\mu_2}{\phi_{\mu_3\ldots\mu_s)\lambda}}^\lambda.
\end{equation}
The expressions are true for $AdS_D$ though we shall explicitly consider the case of $D=5$ only.
It may be shown that this action is invariant under the gauge transformation 
\begin{equation}
\label{gaugetrs}
\phi_{\mu_1\ldots \m_s}\mapsto \phi_{\m_1\ldots \m_s}+\nabla_{(\m_1}\xi_{\m_2\ldots \m_s)}.
\end{equation}
For a consistent description, it turns out that the fields $\phi$ necessarily satisfy  a double-tracelessness constraint $\phi_{\m_1\ldots \m_{s-4}\nu\rho}\,^{\nu\rho}=0$ for $s\geq 4$ and $\xi$ satisfies a tracelessness constraint $\xi_{\m_1\ldots \m_{s-3}\nu}\,^{\nu}=0$. It is then straightforward to evaluate the functional integral
\begin{equation}
\label{pathint}
Z^{\left(s\right)}=\frac{1}{\text{Vol(gauge group)}}\int \left[D\phi_{\left(s\right)}\right]e^{-S\left[\phi_{\left(s\right)}\right]},
\end{equation}
to obtain the partition function as a ratio of one loop determinants evaluated over symmetric transverse traceless (STT) fields
\begin{equation}
\label{determinants}
\ZZ^{(s)}={\left[\det\left(-\square-\frac{\left(s-1\right)\left(3-D-s\right)}{\ell^2}\right)_{(s-1)}\right]^{\frac{1}{2}}\over \left[\det\left(-\square+\frac{s^2+\left(D-6\right)s-2\left(D-3\right)}{\ell^2}\right)_{(s)}\right]^{\frac{1}{2}}}.
\end{equation}
The subscripts indicate that the numerator is evaluated over rank $s-1$ STT fields and the denominator is evaluated over rank $s$ STT fields. The numerator is the ghost determinant that arises from gauge fixing the freedom \eqref{gaugetrs}. These determinants may be evaluated over quotients of AdS space using the techniques of \cite{Gopakumar:2011qs}. In the specific case of AdS$_5$ where we turn on a temperature $\beta$ as well as chemical potentials $\alpha_1$ and $\alpha_2$ for the $SO(4)$ Cartans, we find that the partition function is given by
\begin{equation}
\label{refinedpart}
\log\ZZ^{(s)} = \sum_{m=1}^\infty {1\over m}{e^{-m\beta \left(s+2\right)}\over \vert 1-e^{-m\left(\beta-i\phi_1\right)}\vert ^2 \vert 1-e^{-m\left(\beta-i\phi_2\right)}\vert ^2} \left[\chi^{SO\left(4\right)}_{\left(s,0\right)}-\chi^{SO\left(4\right)}_{\left(s-1,0\right)}e^{-m\beta}\right],
\end{equation}
and $\phi_1=\a_1+\a_2$, $\phi_2=\a_1-\a_2$. The reader will recognize this as the partition function $\tr\left(e^{-\beta\,H-\alpha^i J_i}\right)$ computed in the \textit{grand canonical} ensemble for the conformal primary with highest weights $\left(s+2,\tfrac{s}{2},\tfrac{s}{2}\right)$. It was observed in \cite{Lal:2012ax} that it is possible to formally invert this procedure, i.e., given a character of the conformal algebra, one may infer the spectrum of the corresponding kinetic operator which gives rise to the corresponding grand canonical partition function.

\section{Unitary Irreducible Representations of the $so(2,4)$ Algebra}
\label{app: so24rep}

Fields in AdS$_5$ carry quantum numbers under $so(2,4)$, 
the isometry algebra of AdS$_5$, and fall into its Unitary Irreducible Representations (UIRs). A necessary condition for any AdS/CFT duality is that for every field in AdS$_5$, there is a `dual' operator in the CFT$_4$  carrying the same $so(2,4)$ quantum numbers as the AdS$_5$ field. We therefore review very briefly the set of UIRs of $so(2,4)$. These have been extensively explored in \cite{mack1977all,Dolan:2005wy} and particularly accessible accounts are available in \cite{Minwalla:1997ka,Barabanschikov:2005ri}. 

The starting point is 
the Verma module $\cV(\D,(\ell_1,\ell_2))$ of $so(2,4)$\,. The numbers $\D$ and $(\ell_1,\ell_2)$ label
the irreps of $so(2)\oplus so(4)$ subalgebra carried by the lowest weight state of the module.
Since $so(4)\simeq su(2)\oplus su(2)$, we shall also
use the $su(2)\oplus su(2)$ label $[j_+,j_-]$.
The two labels are simply related by 
$j_\pm=(\ell_1\pm \ell_2)/2$\,.
The Verma module $\cV(\D,(\ell_1,\ell_2))$ is unitary and irreducible when 
$\D$ is greater than the critical dimension $\D_{\ell_1,\ell_2}$\,,
which will be introduced shortly in below.

\begin{itemize}
\item

The UIR belonging to the interior the unitary region of $\D$
is referred as to \emph{long representations}.
They can be realized as 
higher-spin operators in CFT$_4$
 or as massive higher-spin fields in AdS$_5$
with the mass-squared
given by $M^2=[\D(\D-4)-\ell_1-\ell_2]/L^2$
($L$ is the radius of AdS$_5$) \cite{Metsaev:1995re,Metsaev:2004ee,Metsaev:2014sfa}.
\end{itemize}

In the critical cases lying on the boundary of 
the unitary region, the Verma module develops an invariant subspace
and an UIR can be obtained by quotienting the Verma module
with the invariant subspace.
These `critical' representations are again divided into two groups,
\emph{semi-short} and \emph{short}, depending on $(\ell_1,\ell_2)$\,.
\begin{itemize}
\item
The semi-short UIR appears when $\ell_1\neq \pm\ell_2$
and $\D$ reaches its critical value $\D_{\ell_1,\ell_2}=\ell_1+2$\,,
and the UIR is given by the quotient,
\be 
\cD(\ell_1+2,(\ell_1,\pm \ell_2))
=\cV(\ell_1+2,(\ell_1,\pm \ell_2))/\cV(\ell_1+3,(\ell_1-1,\pm \ell_2))\,.
\ee
The semi-short representations 
can be realized as conserved current operator in CFT$_4$
or massless (mixed-symmetry) higher-spin fields in AdS$_5$\,.
\item
The short representation arises when $|\ell_2|=\ell_1$
and $\Delta_{\ell_1,\pm \ell_1} = \ell_1+1$\,. 
The invariant subspace of the Verma module appearing in this case is a semi-short representation, hence again contains an invariant subspace.
Therefore, the UIR is given by a `double' quotient,
\ba
\cD(\ell+1,(\ell,\pm\ell))\eq \cV(\ell+1,(\ell,\pm\ell))/\cD(\ell+2,(\ell,\pm(\ell-1)))\,,
\nn
\cD(\ell+2,(\ell,\pm(\ell-1)))\eq \cV(\ell+2,(\ell,\pm(\ell-1)))/\cV(\ell+3,(\ell-1,\pm(\ell-1)))\,.
\ea
Differently from the long and semi-short representations, the short representations cannot be realized as a propagating AdS$_5$ field
but only as a conformal field operator in CFT$_4$\,.
\end{itemize}

A convenient way to treat the representation spaces, in particular
their tensor products and decompositions, is to use the Lie algebra character.
In case of $so(2,4)$, it is given by the trace,
\be
    \chi_\cR(q,x_+,x_-)=\tr_\cR\left(q^{D}\,{x_+}^{J_+}\,{x_-}^{J_-}\right),
    \label{char cartan}
\ee 
over a representation space $\cR$\,. Here $D$, $J_\pm$ are
the Cartan subalgebra of $so(2,4)$ and 
corresponds to the subalgebra $so(2)$ and $su(2)\oplus su(2)$\,.
Since all the UIRs are given in term of the Verma module.
Their characters can be also expressed by the Verma module character.
It is given by the following simple function,
\begin{equation}\label{long}
\chi_{\D,[j_+,j_-]}\!\left(q,x_+,x_-\right)
=q^{\Delta}\,P(q,x_+,x_-)\,\chi_{j_+}\!\left(x_+\right)\chi_{j_-}\!\left(x_-\right),
\end{equation}
where $\chi_j$ is the character of the spin-$j$ representation of $su(2)$ 
taking the form,
\begin{equation}
\chi_{j}\!\left(x\right)= \frac{x^{j+\frac12}-x^{-j-\frac12}}{x^{\frac12}-x^{-\frac12}}\,,
\end{equation}
and $P(q,x_+,x_-)$ is given by 
\be
	P(q,x_+,x_-)
	=\frac1{\left(1-q\,x_+^{\frac12}\,x_-^{\frac12}\right)
	\left(1-q\,x_+^{-\frac12}\,x_-^{\frac12}\right)
	\left(1-q\,x_+^{\frac12}\,x_-^{-\frac12}\right)
	\left(1-q\,x_+^{-\frac12}\,x_-^{-\frac12}\right)}.
\ee
In the case of long representations, its character
is simply that of Verma module.
About the semi-short and short representations, 
the character is given by the difference,
\ba
\chi_{\cD(\ell_1+2,(\ell_1,\ell_2))}\eq \chi_{\ell_1+2,(\ell_1,\ell_2)}-
\chi_{\ell_1+3,(\ell_1-1,\ell_2)}\,,
\label{semi-short char}\\
\chi_{\cD(\ell+1,(\ell,\pm\ell))}\eq
\chi_{\ell+1,(\ell,\pm\ell)}
-\chi_{\ell+2,(\ell+1,\pm(\ell-1))}+\chi_{\ell+3,(\ell-1,\pm(\ell-1))}\,.
\label{short char}
\ea
It is often useful to work with $(\b,\a_+,\a_-)$ variables
which are related to $(q,x_+,x_-)$ as $q=e^{-\b}$ and $x_\pm=e^{i\,\a_\pm}$. Interpreting the character as a partition function, 
the variables
$\b$ and $\a_\pm$
would correspond to the inverse temperature and two angular chemical potentials.

As a final remark we explicitly evaluate \eqref{short char} for the case of the the boundary singletons that we need. We obtain
\be\label{shortscalar}
\chi_{0}(g) = \chi_{\mathcal{D}(1,(0,0))} = \frac{ e^{\beta}-e^{-\beta }}{4 \left(\cos\alpha_1-\cosh\beta\right) \left(\cos\alpha_2-\cosh \beta \right)}
\ee
for the scalar, 
\be\label{shortmajo}
\chi_{\tfrac12}(g) = \chi_{\mathcal{D}(\frac{3}{2},(\frac{1}{2},\frac{1}{2}))} + \chi_{\mathcal{D}(\frac{3}{2},(\frac{1}{2},-\frac{1}{2}))} = \frac{ \left(e^{\frac{\beta}{2} }-e^{-\frac{\beta}{2}}\right) \cos \frac{\alpha_1}{2}\,\cos \frac{\alpha_2}{2}}{\left(\cos\alpha_1-\cosh\beta\right) \left(\cos\alpha_2-\cosh \beta \right)}
\ee
for the fermion, and
\begin{equation}\label{chi1}
\chi_{1}(g) = \frac{e^{-2 \beta } \left(-2 e^{\beta } \left(\cos\alpha_1 +\cos\alpha_2\right)+e^{2 \beta } (2 \cos\alpha_1 \cos\alpha_2+1)+1\right)}{2 \left(\cos\alpha_1-\cosh\beta\right) \left(\cos\alpha_2-\cosh \beta \right)}
\end{equation}
for the spin-1 field. These expressions are useful for the computations in the main text.

\section{Vector Models}\label{hsrev}

Let us first consider the one-loop vacuum energy of the higher spin gravities in AdS$_5$\,,
which are dual to free vector model CFTs in four dimensions.
In even boundary dimensions, we have more possibilities of free CFT as 
there are infinitely many singleton representations.
In four boundary dimensions, they correspond simply to 
the massless spin $j$ representations.
Here, we consider the $j=0, 1/2$ and  1 cases whose AdS duals
are referred to as the type A, B and C higher spin theories, respectively.
The other CFTs with $j>1$ are rather
exotic as they do not have local stress tensor (hence the dual higher spin theory contains gravity inside).
See \cite{Bae:2016xmv} for the general $j$ cases.

In each of these type A, B, C cases, we have again two different higher spin theories 
depending on their spectrum contains
only even spin fields or
all integer spin fields.
And these models are referred to as minimal or non-minimal
and correspond in boundary to $O(N)$ or $U(N)$ free CFT, respectively. 
Practically they are distinguished whether the two fields in a bilinear operator are symmetric.
The non-minimal model does not enjoy such symmetry whereas the minimal model does so.
This symmetry of the bilinear CFT operators can be simply reflected 
at the level of character.
The space of bilinear operators without any symmetry, hence 
giving the $U(N)$-model operator spectrum, can be obtained from
\be
    \chi_{U(N)}(\b,\a_1,\a_2)=\left[\chi_{sing}(\b,\a_1,\a_2)\right]^2\,.
    \label{U(N)}
\ee 
About the $O(N)$ model, the space of bilinear operators symmetric in two fields can be obtained from
\be
    \chi_{O(N)}(\b,\a_1,\a_2)=\tfrac{1}{2}\left[\chi_{sing}(\b,\a_1,\a_2)^2 + \left(-1\right)^{F_{sing}}\chi_{sing}\left(2\b,2\a_1,2\a_2\right)\right]\,.
    \label{O(N)}
\ee
Here $F_{sing}$ is the fermion number associated with the singleton representation. Following standard conventions, it is 0 for bosonic fields, and 1 for fermionic fields.
By AdS/CFT dualities, the contents in \eqref{U(N)} and \eqref{O(N)} correspond
to AdS$_5$ fields in non-minimal and minimal theories, respectively.


With the above inputs, we will now use the techniques reviewed in Section \ref{CIRZ} to compute the one-loop vacuum energy for higher-spin theories in global AdS$_5$ and the one-loop Casimir energy in thermal AdS$_5$. The singletons for type A, type B and type C dualities have the lowest weights $[\D,j_+,j_-]$,
\be
    {\rm A}\,:\,[1,0,0]\,,
    \qquad
    {\rm B}\,:\,[\tfrac32,\tfrac12,0]
    \oplus [\tfrac32,0,\tfrac12]\,,
    \qquad
    {\rm C}\,:\,
    [2,1,0]\oplus[2,0,1]\,,
\ee 
respectively, and the corresponding characters can be determined from \eqref{short char}. We will first focus on the simpler case of Casimir energies.

\subsection*{Casimir Energies}

Firstly, we note that though singletons do not represent propagating degrees of freedom, formally their partition function may be evaluated through a one-loop determinant in AdS$_5$ and the answer matched with the CFT result. In particular, using the prescription of \eqref{casimirexp} and Laurent expanding the characters $(-1)^{F_{sing}}\chi_{sing}\left(\b,0,0\right)$ about $\b=0$ and picking $-\tfrac12$ times the $\cO\left(\b\right)$ coefficient, we find
\begin{equation}\label{ecsing}
E_{c;(0)} = \frac1{240}\,,\quad E_{c;(\tfrac12)} = \frac{17}{960}\,,\quad E_{c;(1)} = \frac{11}{120}\,,
\end{equation}
where the subscript $(s)$ reminds us that this is the Casimir energy of a spin-$s$ singleton. Next, the Casimir energies of the corresponding bulk non-minimal and minimal higher-spin theories may be found using the partition functions \eqref{U(N)} and \eqref{O(N)} respectively. We find that for the non-minimal version of all three dualities the Casimir energy vanishes \cite{Bae:2016hfy}
\begin{equation}\label{ecnonmin}
E^{\,\text{non-min}}_{c;(A/B/C)} = 0.
\end{equation}
In contrast, for the minimal cases \cite{Bae:2016hfy}
\begin{equation}\label{ecmin}
E^{\,\text{min}}_{c;(A/B/C)} = E_{c;(0/\tfrac12/1)} \quad \text{respectively}\,.
\end{equation}
We now review the argument of \cite{Giombi:2014yra} for interpreting the non-vanishing result as a shift in the dictionary between the bulk coupling constant $g$ and $N$. In particular, it turns out that the total Casimir energy in the boundary theory scales as \cite{Giombi:2014yra} 
\begin{equation}\label{ecbdry}
F_{O(N)\,\text{sing}}\left(\beta\right) = N\,\beta\,E_{sing} + \hat{F}_{O(N)\,\text{sing}}.
\end{equation}
On the bulk side we find a non-vanishing result at one-loop. The bulk answer therefore has the structure
\begin{equation}\label{ecbulk}
\G_{A/B/C,\,\text{min}} = {1\over g} S_{A/B/C,\,\text{min}} +\beta E_{c;(A/B/C)}^{\,\text{min}} +\hat{\cF}_{A/B/C,\,\text{min}}\left(\beta\right) +\ldots,
\end{equation}
where $S_{A/B/C,\,\text{min}}$ is the corresponding classical on-shell action. With $\hat{F}_{O(N)\,\text{sing}} = \hat{\cF}_{A/B/C,\,\text{min}}$, \eqref{ecbdry} and \eqref{ecbulk} are consistent provided
\begin{equation}
g^{-1} = N-1,\quad S_{A/B/C,\,\text{min}} = \beta\, E_{c;(0/\tfrac12/1)}.
\end{equation}
\subsection*{Vacuum Energies}
We now turn to the computation of one-loop vacuum energies in global AdS$_5$. The computations are a bit more invloved in practice but conceptually they are very similar to the Casimir energy computations presented above. The only difference is that instead of expanding the characters and corresponding partition functions, we shall first be computing the functions $f_{\cH|n}$ defined in \eqref{fhn} and Laurent series expanding those to extract the corresponding $\g_{\cH|n}$s defined in \eqref{fin G g}. As in the Casimir energy case, we have picked the computationally most convenient prescription to work in. We remind the reader that all three prescriptions are equivalent for higher-spin partition functions, and emphasize that \textit{a priori} this will not be true for bulk duals of matrix CFTs. We also mention that while we are focussed here on the cases of Higher-Spin/CFT dualities involving singletons carrying spins 0, $\tfrac12$ and 1, the results presented here are valid for more general spin $s$ \cite{Bae:2016xmv}. For the singleton cases, it is straightforward to compute the $f_{sing|n}$ and expand in $\beta$ to extract the coefficients
\be
\gamma_{(s)|2} = \frac{15\,s^4-1}{30}\,,\qquad 
\gamma_{(s)|1} = \frac{6\,s^4-3\,s^2+1}{18}\,, \qquad 
\gamma_{(s)|0} =\frac{s^4-s^2}{2} \,.
\label{gamma S j}
\ee
Finally, summing these three numbers, we obtain the vacuum energy as
\begin{equation}\label{vacjsingleton}
\Gamma^{\sst (1)\,\text{ren}}_{(s)} = \left(-1\right)^{2s}\left(1-\tfrac12\delta_{s,0}\right)\frac{60\,s^4-30\,s^2+1}{45}\,\log R\,,
\end{equation}
where $\left(-1\right)^{2s}$ arises from the possibly fermionic statistics of the singleton. The reader would recognize, for the $s=0,\,\tfrac12,\,1$ instances, the coefficient of the $\log R$ term as the conformal anomaly of the spin-$s$ singleton on $S^4$.

We next turn to the vacuum energy of the non-minimal theory, whose partition function is given by \eqref{U(N)}. Again we may evaluate the $f_{\cH|n}$ and expand in $\beta$ to obtain
\begin{equation}
\begin{split}
\gamma_{(s)|2}^{\text{non-min}} = \frac{2}{105}\,n_s \left(72\,s^4-24\,s^2+1\right),&\quad 
\gamma_{(s)|1}^{\text{non-min}} = {4\over 315}\,n_s\left(60\,s^4-27\,s^2+2\right),\\
\gamma_{(s)|0}^{\text{non-min}} &=\frac{8}{15}\,
n_s\,(s^4-s^2)\,,
\end{split}
\label{gamma type j}
\end{equation}
where $n_s= \tfrac{\left(2s-1\right)2s\left(2s+1\right)}{6}$ is an integer. The terms in \eqref{gamma type j} add to give the full one-loop vacuum energy. Using  \eqref{vacjsingleton}, we find that
\be
\Gamma^{\sst (1)\,\text{ren}}_{A/B/C,\text{non-min}} 
= 2\,n_{0/\frac12/1}\,\Gamma^{\sst (1)\,\text{ren}}_{(0/\frac12/1)}\,.
\label{1lvacjnonmin}
\ee
We thus reproduce the results that the one-loop vacuum energy vanishes for type A and B theories \cite{Giombi:2014iua} and equals twice that of the singleton for the type C theory \cite{Beccaria:2014zma}.

Having thus reproduced the results for the non-minimal dualities, we now turn to the minimal ones, where the thermal partition function is given by \eqref{O(N)}. The contribution to the one-loop vacuum energy of the first term there has already been evaluated in the minimal case, but for the overall factor of $\tfrac12$. It turns out that the second term contributes $\Gamma^{\sst (1)\,\text{ren}}_{(s)}$ to the vacuum energy \cite{Bae:2016rgm,Bae:2016hfy,Bae:2016xmv}. We finally obtain
\begin{equation}
\Gamma^{\sst (1)\,\text{ren}}_{A/B/C,\text{min}} ={1\over 2}\,\Gamma^{\sst (1)\,\text{ren}}_{s,\text{non-min}}+\Gamma^{\sst (1)\,\text{ren}}_{(s)}=
\left[\left(-1\right)^{2s}\,n_s+1\right]
\Gamma^{\sst (1)\,\text{ren}}_{(s)}\,.
\label{min VE}
\end{equation}
This reproduces using CIRZ, the results of \cite{Giombi:2014iua} for the Type A, B dualities and of \cite{Beccaria:2014zma} for the Type C duality.

We conclude this section with a discussion of the proposal of \cite{Giombi:2013fka,Giombi:2014iua} for interpreting the non-zero one-loop answer derived above for the bulk higher-spin theory. For definiteness, we will work with a spin-$0$ field, but the discussion readily generalizes, and we will state the final result for all cases discussed above. Firstly, the logarithmically divergent part of the free energy for a theory with $N$ scalar fields defined on $S^4$ is \cite{Duff:1977ay,Christensen:1978md}
\begin{equation}\label{fcft}
    F_{CFT_4} = {N\over 90}\log \Lambda_{CFT} + \mathcal{O}\left(\Lambda_{CFT}^{0}\right).
\end{equation}
Next, the computation of the one-loop vacuum energy above indicates that the bulk free energy has the expansion
\begin{equation}\label{fads}
    F_{AdS_5} = \left({\cL_0\over g} +{1\over 90}\right)\log R + \mathcal{O}\left(g\right),
\end{equation}
where $\cL_0\,\log R$ is the on-shell action of the AdS$_5$ theory. Then equating \eqref{fcft} with \eqref{fads} while using $\log \Lambda_{CFT} \sim \log R$ and identifying $\cL_0 = \tfrac1{90}$ yields \cite{Giombi:2014iua}
\begin{equation}
 g^{-1} = N-1.
\end{equation}
This discussion generalizes for the Type B and Type C cases also. In summary, for the non-minimal case, we find \cite{Giombi:2014iua,Beccaria:2014zma}
\begin{equation}
 \text{Type A, B:}\quad g^{-1} =N, \qquad \text{Type C:} \quad g^{-1} = N-1,
\end{equation}
while for the minimal case we find \cite{Giombi:2014iua,Beccaria:2014zma}
\begin{equation}
 \text{Type A, B:}\quad g^{-1} =N-1, \qquad \text{Type C:} \quad g^{-1} = N-2.
\end{equation}
The corresponding expressions for spin-$s$ singletons is available in \cite{Bae:2016xmv}.

\bibliographystyle{JHEP}
\bibliography{matrix}

\providecommand{\href}[2]{#2}\begingroup\raggedright\begin{thebibliography}{10}

\bibitem{Maldacena:1997re}
J.~M. Maldacena, \emph{{The Large N limit of superconformal field theories and
  supergravity}}, \href{http://dx.doi.org/10.1023/A:1026654312961}{\emph{Int.
  J. Theor. Phys.} {\bf 38} (1999) 1113--1133},
  [\href{http://arxiv.org/abs/hep-th/9711200}{{\tt hep-th/9711200}}].

\bibitem{Aharony:1999ti}
O.~Aharony, S.~S. Gubser, J.~M. Maldacena, H.~Ooguri and Y.~Oz, \emph{{Large N
  field theories, string theory and gravity}},
  \href{http://dx.doi.org/10.1016/S0370-1573(99)00083-6}{\emph{Phys. Rept.}
  {\bf 323} (2000) 183--386}, [\href{http://arxiv.org/abs/hep-th/9905111}{{\tt
  hep-th/9905111}}].

\bibitem{Polchinski:1995mt}
J.~Polchinski, \emph{{Dirichlet Branes and Ramond-Ramond charges}},
  \href{http://dx.doi.org/10.1103/PhysRevLett.75.4724}{\emph{Phys. Rev. Lett.}
  {\bf 75} (1995) 4724--4727}, [\href{http://arxiv.org/abs/hep-th/9510017}{{\tt
  hep-th/9510017}}].

\bibitem{Klebanov:2002ja}
I.~R. Klebanov and A.~M. Polyakov, \emph{{AdS dual of the critical O(N) vector
  model}}, \href{http://dx.doi.org/10.1016/S0370-2693(02)02980-5}{\emph{Phys.
  Lett.} {\bf B550} (2002) 213--219},
  [\href{http://arxiv.org/abs/hep-th/0210114}{{\tt hep-th/0210114}}].

\bibitem{Gaberdiel:2010pz}
M.~R. Gaberdiel and R.~Gopakumar, \emph{{An AdS$_3$ Dual for Minimal Model
  CFTs}}, \href{http://dx.doi.org/10.1103/PhysRevD.83.066007}{\emph{Phys. Rev.}
  {\bf D83} (2011) 066007}, [\href{http://arxiv.org/abs/1011.2986}{{\tt
  1011.2986}}].

\bibitem{Shenker:2011zf}
S.~H. Shenker and X.~Yin, \emph{{Vector Models in the Singlet Sector at Finite
  Temperature}},  \href{http://arxiv.org/abs/1109.3519}{{\tt 1109.3519}}.

\bibitem{Sundborg:1999ue}
B.~Sundborg, \emph{{The Hagedorn transition, deconfinement and N=4 SYM
  theory}}, \href{http://dx.doi.org/10.1016/S0550-3213(00)00044-4}{\emph{Nucl.
  Phys.} {\bf B573} (2000) 349--363},
  [\href{http://arxiv.org/abs/hep-th/9908001}{{\tt hep-th/9908001}}].

\bibitem{wittenHS}
E.~Witten, \emph{{Spacetime reconstruction}}, {\emph{Conf. in Honor of John
  Schwarz’s 60th Birthday, California Institute of Technology, Pasadena, CA,
  USA, \url{http://theory.caltech.edu/jhs60/witten/1.html}} (Nov. 3-4. 2001) }.

\bibitem{Gross:1988ue}
D.~J. Gross, \emph{{High-Energy Symmetries of String Theory}},
  \href{http://dx.doi.org/10.1103/PhysRevLett.60.1229}{\emph{Phys. Rev. Lett.}
  {\bf 60} (1988) 1229}.

\bibitem{Gross:1988ib}
D.~J. Gross and V.~Periwal, \emph{{String Perturbation Theory Diverges}},
  \href{http://dx.doi.org/10.1103/PhysRevLett.60.2105}{\emph{Phys. Rev. Lett.}
  {\bf 60} (1988) 2105}.

\bibitem{Beisert:2003te}
N.~Beisert, M.~Bianchi, J.~F. Morales and H.~Samtleben, \emph{{On the spectrum
  of AdS / CFT beyond supergravity}},
  \href{http://dx.doi.org/10.1088/1126-6708/2004/02/001}{\emph{JHEP} {\bf 02}
  (2004) 001}, [\href{http://arxiv.org/abs/hep-th/0310292}{{\tt
  hep-th/0310292}}].

\bibitem{Bianchi:2003wx}
M.~Bianchi, J.~F. Morales and H.~Samtleben, \emph{{On stringy AdS(5) x S**5 and
  higher spin holography}},
  \href{http://dx.doi.org/10.1088/1126-6708/2003/07/062}{\emph{JHEP} {\bf 07}
  (2003) 062}, [\href{http://arxiv.org/abs/hep-th/0305052}{{\tt
  hep-th/0305052}}].

\bibitem{Beisert:2004di}
N.~Beisert, M.~Bianchi, J.~F. Morales and H.~Samtleben, \emph{{Higher spin
  symmetry and N=4 SYM}},
  \href{http://dx.doi.org/10.1088/1126-6708/2004/07/058}{\emph{JHEP} {\bf 07}
  (2004) 058}, [\href{http://arxiv.org/abs/hep-th/0405057}{{\tt
  hep-th/0405057}}].

\bibitem{Bianchi:2005ze}
M.~Bianchi, P.~J. Heslop and F.~Riccioni, \emph{{More on La Grande Bouffe}},
  \href{http://dx.doi.org/10.1088/1126-6708/2005/08/088}{\emph{JHEP} {\bf 08}
  (2005) 088}, [\href{http://arxiv.org/abs/hep-th/0504156}{{\tt
  hep-th/0504156}}].

\bibitem{Joung:2012fv}
E.~Joung, L.~Lopez and M.~Taronna, \emph{{Solving the Noether procedure for
  cubic interactions of higher spins in (A)dS}},
  \href{http://dx.doi.org/10.1088/1751-8113/46/21/214020}{\emph{J. Phys.} {\bf
  A46} (2013) 214020}, [\href{http://arxiv.org/abs/1207.5520}{{\tt
  1207.5520}}].

\bibitem{Joung:2012rv}
E.~Joung, L.~Lopez and M.~Taronna, \emph{{On the cubic interactions of massive
  and partially-massless higher spins in (A)dS}},
  \href{http://dx.doi.org/10.1007/JHEP07(2012)041}{\emph{JHEP} {\bf 07} (2012)
  041}, [\href{http://arxiv.org/abs/1203.6578}{{\tt 1203.6578}}].

\bibitem{Joung:2012hz}
E.~Joung, L.~Lopez and M.~Taronna, \emph{{Generating functions of
  (partially-)massless higher-spin cubic interactions}},
  \href{http://dx.doi.org/10.1007/JHEP01(2013)168}{\emph{JHEP} {\bf 01} (2013)
  168}, [\href{http://arxiv.org/abs/1211.5912}{{\tt 1211.5912}}].

\bibitem{Camporesi:1990wm}
R.~Camporesi, \emph{{Harmonic analysis and propagators on homogeneous spaces}},
  \href{http://dx.doi.org/10.1016/0370-1573(90)90120-Q}{\emph{Phys. Rept.} {\bf
  196} (1990) 1--134}.

\bibitem{Camporesi:1993mz}
R.~Camporesi and A.~Higuchi, \emph{{Arbitrary spin effective potentials in
  anti-de Sitter space-time}},
  \href{http://dx.doi.org/10.1103/PhysRevD.47.3339}{\emph{Phys. Rev.} {\bf D47}
  (1993) 3339--3344}.

\bibitem{Camporesi}
R.~Camporesi and A.~Higuchi, \emph{The plancherel measure for p-forms in real
  hyperbolic spaces}, {\emph{Journal of Geometry and Physics} {\bf 15} (1994)
  57--94}.

\bibitem{Camporesi:1994ga}
R.~Camporesi and A.~Higuchi, \emph{{Spectral functions and zeta functions in
  hyperbolic spaces}}, \href{http://dx.doi.org/10.1063/1.530850}{\emph{J. Math.
  Phys.} {\bf 35} (1994) 4217--4246}.

\bibitem{Giombi:2012ms}
S.~Giombi and X.~Yin, \emph{{The Higher Spin/Vector Model Duality}},
  \href{http://dx.doi.org/10.1088/1751-8113/46/21/214003}{\emph{J. Phys.} {\bf
  A46} (2013) 214003}, [\href{http://arxiv.org/abs/1208.4036}{{\tt
  1208.4036}}].

\bibitem{Giombi:2013fka}
S.~Giombi and I.~R. Klebanov, \emph{{One Loop Tests of Higher Spin AdS/CFT}},
  \href{http://dx.doi.org/10.1007/JHEP12(2013)068}{\emph{JHEP} {\bf 12} (2013)
  068}, [\href{http://arxiv.org/abs/1308.2337}{{\tt 1308.2337}}].

\bibitem{Giombi:2014iua}
S.~Giombi, I.~R. Klebanov and B.~R. Safdi, \emph{{Higher Spin
  AdS$_{d+1}$/CFT$_d$ at One Loop}},
  \href{http://dx.doi.org/10.1103/PhysRevD.89.084004}{\emph{Phys. Rev.} {\bf
  D89} (2014) 084004}, [\href{http://arxiv.org/abs/1401.0825}{{\tt
  1401.0825}}].

\bibitem{Giombi:2014yra}
S.~Giombi, I.~R. Klebanov and A.~A. Tseytlin, \emph{{Partition Functions and
  Casimir Energies in Higher Spin AdS$_{d+1}$/CFT$_d$}},
  \href{http://dx.doi.org/10.1103/PhysRevD.90.024048}{\emph{Phys. Rev.} {\bf
  D90} (2014) 024048}, [\href{http://arxiv.org/abs/1402.5396}{{\tt
  1402.5396}}].

\bibitem{Beccaria:2014jxa}
M.~Beccaria, X.~Bekaert and A.~A. Tseytlin, \emph{{Partition function of free
  conformal higher spin theory}},
  \href{http://dx.doi.org/10.1007/JHEP08(2014)113}{\emph{JHEP} {\bf 08} (2014)
  113}, [\href{http://arxiv.org/abs/1406.3542}{{\tt 1406.3542}}].

\bibitem{Beccaria:2014xda}
M.~Beccaria and A.~A. Tseytlin, \emph{{Higher spins in AdS$_{5}$ at one loop:
  vacuum energy, boundary conformal anomalies and AdS/CFT}},
  \href{http://dx.doi.org/10.1007/JHEP11(2014)114}{\emph{JHEP} {\bf 11} (2014)
  114}, [\href{http://arxiv.org/abs/1410.3273}{{\tt 1410.3273}}].

\bibitem{Beccaria:2014zma}
M.~Beccaria and A.~A. Tseytlin, \emph{{Vectorial AdS$_5$/CFT$_4$ duality for
  spin-one boundary theory}},
  \href{http://dx.doi.org/10.1088/1751-8113/47/49/492001}{\emph{J. Phys.} {\bf
  A47} (2014) 492001}, [\href{http://arxiv.org/abs/1410.4457}{{\tt
  1410.4457}}].

\bibitem{Beccaria:2014qea}
M.~Beccaria, G.~Macorini and A.~A. Tseytlin, \emph{{Supergravity one-loop
  corrections on AdS$_7$ and AdS$_3$, higher spins and AdS/CFT}},
  \href{http://dx.doi.org/10.1016/j.nuclphysb.2015.01.014}{\emph{Nucl. Phys.}
  {\bf B892} (2015) 211--238}, [\href{http://arxiv.org/abs/1412.0489}{{\tt
  1412.0489}}].

\bibitem{Giombi:2016pvg}
S.~Giombi, I.~R. Klebanov and Z.~M. Tan, \emph{{The ABC of Higher-Spin
  AdS/CFT}},  \href{http://arxiv.org/abs/1608.07611}{{\tt 1608.07611}}.

\bibitem{Gunaydin:2016amv}
M.~Gunaydin, E.~D. Skvortsov and T.~Tran, \emph{{Exceptional F(4) Higher-Spin
  Theory in AdS(6) at One-Loop and other Tests of Duality}},
  \href{http://arxiv.org/abs/1608.07582}{{\tt 1608.07582}}.

\bibitem{Pang:2016ofv}
Y.~Pang, E.~Sezgin and Y.~Zhu, \emph{{One Loop Tests of Supersymmetric Higher
  Spin $AdS_4/CFT_3$}},  \href{http://arxiv.org/abs/1608.07298}{{\tt
  1608.07298}}.

\bibitem{Skvortsov:2017ldz}
E.~D. Skvortsov and T.~Tran, \emph{{AdS/CFT in Fractional Dimension and Higher
  Spin Gravity at One Loop}},  \href{http://arxiv.org/abs/1707.00758}{{\tt
  1707.00758}}.

\bibitem{Giombi:2016ejx}
S.~Giombi, \emph{{TASI Lectures on the Higher Spin - CFT duality}},
  \href{http://arxiv.org/abs/1607.02967}{{\tt 1607.02967}}.

\bibitem{Barabanschikov:2005ri}
A.~Barabanschikov, L.~Grant, L.~L. Huang and S.~Raju, \emph{{The Spectrum of
  Yang Mills on a sphere}},
  \href{http://dx.doi.org/10.1088/1126-6708/2006/01/160}{\emph{JHEP} {\bf 01}
  (2006) 160}, [\href{http://arxiv.org/abs/hep-th/0501063}{{\tt
  hep-th/0501063}}].

\bibitem{Newton:2008au}
T.~H. Newton and M.~Spradlin, \emph{{Quite a Character: The Spectrum of
  Yang-Mills on S3}},
  \href{http://dx.doi.org/10.1016/j.physletb.2009.01.044}{\emph{Phys. Lett.}
  {\bf B672} (2009) 382--385}, [\href{http://arxiv.org/abs/0812.4693}{{\tt
  0812.4693}}].

\bibitem{Bae:2016rgm}
J.-B. Bae, E.~Joung and S.~Lal, \emph{{One-loop test of free SU(N ) adjoint
  model holography}},
  \href{http://dx.doi.org/10.1007/JHEP04(2016)061}{\emph{JHEP} {\bf 04} (2016)
  061}, [\href{http://arxiv.org/abs/1603.05387}{{\tt 1603.05387}}].

\bibitem{Bae:2016hfy}
J.-B. Bae, E.~Joung and S.~Lal, \emph{{On the Holography of Free Yang-Mills}},
  \href{http://dx.doi.org/10.1007/JHEP10(2016)074}{\emph{JHEP} {\bf 10} (2016)
  074}, [\href{http://arxiv.org/abs/1607.07651}{{\tt 1607.07651}}].

\bibitem{Bae:2016xmv}
J.-B. Bae, E.~Joung and S.~Lal, \emph{{A Note on Vectorial AdS$_5$/CFT$_4$
  Duality for Spin-$j$ Boundary Theory}},
  \href{http://dx.doi.org/10.1007/JHEP12(2016)077}{\emph{JHEP} {\bf 12} (2016)
  077}, [\href{http://arxiv.org/abs/1611.00112}{{\tt 1611.00112}}].

\bibitem{Bae:2017spv}
J.-B. Bae, E.~Joung and S.~Lal, \emph{{One-Loop Free Energy of Tensionless Type
  IIB String in AdS$_5\times$S$^5$}},
  \href{http://arxiv.org/abs/1701.01507}{{\tt 1701.01507}}.

\bibitem{Gibbons:2006ij}
G.~W. Gibbons, M.~J. Perry and C.~N. Pope, \emph{{Partition functions, the
  Bekenstein bound and temperature inversion in anti-de Sitter space and its
  conformal boundary}},
  \href{http://dx.doi.org/10.1103/PhysRevD.74.084009}{\emph{Phys. Rev.} {\bf
  D74} (2006) 084009}, [\href{http://arxiv.org/abs/hep-th/0606186}{{\tt
  hep-th/0606186}}].

\bibitem{Camporesi:1995fb}
R.~Camporesi and A.~Higuchi, \emph{{On the Eigen functions of the Dirac
  operator on spheres and real hyperbolic spaces}},
  \href{http://dx.doi.org/10.1016/0393-0440(95)00042-9}{\emph{J. Geom. Phys.}
  {\bf 20} (1996) 1--18}, [\href{http://arxiv.org/abs/gr-qc/9505009}{{\tt
  gr-qc/9505009}}].

\bibitem{Diaz:2007an}
D.~E. Diaz and H.~Dorn, \emph{{Partition functions and double-trace
  deformations in AdS/CFT}},
  \href{http://dx.doi.org/10.1088/1126-6708/2007/05/046}{\emph{JHEP} {\bf 05}
  (2007) 046}, [\href{http://arxiv.org/abs/hep-th/0702163}{{\tt
  hep-th/0702163}}].

\bibitem{Bianchi:2006ti}
M.~Bianchi, F.~A. Dolan, P.~J. Heslop and H.~Osborn, \emph{{N=4 superconformal
  characters and partition functions}},
  \href{http://dx.doi.org/10.1016/j.nuclphysb.2006.12.005}{\emph{Nucl. Phys.}
  {\bf B767} (2007) 163--226}, [\href{http://arxiv.org/abs/hep-th/0609179}{{\tt
  hep-th/0609179}}].

\bibitem{Polyakov:2001af}
A.~M. Polyakov, \emph{{Gauge fields and space-time}},
  \href{http://dx.doi.org/10.1142/S0217751X02013071}{\emph{Int. J. Mod. Phys.}
  {\bf A17S1} (2002) 119--136},
  [\href{http://arxiv.org/abs/hep-th/0110196}{{\tt hep-th/0110196}}].

\bibitem{Aharony:2003sx}
O.~Aharony, J.~Marsano, S.~Minwalla, K.~Papadodimas and M.~Van~Raamsdonk,
  \emph{{The Hagedorn - deconfinement phase transition in weakly coupled large
  N gauge theories}},
  \href{http://dx.doi.org/10.4310/ATMP.2004.v8.n4.a1}{\emph{Adv. Theor. Math.
  Phys.} {\bf 8} (2004) 603--696},
  [\href{http://arxiv.org/abs/hep-th/0310285}{{\tt hep-th/0310285}}].

\bibitem{Nishioka:2008gz}
T.~Nishioka and T.~Takayanagi, \emph{{On Type IIA Penrose Limit and N=6
  Chern-Simons Theories}},
  \href{http://dx.doi.org/10.1088/1126-6708/2008/08/001}{\emph{JHEP} {\bf 08}
  (2008) 001}, [\href{http://arxiv.org/abs/0806.3391}{{\tt 0806.3391}}].

\bibitem{Jaroszewicz:1991dd}
T.~Jaroszewicz and P.~S. Kurzepa, \emph{{Polyakov spin factors and Laplacians
  on homogeneous spaces}},
  \href{http://dx.doi.org/10.1016/0003-4916(92)90286-U}{\emph{Annals Phys.}
  {\bf 213} (1992) 135--165}.

\bibitem{Gopakumar:2011qs}
R.~Gopakumar, R.~K. Gupta and S.~Lal, \emph{{The Heat Kernel on $AdS$}},
  \href{http://dx.doi.org/10.1007/JHEP11(2011)010}{\emph{JHEP} {\bf 11} (2011)
  010}, [\href{http://arxiv.org/abs/1103.3627}{{\tt 1103.3627}}].

\bibitem{Lal:2012ax}
S.~Lal, \emph{{CFT(4) Partition Functions and the Heat Kernel on AdS(5)}},
  \href{http://dx.doi.org/10.1016/j.physletb.2013.10.043}{\emph{Phys. Lett.}
  {\bf B727} (2013) 325--329}, [\href{http://arxiv.org/abs/1212.1050}{{\tt
  1212.1050}}].

\bibitem{mack1977all}
G.~Mack, \emph{{All Unitary Ray Representations of the Conformal Group SU(2,2)
  with Positive Energy}},
  \href{http://dx.doi.org/10.1007/BF01613145}{\emph{Commun. Math. Phys.} {\bf
  55} (1977) 1}.

\bibitem{Dolan:2005wy}
F.~A. Dolan, \emph{{Character formulae and partition functions in higher
  dimensional conformal field theory}},
  \href{http://dx.doi.org/10.1063/1.2196241}{\emph{J. Math. Phys.} {\bf 47}
  (2006) 062303}, [\href{http://arxiv.org/abs/hep-th/0508031}{{\tt
  hep-th/0508031}}].

\bibitem{Minwalla:1997ka}
S.~Minwalla, \emph{{Restrictions imposed by superconformal invariance on
  quantum field theories}}, {\emph{Adv. Theor. Math. Phys.} {\bf 2} (1998)
  781--846}, [\href{http://arxiv.org/abs/hep-th/9712074}{{\tt
  hep-th/9712074}}].

\bibitem{Metsaev:1995re}
R.~R. Metsaev, \emph{{Massless mixed symmetry bosonic free fields in
  d-dimensional anti-de Sitter space-time}},
  \href{http://dx.doi.org/10.1016/0370-2693(95)00563-Z}{\emph{Phys. Lett.} {\bf
  B354} (1995) 78--84}.

\bibitem{Metsaev:2004ee}
R.~R. Metsaev, \emph{{Mixed symmetry massive fields in AdS(5)}},
  \href{http://dx.doi.org/10.1088/0264-9381/22/13/016}{\emph{Class. Quant.
  Grav.} {\bf 22} (2005) 2777--2796},
  [\href{http://arxiv.org/abs/hep-th/0412311}{{\tt hep-th/0412311}}].

\bibitem{Metsaev:2014sfa}
R.~R. Metsaev, \emph{{Mixed-symmetry fields in AdS(5), conformal fields, and
  AdS/CFT}}, \href{http://dx.doi.org/10.1007/JHEP01(2015)077}{\emph{JHEP} {\bf
  01} (2015) 077}, [\href{http://arxiv.org/abs/1410.7314}{{\tt 1410.7314}}].

\bibitem{Duff:1977ay}
M.~J. Duff, \emph{{Observations on Conformal Anomalies}},
  \href{http://dx.doi.org/10.1016/0550-3213(77)90410-2}{\emph{Nucl. Phys.} {\bf
  B125} (1977) 334}.

\bibitem{Christensen:1978md}
S.~M. Christensen and M.~J. Duff, \emph{{New Gravitational Index Theorems and
  Supertheorems}},
  \href{http://dx.doi.org/10.1016/0550-3213(79)90516-9}{\emph{Nucl. Phys.} {\bf
  B154} (1979) 301}.

\end{thebibliography}\endgroup
\end{document}